%% file: Main.tex
\documentclass[final,5p,times,twocolumn]{elsarticle}

\newcounter{insight}[section]
\newenvironment{insight}[1]
{\refstepcounter{insight}
 \begin{tcolorbox}\textbf{Key insights \theinsight: \textit{(#1)}}%
}%
{
  \end{tcolorbox}\ignorespacesafterend
} 

\usepackage{amssymb}
\usepackage[hidelinks]{hyperref}
\usepackage{booktabs}
\usepackage{multirow}
\usepackage[table,xcdraw]{xcolor}
\usepackage[normalem]{ulem}
\useunder{\uline}{\ul}{}
\usepackage{longtable}
\usepackage{geometry}
\usepackage{array}
\usepackage{lipsum}
\usepackage{tcolorbox}
\usepackage{fontawesome5}

\newcolumntype{C}[1]{>{\centering\arraybackslash}p{#1}}
\newcolumntype{L}[1]{>{\raggedright\arraybackslash}p{#1}}

\newcommand{\newtext}[1]{{#1}}
\newcommand{\oldtext}[1]{}

\journal{Journal of Systems and Software}

\begin{document}

\begin{frontmatter}

\title{Why do women pursue a PhD in Computer Science?}

\author[RWTH]{Erika Ábrahám}
\author[NOVA]{Miguel Goulão}
\author[MATF]{Milena Vujo\v{s}evi\'c Jani\v{c}i\'c}
\author[TUDublin]{Sarah Jane Delany}
\author[IUSA]{Amal Mersni}
\author[NURE]{Oleksandra Yeremenko}
\author[IUSA]{Ozge Buyukdagli}
\author[UCA]{Karima Boudaoud}
\author[BAMBERG]{Caroline Oehlhorn}
\author[BAMBERG]{Ute Schmid}
\author[RWTH]{Christina Büsing}
\author[RWTH]{Helen Bolke-Hermanns}
\author[RWTH]{Kaja Köhnle}
\author[ISEL]{Matilde Pato}
\author[ADYU]{Deniz Sunar Cerci}
\author[KIT]{Larissa Schmid}

\affiliation[RWTH]{organization={Department of Computer Science, RWTH Aachen University},
            city={Aachen},
            country={Germany}}

\affiliation[NOVA]{organization={Department of Computer Science, NOVA School of Science and Technology, NOVA LINCS}, 
            city={Lisboa},
            country={Portugal}}

\affiliation[MATF]{organization={Department of Computer Science, Faculty of Mathematics, University of Belgrade}, 
            city={Belgrade},
            country={Serbia}}

\affiliation[TUDublin]{organization={School of Computer Science, Technological University Dublin},
            city={Dublin},
            country={Ireland}}

\affiliation[IUSA]{organization={Department of Engineering, Faculty of Engineering and Natural Sciences, International University of Sarajevo}, 
            city={Sarajevo},
            country={Bosnia and Herzegovina}}

\affiliation[NURE]{organization={V.V. Popovskyy Department of Infocommunication Engineering, Faculty of Infocommunications, Kharkiv National University of Radio Electronics}, 
            city={Kharkiv},
            country={Ukraine}}            
 \affiliation[UCA]{organization={Université Côte d'Azur, Laboratoire I3S - CNRS},
            city={Sophia Antipolis},
            country={France}}  

\affiliation[BAMBERG]{organization={Faculty Information Systems and Applied Computer Sciences, University of Bamberg},
            city={Bamberg},
            country={Germany}}

\affiliation[ISEL]{organization={Department of Electronical Engineering, Telecommunications and Computers, ISEL, Lisbon School of Engineering, Instituto Politécnico de Lisboa}, 
            city={Lisboa},
            country={Portugal}}

\affiliation[ADYU]{organization={Department of Physics, Faculty of Arts and Sciences},
            city={Adiyaman},
            country={Turkiye}}

\affiliation[KIT]{organization={Department of Computer Science, Karlsruhe Institute of Technology},
            city={Karlsruhe},
            country={Germany}}

\begin{abstract}
\paragraph{Context}

Computer science, even now, attracts a small number of women, and the proportion of women in the field decreases through advancing career stages. Consequently, few women progress to PhD studies in computer science after completing master's studies. Empowering women at this stage in their careers is essential, not just for equality reasons, but to unlock untapped potential for society, industry and academia. 

\paragraph{Objective}
This paper aims to identify students’ career assumptions and information related to PhD studies \newtext{focused}  \oldtext{with a focus} on gender-based differences. We propose a program to inform female master students about PhD studies that explains the process, clarifies misconceptions, and alleviates concerns.

\paragraph{Method}
An extensive survey was conducted to identify factors that encourage and discourage students from undertaking PhD studies. The analysis identified statistically significant differences between those who undertook PhD studies and those who didn't, as well as statistically significant gender differences. A catalogue of questions to initiate discussions with potential PhD students which allowed them to explore these factors was developed. These were structured into a \textit{Women's Career Lunch} program where students can explore and discuss the benefits of PhD study. 

\paragraph{Results}
Encouraging factors toward PhD study include interest and confidence in research arising from a research involvement during earlier studies; enthusiasm for and self-confidence in computer science in addition to an interest in an academic career; encouragement from external sources; and a positive perception towards PhD studies which can involve achieving personal goals. 
Discouraging factors include uncertainty and lack of knowledge of the PhD process, a perception of lower job flexibility, and the requirement for long-term commitment.
Gender differences highlighted that female students who pursue a PhD have less confidence in their technical skills than males but a higher preference for interdisciplinary areas. 
Female students are less inclined than males to perceive the industry as offering better job opportunities and more flexible career paths than academia.  

\paragraph{Conclusions}
The insights collected from the survey facilitated the development of a \newtext{questions catalogue} \oldtext{catalogue of questions that were} structured into the \textit{Women Career Lunch} program to help students make a more informed decision concerning whether they should pursue a PhD in computer science. Localised versions of this program, in 8 languages, were created to support its adoption in different countries and assist in mitigating the female under-representation challenge.
\end{abstract}

\begin{highlights}
\item A comprehensive survey with more than 500 participants on the factors supporting, or deterring students from pursuing a PhD in Computer Science and how these factors vary by gender.
\item A characterisation of students' career assumptions and information needs concerning PhD studies in Computer Science.
\item A catalogue of questions aimed at sparking constructive discussion on the option to pursue a PhD in Computer Science.
\item A program targeted at female students to help them make more informed decisions concerning PhD studies in Computer Science.
\end{highlights}


\end{frontmatter}

\textit{In loving memory of Bara Buhnova and Ivica Crnkovic.}

\input{01-Introduction}

\input{02-Background}

\input{03-RelatedWork}

\input{04-Survey}

\input{05-Discussion}

\input{06-CareerLunch}

\input{07-Conclusions}

\input{08-DataAvailability}

\paragraph{Acknowledgements}
This work was supported by COST Action CA19122 - EUGAIN (European Network for Gender Balance in Informatics).
The authors also acknowledge the financial support by the Ministry of Science, Technological Development, and Innovation of the Republic of Serbia, grant number 451-03-47/2023-01/ 200104, the support from UID/04516/NOVA Laboratory for Computer Science and Informatics (NOVA LINCS) with the financial support of FCT.IP, and by the pilot program Core Informatics at KIT (KiKIT) of the Helmholtz Association (HGF). 

\bibliographystyle{elsarticle-num} 
\bibliography{references}

\appendix
\include{AppendixModule5}

\end{document}

%% file: 01-Introduction.tex
\section{Introduction}
\label{sec:introduction}

Computer science attracts less women than men. In addition, the percentage of women decreases with \newtext{the advancement of their careers.} \oldtext{advancing career stage.}
This poses a severe loss of opportunities, as the significance of women in computer science cannot be overemphasised. Diversity catalyses innovation and progress in the technological sector. Women supply distinctive perspectives, experiences, and problem-solving methods to the industry, making it more diverse and all-encompassing.
Additionally, as technology continues to shape our world, women in computer science research play an essential role in addressing crucial societal issues and driving groundbreaking advancements.

\newtext{Across Europe, numerous activities have been launched to counteract this so-called ``leaky pipeline'' phenomenon.}
\oldtext{To counteract this so-called ``leaky pipeline'' phenomenon, numerous activities have been launched across Europe.}
For example, Informatics Europe created the Working Group \emph{Women in Informatics Research and Education (WIRE)} to support, connect, and structure these activities. One activity rooted in WIRE has been the launch of the annual WIRE Workshop, offering a platform for networking and exchange. 

The surprisingly large interest in the WIRE Workshop has shown the need for more \newtext{additional} support \oldtext{in this direction}. This gave the motivation to extend the WIRE Working Group to a larger European effort, resulting in the European COST Action \emph{European Network for Gender Balance in Informatics (EUGAIN)} (https://eugain.eu). The objectives of EUGAIN are, beyond those of WIRE, to create a European platform for networking and exchange and to provide guidelines and recommendations for colleagues working at the forefront of the efforts for gender balance in computer science.

Beyond bachelor and master studies, PhDs in computer science cultivate a more robust, well-rounded, and equitable future for technology.
Embracing diversity and empowering women in computer science at the doctoral level is not only an issue of equality but also a way to unlock untapped potential and establish a brighter future for the industry and society.

However, with the transition from master's to PhD studies, we observe a weakening of the female presence. In Europe, an estimated 25.5\% of the master's students were female, while 24.1\% of PhD students were female in 2021/2022~\cite{InformaticsEuropeDataPortal}. This gap varies somewhat depending on the country and institution.  
For example, at RWTH Aachen University in the year 2022, 20\% of the master's students were female, but only 19\% of the PhD students. This gap was significantly wider at the Department for Computer Science, Faculty of Mathematics, University of Belgrade, where there were 41\% female students at the master's level but only 14\% at the PhD level. 
A key goal of groups like Informatics Europe, EUGAIN, and others concerned with equality and inclusion is to increase the number of female PhD students and facilitate them \oldtext{to complete} \newtext{in completing} their studies successfully.
To achieve this goal, we developed a \emph{program to inform female master's students about PhD studies}. 
This program has been constructed to be executable at interested computer science departments without intensive preparation, requiring little time and effort and a small financial budget, but still having a substantial impact. For maximal \oldtext{impact} \newtext{effect}, the program is also transferable to other STEM areas.

This paper reports on this work, explaining our starting objectives, methodology to achieve \oldtext{those objectives} \newtext{them}, and the developed program itself.

The goal of this program is not to convince female students to start a PhD but to \oldtext{inform them in a way that they can} \newtext{help them} make an informed decision that best fits their interests and abilities.
To achieve this goal, we needed to identify why fewer women than men choose PhD studies.
Once we had explored the reasons, we needed an appropriate way to convey relevant information to the female students.
We have to counteract wrong assumptions and uncertainties to equip the involved students with the knowledge \oldtext{that} they need to identify their \oldtext{individual} optimal career preferences.

Thus, we had the following objectives:
\begin{enumerate}
\item[O1] Identify which aspects students consider when they decide on their career path after their master's studies, with a focus on gender-based differences. Identify areas of missing knowledge, wrong assumptions, and uncertainties which might hinder master's students from starting a PhD for inappropriate reasons. 
\item[O2] Identify \oldtext{which information} \newtext{what} needs to be conveyed to female students to improve their decision-making.
\item[O3] Develop a program to convey \oldtext{the identified} \newtext{this} information to female students \newtext{in an engaging and practical way} \oldtext{engagingly and practically}.
\end{enumerate}

To achieve these goals, we used the following methodology:
\begin{itemize}
  \item[M1] To achieve objective O1, we set up an international survey, targeting primarily European countries but open to respondents worldwide. 
The survey was designed in such a way that the statistical evaluation of answers to the survey should facilitate answering the following research questions:
\begin{description}
    \item[RQ1:] What main supporting factors encourage enrolment in a PhD program?    
    \item[RQ2:] What main blocking factors discourage enrolment in a PhD program?
    \item[RQ3:] To what extent do these factors differ by gender?
\end{description}

\item[M2] To achieve objective O2, the answers to the research questions RQ1-RQ3 were used to design a catalogue of questions to prime a discussion with master students about PhD studies, where this discussion will facilitate conveying the relevant knowledge, strengthening supporting factors, and weakening blocking factors, particularly to female students.

\item[M3] To achieve objective O3, we designed a program named \emph{Women's Career (WoCa) Lunch}. The program invites female master's students whose grades are strong enough to consider a PhD to 8 gatherings in small groups. Each module lasts one hour during a common lunch, which gave the program its name. During lunch, the students can chat with (typically female) guests who have completed their PhDs in computer science and work in companies and universities. This format creates a safe space where prospective female PhD students can be informed about PhD studies, using the questions from the catalogue resulting from M2, where each meeting has a specific topical focus. 

\end{itemize}

The rest of the paper describes the above steps in detail. We start with some background in Section \ref{sec:background}, followed by a discussion on related work in Section \ref{sec:relatedWork}. Section \ref{sec:survey} describes how we created, distributed, and evaluated the survey. Section \ref{sec:Results} presents the survey results, which are then discussed in Section \ref{sec:discussion}. Section \ref{sec:catalogue} describes the question catalogue, which we derived from the survey's results. Section \ref{sec:careerlunch} describes the Women Career Lunch program and its pilot executions. We conclude the paper with some observations in Section \ref{sec:conclusions}.

%% file: 02-Background.tex
\section{Background}
\label{sec:background}

Historically, women have been notably underrepresented in STEM disciplines and continue to be outnumbered by men~\cite{whysofew}. 
Physics, engineering, and computer science exhibit the most pronounced disparities -- with only 20\% of bachelor's degrees earned by women in these fields in 2006~\cite{whysofew}.
The percentage of computer science bachelor's degrees earned by women in the US has even dropped since 1986~\cite{whysofew}, emphasising that progress towards gender parity should not be taken for granted. 
Additionally, there is evidence that even if women are majoring in computer science, only 38\% of them go on to work in computer-related jobs~\cite{STEMequity}. 
This results in women making up only 26\% of \oldtext{professionals in computing} \newtext{computing professionals} -- around the same percentage as almost 60 years ago~\cite{solvingequation}. 

This is in line with the metaphor of the \emph{leaky pipeline} in STEM and, more specifically, within the domain of computer science. It describes a pattern where women and underrepresented groups start their education and careers but face challenges at each step \newtext{leading} \oldtext{that lead} to fewer of them continuing. These challenges, like gender bias and workplace culture~\cite{corallo2022,Andre2022GenderMSc,rowe2021realities,wagner2020challenges}, cause a gradual drop in their participation~\cite{theorizingprogress}. While \newtext{many} \oldtext{a large number of} female students start in undergraduate courses worldwide, they do not continue to further computer science post-graduate studies~\cite{MiliBarkHend200627}. 
The leaky pipeline metaphor is a simple and illustrative way to discuss attrition. However, it is crucial to recognise its limitations regarding, e.g., the normative educational pathway it suggests and the exclusion of interdisciplinary fields~\cite{vitores2016trouble,soe2008s}. 

To quantify the number of female students at each stage, we \oldtext{show} \newtext{use} the data collected over the past decade by Informatics Europe,\oldtext{. 
The data is} available from their data portal~\cite{InformaticsEuropeDataPortal}.

\begin{figure}
    \centering
    \includegraphics[width=0.5\textwidth]{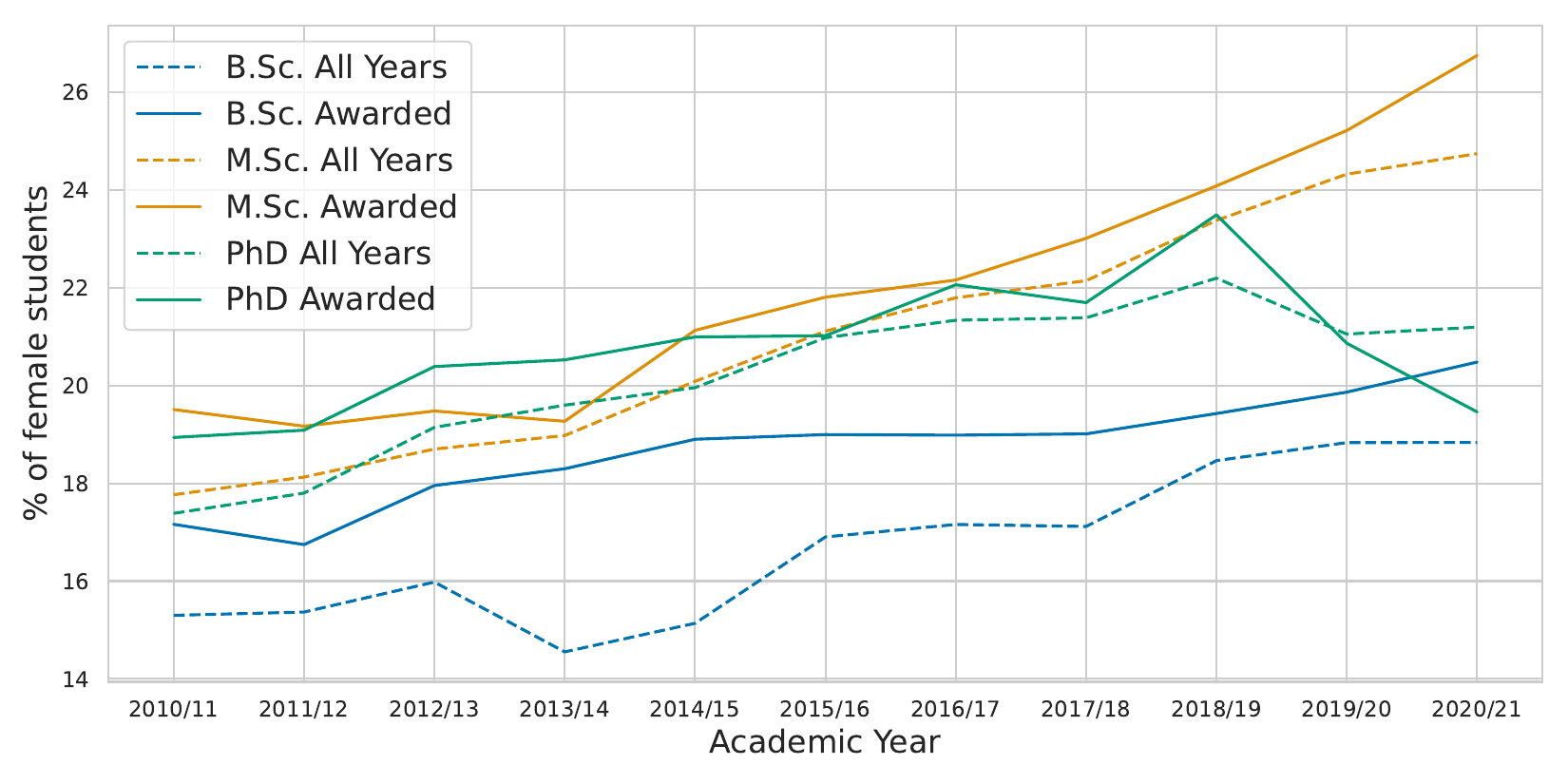}
    \caption{Percent of female computer science students within Europe~\cite{InformaticsEuropeDataPortal}}
    \label{fig:background:percent-female}
\end{figure}

\oldtext{This comprehensive data sheet encompasses information concerning the participation of women in computer science bachelor, master, and PhD programs across all European countries.} 
Figure~\ref{fig:background:percent-female} shows a snapshot of the \oldtext{landscape} \newtext{participation of women in computer science degree programs across all European countries} \oldtext{based on data} from 2010/11 to 2020/21. 
Firstly, the overall percentage of women at all stages of academic progression remains consistently low, with a maximum of just over a quarter. Despite the significant strides in promoting gender diversity in recent years, there is still a notable gender gap. \oldtext{in the field.}
More notably, the data illustrates a significant difference in the percentages of women enrolled in bachelor, master, and PhD programs versus those awarded the respective degrees. It is encouraging to observe that a larger proportion of women are awarded the degree compared to those who enrol in these programs, indicating that \oldtext{once women enter these programs} they tend to excel \newtext{once women enter these programs}.
However, the critical observation is the disparity in trends between master and PhD programs. While there has been a steady increase in \oldtext{the number of} female students enrolling in computer science master programs, the same does not hold for PhD programs. After reaching a peak in 2018/19, the number of female PhD graduates started to decline. 
Furthermore, the data reveals a stagnation in the percentage of female PhD students, especially from 2015/16 onward. This stagnation is alarming, considering that there is a growing pool of female master graduates who could potentially transition to PhD programs. In other words, there is an untapped resource of talented women who have completed their master degrees but are not proceeding to PhD studies. 

Statistics show that the stagnation in the percentage of female PhD students in computer science is not limited to Europe. For example, the Taulbee survey~\cite{taulbee2020} shows that 23 to 25\% of enrolled PhD students were women, while a little over 30\% of master degrees were awarded to women in the US. 
It is worth noting that while this stagnation is prevalent in Western countries~\cite{vitores2016trouble}, it is less commonplace in other places~\cite{Varma2010,lagesen2008-malaysia}. However, also in Europe, there are certain countries (e.g. Romania) where the average percentage of PhDs awarded to women since 2014/2015 is 43.3\%, with an average of 35.5\% of enrolled female students, following an average of 39.3\% MScs awarded and 35.7\% of enrolled female students, albeit with a relatively low ratio of PhDs awarded per million inhabitants~\cite{InformaticsEuropeDataPortal}.

In an era where technology plays an ever-increasing role in shaping our world, it is crucial to ensure diversity in the individuals who contribute to the development and advancement of computer science. 
A key milestone in the academic journey of computer science students is the transition from a masters degree to a PhD program. This transition not only opens doors to advanced research opportunities but also \newtext{creates} \oldtext{leads to the creation of} future educators, researchers, and industry leaders. 
However, little is known about the underlying reasons behind the limited transition of women from masters programs in computer science to enrolment in PhD programs. Assessing the reasons behind their choice holds both academic and practical importance. A comprehensive understanding of the multifaceted factors that influence women's decisions in this context not only contributes to the academic discourse but also informs evidence-based strategies aimed at promoting gender diversity in PhD programs. 

%% file: 03-RelatedWork.tex
\section{Related Work}
\label{sec:relatedWork}

Research into \newtext{the} challenges women face in computer science (CS) education \cite{yates2022female,beyer2014women,beyer2003gender} has investigated \oldtext{influential} \newtext{the} factors \newtext{influencing} \oldtext{on women's} \newtext{their} decisions \oldtext{to choose} \newtext{to pursue} this field \oldtext{to} \newtext{of} study \cite{pantic2020retention,pantic2019factors,genut2019factors}, as well as dropout reasons  \cite{beaubouef2005high,pappas2016investigating}. While these studies have offered strategies to bolster diversity, equity, and inclusivity (DEI) at the undergraduate level \cite{roberts2002encouraging,cohoon2003must,lagesen2007strength}, a distinct research gap remains concerning the transition from master to  PhD studies. 
\oldtext{The enrolment phase is a critical stage in the academic journey. Before delving into retention approaches, it is essential to understand the nuances of enrolment, as this can provide valuable insights and help address potential misconceptions or barriers from the outset.} \newtext{This transition phase is a critical stage in the academic journey, and understanding its nuances is essential before addressing retention approaches. Understanding the transition phase provides valuable insights and helps address potential misconceptions or barriers from the outset. } 

\begin{figure*} [htb!]
    \centering
    \includegraphics[width=\textwidth]{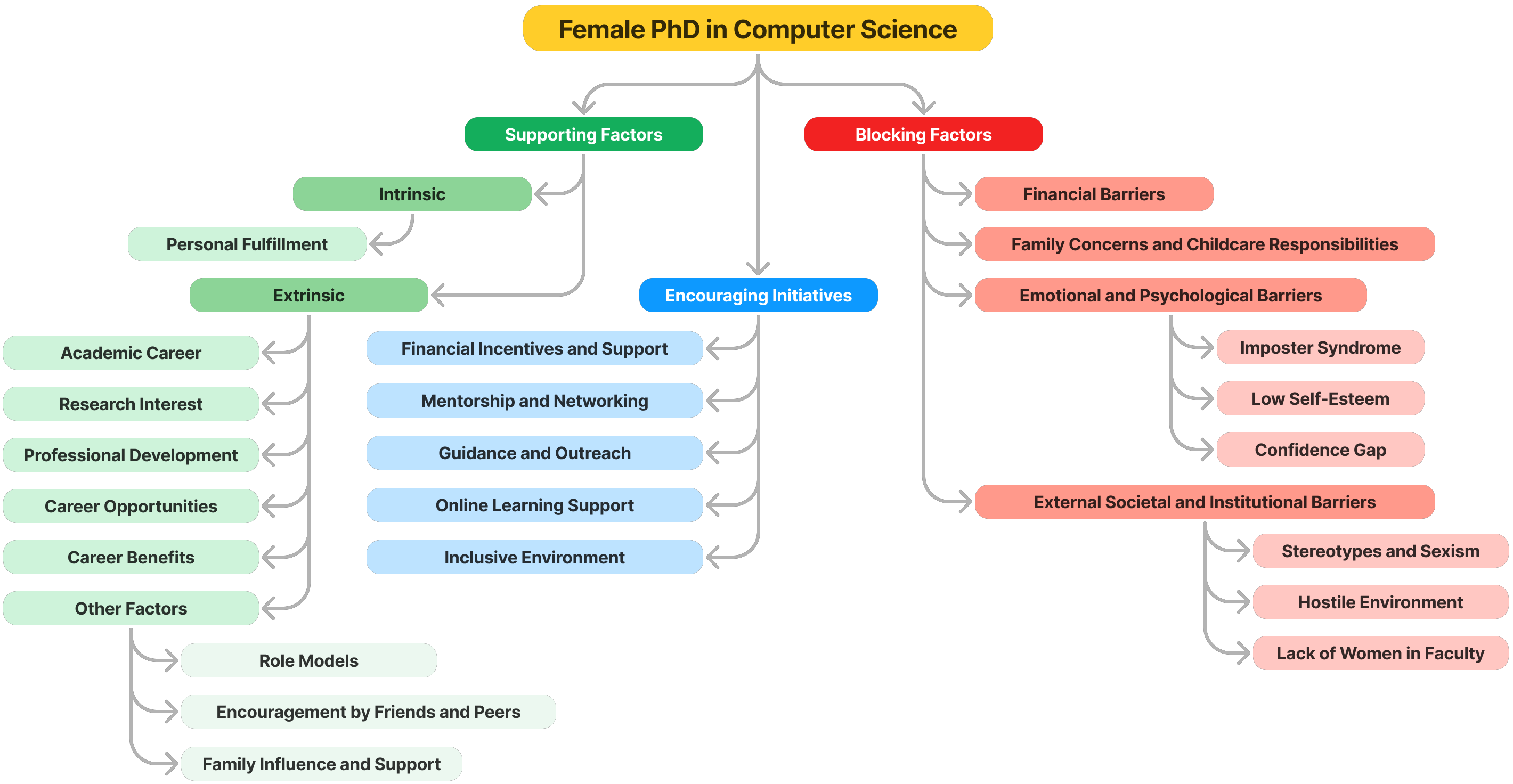}
   \caption{\ Primary factors and initiatives related to female enrolment in CS doctoral programmes.}
    \label{fig:Primaryfactorsandinitiatives}
\end{figure*}

While a significant portion of the existing literature on women PhD candidates \oldtext{mainly} focuses \newtext{mainly} on retention and attrition dynamics \cite{Andre2022GenderMSc,alfermann2021should,wollast2023facing}, \newtext{we highlight in Figure \ref{fig:Primaryfactorsandinitiatives} }
\oldtext{highlighting the driving forces behind their initial decision to enrol is also crucial. 
Our study sheds light on these factors, enriching the existing knowledge and expanding the scope of research on the topic.
outlines} the primary factors and initiatives \newtext{relating} \oldtext{related} to female enrolment in CS doctoral programmes. \oldtext{to provide a visual overview of the major themes addressed in this section.}

\subsection{\textbf{The main supporting factors encouraging enrolment in a PhD programme}}

 Predominantly, existing studies tend to focus on broader disciplines such as STEM \cite{diogo2022tales}, engineering \cite{london2014motivations,tarvid2014motivation}, and computing \cite{tarvid2014motivation,moreno2013motivations} rather than exclusively on computer science. 
 \oldtext{This research works do not solely target female participants but} 
 Data from these studies indicate \oldtext{diverse} \newtext{varying} levels of female involvement. 
 \oldtext{Specifically, some} \newtext{Some studies} have observed \oldtext{a noteworthy representation of females} \newtext{noteworthy female representation} in their interviews and questionnaires \cite{diogo2022tales,tarvid2014motivation}, while others have reported typical levels of participation \cite{london2014motivations,moreno2013motivations}. These findings provide valuable insights into \oldtext{females'} \newtext{women’s} perspectives and initial motivations for enrolment. \newtext{The study by \cite{motogna2022retaining} is of particular interest, as it explains factors encouraging women} \oldtext{Of particular interest is the study by \cite{motogna2022retaining}, which explains factors encouraging females} to undertake doctoral degrees in computer science, though not confined solely to the enrolment phase.  
 
 Self-determination Theory (SDT) and \newtext{its subsidiary} Cognitive Evaluation Theory \newtext{(CET)}\oldtext{(a sub-theory within SDT)} suggest \newtext{that} needs for competence, autonomy, and relatedness impact both intrinsic \newtext{motivations }(\oldtext{i.e.,} \newtext{such as} personal fulfilment) and extrinsic \newtext{motivations}\oldtext{ (i.e.,} \newtext{including} academic career goals, professional development, \newtext{and} career benefits and opportunities\oldtext{, and other external drivers) motivations} \cite{ryan2000self}. SDT suggests that while intrinsic motivation stems from genuine interest, the intensity of extrinsic motivation varies from reward-driven to aligning closely with personal values, often leading to enhanced performance and well-being \cite{ryan2000self,ryan2000intrinsic}. 
 
Women's motivations for embarking on doctoral studies are multifaceted. In pursuit of personal fulfilment, they are often driven by an innate joy of exploration and a deep-seated desire to understand complex subjects. This passion goes hand in hand with aspirations \oldtext{of} \newtext{for} personal growth and integrating personal interests into one's professional journey via doctoral research \cite{diogo2022tales,london2014motivations,tarvid2014motivation,moreno2013motivations,motogna2022retaining}. Alongside these motivations, \oldtext{there} is an ambition to achieve \oldtext{remarkable} \newtext{significant} milestones that enhance their career and \newtext{contribute to} global scientific advancements \cite{diogo2022tales,motogna2022retaining}. Elevating one's social standing and gaining enriching learning experiences also play \oldtext{an} important roles \cite{moreno2013motivations}. Notably, research topic selection is \oldtext{influenced significantly}\newtext{heavily influenced} by curiosity and an overarching desire to add value to collective knowledge \cite{tarvid2014motivation,moreno2013motivations}.

\oldtext{Aspiring for an academic career, students} \newtext{Students looking for an academic career} see a PhD as essential for their future \oldtext{in academia}. For some, this commitment to \oldtext{the academic realm} \newtext{academia} has been evident since high school, even when presented with \oldtext{other} career opportunities in \oldtext{the}industry. They are \oldtext{intensely} interested in \oldtext{academic responsibilities, especially} teaching and prefer the university environment over \oldtext{industry}\newtext{corporate settings} \cite{moreno2013motivations}. \oldtext{While} \newtext{For} some PhD students \oldtext{may have a measured enthusiasm towards academia, building} \newtext{the work-life balance of} an academic career is compelling, \oldtext{especially given the work-life balance it can offer. This includes} \newtext{offering} part-time roles, flexible schedules, and a conducive work environment \cite{motogna2022retaining}. 

\oldtext{Research interest} \newtext{An interest in research} is crucial for many PhD students. Doctoral studies \oldtext{not only} deepen students' research capabilities, \oldtext{but also} introduce them to new social networks and refine their writing skills \cite{tarvid2014motivation}. \oldtext{The exposure to research during their master programs, positive experiences and academic endorsements during this period}\newtext{Exposure to research, positive experiences, and academic endorsements during master programmes} can significantly influence students' decision to advance to a PhD \cite{moreno2013motivations}. While the demanding nature of research activities can sometimes diminish enthusiasm \cite{motogna2022retaining}, these positive factors help \oldtext{ counter and} sustain the interest in \oldtext{going to a PhD}\newtext{pursuing a PhD} \cite{london2014motivations}.

Motivated by professional development, many see a PhD as a strategic investment to enhance career potential, with or without a clear career direction. In contrast, some aim to excel in their chosen fields, viewing doctoral study as \newtext{a means of} refining skills and meeting industry expectations \cite{moreno2013motivations}. Pursuing an engineering PhD, in particular, is often associated with the aspiration to align more closely with evolving industry standards and demands \cite{london2014motivations}. Additionally, many advanced roles in scientific research, academia, \oldtext{or}\newtext{and} industry require a PhD. This advanced qualification equips graduates with deeper knowledge and superior problem-solving skills, setting them apart from peers with lesser qualifications \cite{diogo2022tales,tarvid2014motivation, moreno2013motivations}. For those with significant industrial experience, the PhD is often perceived as a tool for professional exploration, offering the prospect of diverse and potentially more challenging roles within industry \cite{diogo2022tales,london2014motivations}. Moreover, those feeling limited or unfulfilled in their current positions might be attracted to a PhD to further their professional development and open new career pathways \cite{tarvid2014motivation, moreno2013motivations} \oldtext{.
For many students, the allure of a PhD lies in the promising career opportunities it provides. Universities often ensure employment upon graduation, offering stability in an unpredictable job market. While some are drawn to the programme for the chance to engage in profound research, others view \oldtext{seeing} it as a strategic move in a competitive career landscape \cite{moreno2013motivations}. Students frequently see a PhD as a stepping stone to better employment opportunities. This prospect significantly}
\newtext{which can specifically} influence female computer science students' decision to undertake doctoral studies \cite{motogna2022retaining}.

A PhD is a gateway to diverse roles beyond academia, spanning sectors like business, government, and non-profits. Many align their research focus with real-world applications, preparing themselves for entrepreneurial and managerial positions outside universities \cite{diogo2022tales}. 
Doctoral studies offer significant career benefits. In a labour market crowded with bachelor's and master's graduates, a doctoral degree \oldtext{distinguishes individuals, enhancing their career prospects, improving job security }\newtext{is a distinction that enhances career prospects,improves job security and can promise higher salaries}  \cite{london2014motivations,tarvid2014motivation}. \oldtext{Moreover, it promises higher salaries and meets the increasing demand of employers \cite{tarvid2014motivation}.}

Beyond direct career benefits, a significant motivation to pursue a PhD is the influence of successful role models. This motivation becomes stronger when observing the accomplishments of other PhD holders in the labour market, fuelling the desire for similar success and recognition \cite{tarvid2014motivation,motogna2022retaining}. 

In line with \oldtext{Cognitive Evaluation Theory}\newtext{CET}, encouragement from peers, friends,\oldtext{and, notably,} faculty members, and supervisors is a vital factor driving the decision to embark on a PhD journey  \cite{diogo2022tales,london2014motivations,tarvid2014motivation,moreno2013motivations,motogna2022retaining}. This choice \oldtext{often has its roots}\newtext{is usually rooted} in familial influences, \oldtext{mainly}\newtext{especially} when parents hold degrees. Exposure to \oldtext{such} academic environments can foster early aspirations, leading some to engage in research during their undergraduate years. Family advice and support further bolster the decision to undertake advanced studies. Furthermore, factors like students' academic achievements, aspirations, and career goals, which their parents' educational background can shape, provide added momentum towards pursuing a PhD \cite{moreno2013motivations,mccallum2012understanding,mullen2003goes}.

\subsection{\textbf{The main blocking factors discouraging enrolment in a PhD programme}}

Enrolling on a PhD programme is a \oldtext{progression through}\newtext{significant milestone in} educational life, \oldtext{culminating in a decision to start an advanced degree} \newtext{beginning with the decision to pursue an advanced degree}. Yet, female candidates may \oldtext{refrain from} \newtext{be reluctant to take} this decision due to their gendered experiences.  This begins at home, \oldtext{with little girls being encouraged} \newtext{where girls are often encouraged} to play with dolls while \oldtext{steering boys} \newtext{boys are steered} towards video and computer games, \oldtext{breeding} \newtext{facilitating} an early familiarity with the technical field \cite{yates2022female,denisco2017state}. Later, \oldtext{in post-secondary education, the gendered home life and classroom already lead to a shrunken pool of females} \newtext{these gendered experiences at home and in the school contribute to a reduced pool of women} willing and able to enter the CS field at the bachelor and master levels \cite{yates2022female,denisco2017state}. 
At the PhD enrolment stage, women may face additional limiting factors \newtext{ including financial constraints and emotional and psychological hurdles} impacting this decision. 
\oldtext{These include financial constraints, emotional and psychological hurdles such as the impostor syndrome, and \oldtext{other} external societal and institutional barriers.} 
Financial barriers \oldtext{encompass the female propensity} \newtext{include the tendency for females} to finance their studies from personal funds, \oldtext{whereas}\newtext{while} male candidates are more likely to receive salaried positions \cite{moskal2002female}. \oldtext{, the need to provide for their children, and other financial concerns} \newtext{Women also need to provide for their children, among other financial concerns \cite{motogna2022retaining}}.  
\oldtext{Women find i} \newtext{It is often} difficult to find affordable daycare \oldtext{nor can they}, \newtext{ and mothers cannot}  expect their \oldtext{child} \newtext{children} to subsist on low-quality food for long periods, \oldtext{thus often having to make a difficult choice}  \newtext{usually forcing them to choose } between their educational aspirations and raising a family \cite{Andre2022GenderMSc,rowe2021realities}.

\oldtext{For} \newtext{Among} women in STEM \oldtext{areas} \newtext{fields}, there is a notable \oldtext{manifestation} \newtext{presence} of \oldtext{the} impostor syndrome, \oldtext{. This syndrome is especially evident when they transition into technical fields} \newtext{especially when transitioning into technical areas}. Even \oldtext{those} women actively working in the sector \oldtext{might} \newtext{may feel they} need additional qualifications to \oldtext{feel they} \newtext{to} compare favourably with their male colleagues. Such beliefs often stem from deeply ingrained self-doubts, leading them to question their abilities and potential \oldtext{success in their career} \newtext{career success} \cite{denisco2017state,Hyrynsalmi2019The}.
From a young age, many women grapple with feelings of low self-esteem \cite{happe2021frustrations,adamecz2023overconfident}, \oldtext{. This self-perception makes} \newtext{making} them self-critical \oldtext{, often} \newtext{and prone to} questioning their competence even when they possess the necessary skills. Women tend to doubt their qualifications when faced with challenging or ambiguous job descriptions in technical roles, \oldtext{. This tendency to underestimate their abilities} \newtext{which} can leave them feeling unsuccessful \oldtext{, pushing} \newtext{and push} them away from their studies or career transitions, even when there is no actual performance deficit \cite{Hyrynsalmi2019The}. 

\oldtext{Added} \newtext{In addition} to \oldtext{the} pre-existing gender biases, societal dynamics further widen the confidence gap among women in technical domains. The observable overconfidence displayed by some men in these areas contrasts starkly with many women's self-doubt \cite{Hyrynsalmi2019The, adamecz2023overconfident}. In programming courses, for example, men with little to no experience \oldtext{still} \newtext{often} believe themselves \newtext{to be} equal \newtext{competitors} to women with much higher experience levels \cite{denisco2017state}. Age biases and competence questions further intensify women's sense of inadequacy. \oldtext{For instance, they} \newtext{Women} often express concerns about being perceived as ``too old'' for the labour market upon graduation or considering a career change. Some, \oldtext{females,} especially those in their thirties and forties, feel that potential employers might overlook their vast skills and experience in favour of younger candidates \cite{Hyrynsalmi2019The}. Moreover, even in female perception, leadership roles in technical fields are still predominantly associated with males, further contributing to the confidence gap \cite{Hyrynsalmi2019The, Wang2019Implicit}.
 
Stereotypes have shaped gender roles, impacting women's experiences in academia and technology. 
The perception that being a ``nerd'' or ``geek'' is unattractive and that computers are mainly for boys negatively impacts young girls' willingness to study computer science \cite{stross2008has}. From post-secondary education to advanced studies \cite{cohoon2009sexism}, sexism, real or perceived, also plays a role \oldtext{ as} \newtext{with} boys and men \oldtext{are} implicitly believed to possess higher competence levels in STEM \cite{denisco2017state,cohoon2009sexism}.

Women in computer science frequently encounter explicit sexism from male peers and professors. \oldtext{Whether in} \newtext{From} competitions \oldtext{or} \newtext{to} securing internships at top tech companies, their achievements are often undermined by male counterparts who suggest their success is due to gender-based quotas or a push for workplace diversity. Some male students resort to derogatory comments related to natural female processes \cite{denisco2017state}. Additionally, male professors might harbour lower expectations for female students, make inappropriate remarks about their personal lives \cite{denisco2017state} or ignore questions from female students \cite{cohoon2009sexism}.
\oldtext{Even} \newtext{Al}though explicit sexism might be less prevalent in CS doctoral programmes, its presence remains \oldtext{so} detrimental, \oldtext{that it can contribute} \newtext{contributing} to a higher attrition rate among female doctoral candidates \cite{cohoon2009sexism}.

Closely related to sexism, hostility in the academic environment can deter women from starting or continuing their doctoral studies. Women tend to perform better in a collaborative and supportive environment, where group success and valued work are more important than individual victories and self-focused learning \cite{yates2022female}. However, classrooms, especially in CS, often cater to a male-centric perspective \cite{yates2022female}, making them feel unwelcoming to many \oldtext{females} \newtext{women} \cite{cohoon2003must}.
Although many institutions have anti-discrimination policies, these measures are \newtext{often} not enough to combat negativity and toxicity in the institutional culture. Such a climate can intensify women's \oldtext{considerations} \newtext{inclinations} to abandon their academic pursuits \cite{Andre2022GenderMSc}.
Dismissive attitudes and a lack of support from supervisors or peers further \oldtext{heighten} \newtext{exacerbate} \oldtext{the} \newtext{this} sentiment \cite{yates2022female}. 

A lack of female faculty \oldtext{in male-dominated fields} \newtext{members} might discourage young women from pursuing advanced careers in \oldtext{these areas} \newtext{in male-dominated fields}. They might struggle to imagine themselves as potential mentors or professors \cite{denisco2017state}.
Not seeing \oldtext{other} women \oldtext{both} at their level and in positions of authority may \oldtext{cause women to feel more isolated. This could present} \newtext{lead to feelings of isolation, presenting} a challenge in terms of motivation due to the need for a more complex integration of social and academic aspects \cite{Andre2022GenderMSc}.

\subsection{\textbf{Initiatives to Encourage Female Enrolment in PhD Computer Science programmes}}
The various initiatives to support female students in \oldtext{PhD} Computer Science \newtext{PhD} programmes are multifaceted. Table \ref{tab:rel_work_initiatives} offers a concise summary of these initiatives, which we have categorised under five groupings 
(i) financial incentives and support, (ii) mentorship and networking, (iii) guidance and outreach, (iv) online learning support, and (v) inclusive environment.

\begin{table*}[!htbp]
\centering
\caption{Overview of Enrolment Initiatives for Females in PhD Computer Sciences programmes}
\label{tab:rel_work_initiatives}
\resizebox{\textwidth}{!}{%
\begin{tabular}{@{}llll@{}}
\toprule
 \textbf{Initiative Category} & \textbf {Ref}& \textbf{Description}& \textbf{Examples} \\ 
  \\\midrule

 Financial Incentives and  & \cite{lockett2023} & Grants, scholarships, and fellowships for & Google PhD fellowship programme, Microsoft Research Women’s  \\
 Support & & PhD women in CS & Fellowship \\
   
  Mentorship and Networking & \cite{informaticseu2023,craGradCohort}& Networking opportunities with successful  & CRA-WP, Women in ML, ACM-W,ACM-W Europe\\
 & & peers and role models, mentoring programs &  \\
 
 Guidance and Outreach & \cite{Andre2022GenderMSc,informaticseu2023,CSBROWN} & programmes educating potential applicants & PhD orientation at Stanford's Computer Science Department,\\
 && about the PhD process, by providing feedback   & Application Feedback programme for underrepresented applicants \\
 && and guidance & at Brown University \\
  
Online Learning Support & \cite{CS_org,munoz2016using} & Online doctoral programmes, remote meetings  &  Online or on campus Computer Sciences PhD \\
&& with supervisors & at the University of South Carolina, College of Engineering and Computing  \\ 

 Inclusive Environment & \cite{rubegni2023owning,jaccheri2020gender} & Anti-discrimination policies and support & NTNU's gender-sensitive policies, Chalmers University of Technology \\
&&  to make the academic environment more & Genie -- Gender Initiative for Excellence   \\
&& welcoming for females & \\

\bottomrule
\end{tabular}%
}

\end{table*}

Financial challenges may discourage female students from enrolling in PhD programmes in various fields, including computer science \oldtext{, as discussed in} \cite{lindner2020,horta2018phd}. 
Many \textbf{financial incentives and support} \oldtext{in terms of} \newtext{such as} scholarships, research grants, and, in some cases, stipends have been introduced to cover tuition fees and other costs \oldtext{and were} explicitly dedicated to female students \cite{lockett2023}. This \newtext{financial assistance} can \oldtext{further} \newtext{increase} \oldtext{motivate} \newtext{motivation} and significantly improve women's PhD opportunities. 

\oldtext{On top of} \newtext{In addition to} financial aid, institutions have actively \oldtext{attempted} \newtext{sought} to increase female enrolment in PhD programmes through various other campaigns and support programmes \cite{Andre2022GenderMSc,informaticseu2023}. Among these, \textbf{mentorship programmes and networking opportunities} \cite{informaticseu2023,craGradCohort} are essential tools to \oldtext{clear any doubts or concerns students might have before starting} \newtext{address students' doubts or concerns before starting} their PhD journey. By organising events that bring together prospective PhD candidates with current female doctoral students or professors who might become potential mentors, institutions provide a platform where experiences can be shared. \oldtext{and a clear view of future.} \newtext{Future} challenges and opportunities can be \newtext{openly} discussed. Such interactions empower attendees to ask questions, \oldtext{and} fully understand the PhD process \cite{Andre2022GenderMSc,fisher2019structure}, \oldtext{thus making} \newtext{and make} informed decisions, \oldtext{and fostering a sense of belonging to the academic community}. Additionally, celebrating female PhD graduates' successes, as done by Stony Brook's Computer Science Department \cite{stonyBrook}, offers attendees an opportunity to network and be inspired by the \oldtext{successes} \newtext{achievements} of their peers.

Furthermore, institutions often host \textbf{guidance and outreach} initiatives, such as specific PhD orientation activities. For instance, Stanford University's Computer Science Department \cite{standfordCS} organises events designed to answer students' questions, explain the details of the programme, and highlight post-PhD career opportunities \cite{craGradCohort}. Similarly, Brown University's Computer Science Department has a specialised PhD application programme \cite{CSBROWN} dedicated to underrepresented groups, including women, ensuring a smooth application process. These tailored initiatives are crucial in helping female students reach their academic goals. 

Following the 2019 pandemic, there has been an accelerated interest in \textbf{online learning supports} that might aid in recruiting female doctoral students. \oldtext{The o} \newtext{O}nline doctoral programmes in information technology \cite{CS_org} highlight this increased visibility. Such \newtext{programme} designs offer advantages for women juggling family responsibilities. A study \cite{munoz2016using} presented the benefits of online education for women with limited resources, such as remote meetings with supervisors, suggesting its potential to support greater female participation in PhDs. 

European universities have developed projects, policies, and strategies to promote an \textbf{inclusive environment} and increase women's participation in CS \cite{rubegni2023owning,jaccheri2020gender}. The Norwegian University of Science and Technology (NTNU) stands out for its gender-sensitive policies and anti-discrimination awareness programmes \cite{ntnu_initiative}. Beyond policy adjustments, some institutions, like Koç University, have adopted gender-responsive pedagogical practices and incorporated gender-equitable design principles into their teaching materials to cultivate a lasting positive impact \cite{koc_initiative}.

Retaining female students in computing is as important as recruiting them. Robust retention initiatives \oldtext{not only} ensure that women continue their doctoral journey \oldtext{but also} \newtext{and} position them as role models for potential entrants. Witnessing the successes and perseverance of their peers can inspire and motivate more women to consider and pursue a PhD in the field. Ongoing mentorship and training \cite{eugain_workshop2022,eugain_workshop2023}, early exposure to computing topics through workshops and hackathons \cite{de2018encouraging}, supportive networks for women at all academic stages \cite{rubegni2023owning}, and skills-enhancing training are essential. It bears reiterating that the continued presence of female role models in computing encourages others to follow similar paths.

%% file: 04-Survey.tex
\section{Survey \& Approach}
\label{sec:survey}

\oldtext{The objective of our survey} \newtext{Our goal} was to identify the factors that students consider when they decide on their career path after their master's studies, with a particular focus on gender-based differences. This section describes the instrument design, the data collection approach, the data cleaning performed on the responses received and the demographics of the valid responses received.

\subsection{Instrument Design}
\label{subsec:InstrumentDesign}

Our survey instrument includes 58 questions (see \tablename~ \ref{tab:QuestionnaireSummary}) used to collect data on:

\begin{itemize}
	\item \textbf{Demographics (Q1-Q7).} These were aimed at finding out and collecting basic demographic data (enrolment country, nationality, degree, progress, gender).
	\item \textbf{Respondents' background as students (Q8-Q21).} These questions were aimed at profiling our respondents concerning their academic trajectory, providing insights about their experiences and preferences.
	\item \textbf{Intentions concerning the possibility of pursuing a PhD (Q22-Q46).} 
	These questions include not only the intentions, but also respondents' perceptions on the factors that are in favour of, or detrimental to, enrolling in a PhD. Those factors include personal factors (e.g. family related), as well as more structural factors (e.g. related to career prospects, or financial security).
	\item \textbf{Respondents' Self-perception (Q47-Q58).} The last group is mostly focused on \oldtext{the perceptions respondents might have} \newtext{respondents' perceptions} of their strengths and weaknesses \oldtext{when} compared to their peers.
\end{itemize}

We used a mixture of open and closed questions, with a clear predominance for the latter (49 out of 58):
\begin{itemize}
	\item \textbf{Likert.} We have 39 5-point \textit{\textbf{Likert}-scale} questions, where the options are \textit{definitely yes}. \textit{rather yes}, \textit{neutral}, \textit{rather not}, and \textit{not at all}. We also included a \textit{does not apply} option in the Likert-scale questions so respondents would not randomly choose one of the options if for some reason the question did not apply to them. \newtext{The variables collected via these questions are \textbf{ordinal}.}
	\item \textbf{SC.} We have 5 \textit{\textbf{S}ingle \textbf{C}hoice} questions that require a \textit{yes} or \textit{no} answer. \newtext{The variables collected via these questions are \textbf{nominal}.}
	\item \textbf{SMC.} We have 4  \textit{\textbf{S}imple \textbf{M}ultiple \textbf{C}hoice} questions. These questions allow the respondent to choose only one of the provided answer alternatives. They include a special \textit{Other} alternative, to allow the respondent to answer in open free text if none of the available alternatives is a good match for the respondent's answer. \newtext{The variables collected via these questions are \textbf{nominal}.}
	\item \textbf{MMC.} We have 2 \textit{\textbf{M}ulti-select \textbf{M}ultiple \textbf{C}hoice} questions. In one of them, we added the \textit{Other} option, to provide a free text alternative to our respondents, if necessary. \newtext{The variables collected via these questions are \textbf{nominal}.}
	\item \textbf{Open.} We have 8 \textit{\textbf{Open}} questions, devoted to the collection of additional comments or insights the respondents might want to share with us. \newtext{The variables collected via these questions are \textbf{open text}.}
\end{itemize}

\begin{table*}[!htbp]
	\centering
	\caption{Survey questions - condensed excerpt.}
	\label{tab:QuestionnaireSummary}
		\begin{footnotesize}
			\begin{tabular}{@{}llll@{}}
				\toprule
				\textbf{Part} & \textbf{Id} & \textbf{Question} & \textbf{Type} \\ \midrule
				\multirow{7}{*}{\rotatebox[origin=c]{90}{I. Personal Info}} & Q1 & Please name the country in which the university you are/were enrolled is located. & SMC+Other \\
				& Q2 & Please name your nationality. & SMC+Other \\
				& Q3 & Please indicate your highest educational rank: & SMC+Other \\
				& Q4 & Please name the study program in which you are/were enrolled: & Open \\
				& Q5 & Please indicate your current occupation: & MMC+Other \\
				& Q6 & What is your gender? & SMC+Other \\
				& Q7 & Do you have any further comments? & Open \\\midrule
				\multirow{18}{*}{\rotatebox[origin=c]{90}{II. Questions on Own Studies}} & Q8 & Did you have any computer science courses during high school? & SC \\
				& Q9 & Are/Have you been personally supervised by a researcher, e.g. by a PhD student or a teaching assistant & SC \\
				&  & for a seminar or project work? & \\
				& Q10 & Are/Have you been involved in research projects during your studies, e.g. in terms of being involved in & SC \\
				&  & research at a department/chair or in a larger scientific project? &  \\
				& Q11 & Are/Have you been personally supervised by a research leader, e.g. by a professor or lecturer for a seminar & SC \\
				&  & or project work? & \\
				&  & \textbf{Please rate each of the following statements on the scale displayed.}& \\
				& Q12 & Do you feel enthusiasm for computer science? & Likert \\
				& Q13 & Do you enjoy your studies (without considering the pandemic)? & Likert \\
				& Q14 & Would you say that you have good grades? & Likert \\
				& Q15 & Would you say that you are successful in your studies? & Likert \\
				& Q16 & Do you feel self-confident in interaction with other students? & Likert \\
				& Q17 & Did/Do you regularly participate in study groups during your studies? & Likert \\
				& Q18 & Do you like theoretical computer science subjects (esp. logic and maths)? & Likert \\
				& Q19 & Do you like practical computer science subjects (esp. programming)? & Likert \\
				& Q20 & Do you like interdisciplinary areas of computer science? & Likert \\
				& Q21 & Do you have any further comments? & Open \\\midrule
				\multirow{27}{*}{\rotatebox[origin=c]{90}{III. Questions on PhD (intentions)}} &  & \textbf{Please evaluate each of the following statements concerning your current situation on the scale displayed.} &  \\
				& Q22 & When being a Bachelor student, I plan(ned) to enrol in master studies. & Likert \\
				& Q23 & I have thought about doing a PhD. & Likert \\
				& Q24 & I plan to do a PhD. & Likert \\
				& Q25 & I am aware of the aims and contents of PhD studies. & Likert \\
				& Q26  & I am aware of the requirements for starting a PhD. & Likert \\
				& Q27 & I have been encouraged to do a PhD. & Likert \\
				& Q28 & After completing my current studies, I am considering moving abroad for a while. & Likert \\
				& Q29 & My family encourages me to do a PhD. & Likert \\
				& Q30 & I was motivated by industrial contacts to do a PhD. & Likert \\
				&  & \textbf{Please evaluate each of the following arguments in favour of enrolling in a PhD on the scale displayed.} &  \\
				& Q31 & The aims and contents of PhD studies suit me. & Likert \\
				& Q32 & I believe I meet the requirements for getting a PhD position. & Likert \\
				& Q33 & A PhD would improve my future career. & Likert \\
				& Q34 & Pursuing a PhD would be fun. & Likert \\
				& Q35 & Better gender balance in computer science would increase my interest in PhD studies. & Likert \\
				&  & \textbf{Please evaluate each of the following arguments against enrolling in a PhD on the scale displayed.} &  \\
				& Q36 & I am worried that PhD studies would negatively impact my future family plan (getting married, having kids,$\ldots$). & Likert \\
				& Q37 & I have concerns that during my PhD studies, I would not have enough money to support my lifestyle. & Likert \\
				& Q38 & I have better job opportunities in the industry than at university. & Likert \\
				& Q39 & I am hesitant to start a PhD because of many factors and circumstances (e.g. career opportunities, flexibility, & Likert \\
				&  & working environment,$\ldots$) are unclear to me. & \\
				& Q40 & I think the industry offers more flexible career opportunities compared to the structured path of PhD studies. & Likert \\
				& Q41 & I am worried about the impact on my future career if I start but do not complete my PhD studies. & Likert \\
				& Q42 & I have reservations about making long-term job commitments for the whole PhD duration. & Likert \\
				& Q43 & I am afraid of being one of only a few females in a PhD group. & Likert \\
				& Q44 & Do/Did you participate in any programs or activities that informed you about PhD studies? & SC \\
				& Q45 & Please describe these programs or activities shortly. & Open \\
				& Q46 & Do you have any further comments concerning PhD intentions? & Open \\\midrule
				
				\multirow{13}{*}{\rotatebox[origin=c]{90}{
						IV. Questions on self-reflection}} &  & \textbf{Please evaluate your strengths as compared to your peers:} &  \\
				& Q47 & I have the competencies to succeed in computer science. & Likert \\
				& Q48 & I have strong programming skills. & Likert \\
				& Q49 & I possess theoretical understanding. & Likert \\
				& Q50 & I have self-confidence. & Likert \\
				& Q51 & I embrace problem-oriented learning. & Likert \\
				& Q52 & I am driven by curiosity to learn. & Likert \\
				& Q53 & I am successful in collaborating with others. & Likert \\
				& Q54 & I love computer science. & Likert \\
				& Q55 & Do you have any insights on what encourages students to start a PhD in computer science? & Open \\
				& Q56 & Do you have any insights on what discourages students from starting a PhD in computer science? & Open \\
				& Q57 & Do you have any further comments? & Open \\
				& Q58 & How did you find out about this survey? & MMC \\ \bottomrule
			\end{tabular}%
	\end{footnotesize}
\end{table*}

We conducted a pilot session with colleagues (all academics) to solicit their feedback, during one of our COST Action meetings, before launching the survey. We reviewed the whole questionnaire to detect clarity issues, problematic questions (e.g. overloaded questions), and any other structural problems. We incorporated this feedback. We also asked participants to fill in the questionnaire so we could assess how long it took them to do so and incorporate that information in the questionnaire invitation letter. 

\subsection{Data Collection}

The target population for this study was computer science students, researchers, and practitioners in Europe. In the case of participants who already have a PhD, they were encouraged to reflect on their experience.  Although this study was open to respondents from countries outside of Europe, we focused our respondents' recruitment efforts on Europe, with all its cultural diversity.

We used an \textit{open invitation} strategy~\cite{wagner2020challenges}. We advertised our survey through several channels, including mailing lists, personal contacts, classes and seminars at Universities, conferences, workshops, and summer schools. We leveraged the contacts of the COST Action members \oldtext{, as well as} \newtext{and} those of Informatics Europe to reach a broad audience of potential respondents. As a key focus of our research is gender inclusion in Informatics, we were particularly interested in gaining insights from female participants. We specifically targeted potential female respondents in our contacts, as several of our research questions depended on having \oldtext{a large number of} \newtext{many} female respondents. We also wanted to have answers from respondents of other genders, so we could find out which \oldtext{of the} candidate factors are \oldtext{, indeed,} are gender-related.

The responses include answers from people in different \newtext{life} stages \oldtext{in their lives}, ranging from undergraduate and graduate students \oldtext{who may be} considering pursuing a PhD to researchers and practitioners who may have completed a PhD or decided against doing one.

\begin{figure}[htb!]
	\centering
	{\includegraphics[width=1.0\columnwidth]{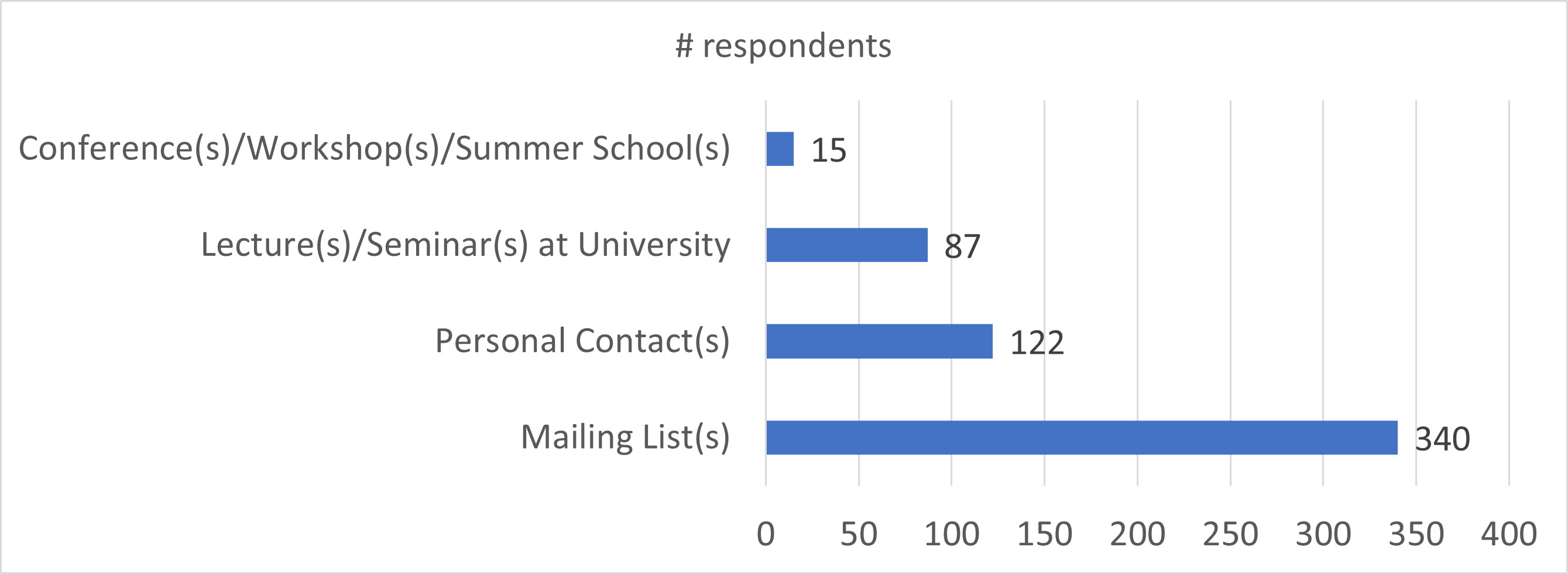}}
	\caption{How participants found about this study}
	\label{fig:HowDidYouFindOutThisSurvey}
\end{figure}

Figure \ref{fig:HowDidYouFindOutThisSurvey} summarises how our respondents found out about the survey. Several respondents learned about the survey via more than one channel.

We targeted European countries, as this survey was primarily aimed at understanding the phenomenon in the European context. That said, the survey link was public and it did reach people from other continents (e.g. we had several respondents from universities in China). We decided to keep answers from other origins, given the mobility of PhD students. It is common to find non-European PhD students in European Universities.

Concerning sample size, we used Yamane's equation~\cite{yamane1973statistics}, as suggested in~\cite{wagner2020challenges}), to compute a suitable sample size $n$:
\begin{equation}
	n = \frac{N}{1 + N \cdot e^{2}}
\end{equation}

With an estimated 28.7 million developers worldwide by 2024~\cite{SoftwareEngineersPopulation}, this would mean around 400 participants, for a precision set to 0.05. This is a conservative estimate, as the number of potential PhD candidates is \oldtext{surely} much lower, \oldtext{, and more so} \newtext{especially} if we focus on Europe.

\begin{equation}
	n = \frac{28,700,000}{1 + 28,700,000 \cdot 0.05^{2}} = 400
\end{equation}

We did not include any localisation for our survey. The same survey version was available for all respondents, regardless of their geographic location.

We did not offer any financial or personalised incentive for participation. The lack of gender balance in Informatics is noticeable and this was enough to attract a significant number of participants.

The data collection process was prolonged in time, due to difficulties in convincing enough people to offer their valuable time for research. We started disseminating the survey in October 2021, mostly via contacts made by the authors, leveraging their networks. By June 2022 we had 129 respondents, mostly from Germany and Portugal. We then increased our efforts to bolster the number of respondents and the diversity of their locations, by encouraging other EUGAIN members to disseminate the questionnaire including Informatics Europe in that process. By February 2023, we had 506 responses. While we were doing some preliminary data analysis, we reiterated our requests for help from EUGAIN members and intensified our dissemination campaign. We finished our data collection on May 30, 2023, with 867 responses from 53 countries.

\subsection{Data cleaning and transformation}
We filtered out all responses that did not meet our inclusion criteria. In particular, we excluded responses that did not answer our research questions (e.g. respondents only answered basic demographic questions).
Of the total 867 responses to our survey, 280 answers contained no relevant information concerning our research questions, leaving us with 587 valid responses. 

We sanitised some answers (e.g. we transformed empty answers to the question \textit{``What is your gender?''} into \textit{``I prefer not to disclose.''} and did the same for a respondent who answered \textit{``Other''}, but then did not fill in the form with any value for \textit{``Other''} - in practice, the respondent preferred not to disclose the gender, as well).

We computed an extra variable to encode whether the respondent has a PhD, is in the process of studying for a PhD, or plans to do a PhD, or if the person has no PhD and no plans to obtain one. We refer to these values in our data analysis tables and figures as \textit{``PhD''} \textit{vs.} \textit{``No PhD''}, respectively.

Finally, we moved free-text responses to a separate file, to preserve anonymity.

\subsection{Survey Response Demographics }

Of the 587 valid responses, 309 respondents identified as \textit{female}, 267 identified as \textit{male}, and 4 respondents reported their gender as \textit{other} (1 \textit{female-to-male transgender} and 3 \textit{non-binary} respondents, one of which also identified as \textit{queer} and \textit{agender}. 7 respondents \textit{did not disclose} their gender. 

\begin{figure}[htb!]
	\centering
	{\includegraphics[width=1.0\columnwidth]{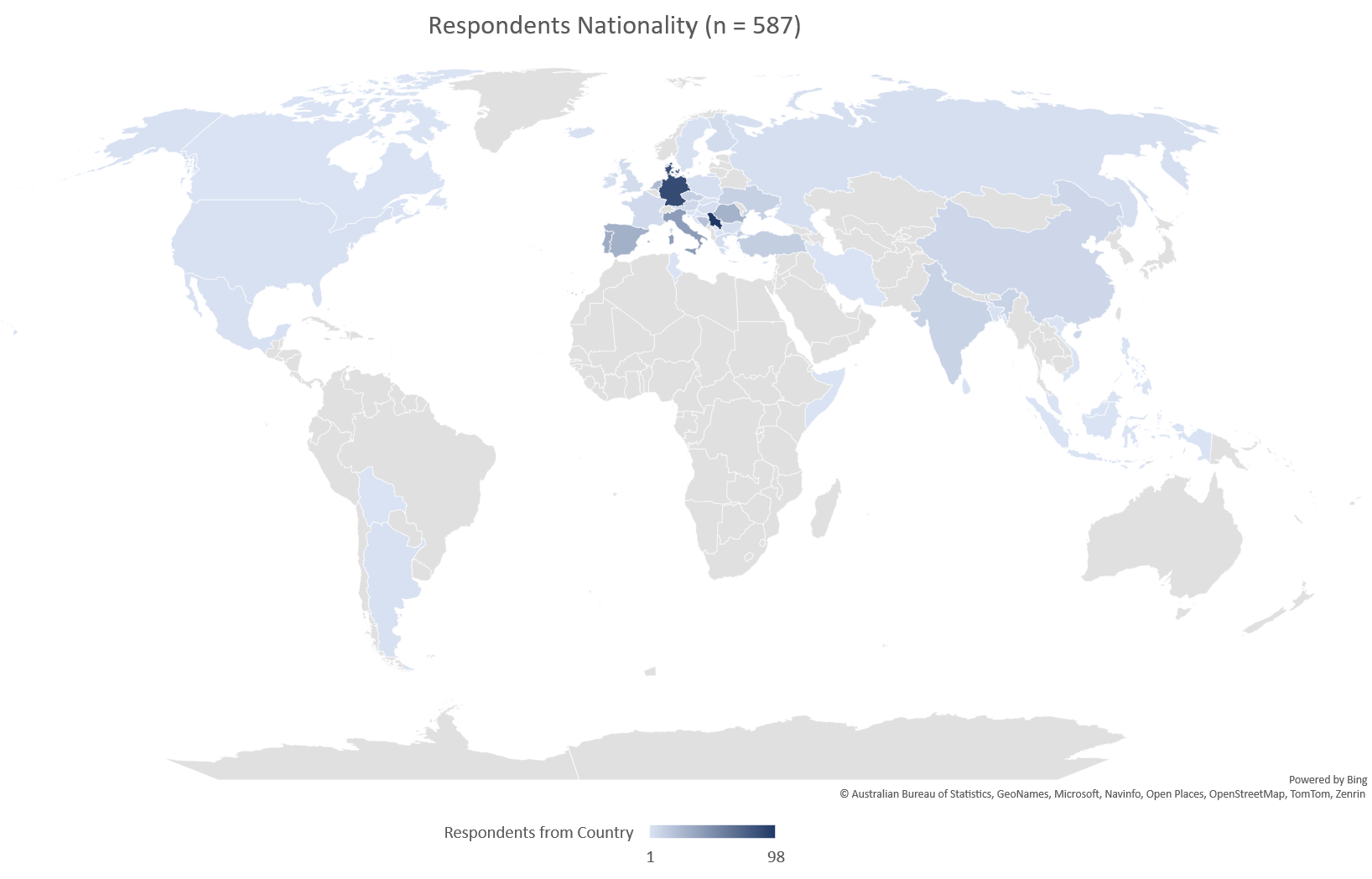}}
	\caption{Respondents nationality}
	\label{fig:RespondentsNationality}
\end{figure}

We received responses from people from 53 nationalities. Although our target population were people working in Europe, the questionnaire was available to anyone with the link and publicly announced at an international level, so we also got some answers from other continents (e.g. with 10 participants from India and 8 from China). The countries with the most responses were Serbia (98), Germany (87) and Denmark (86). Figure \ref{fig:RespondentsNationality} summarises respondents' nationalities. 

\begin{figure}[htb!]
	\centering
	{\includegraphics[width=0.60\columnwidth]{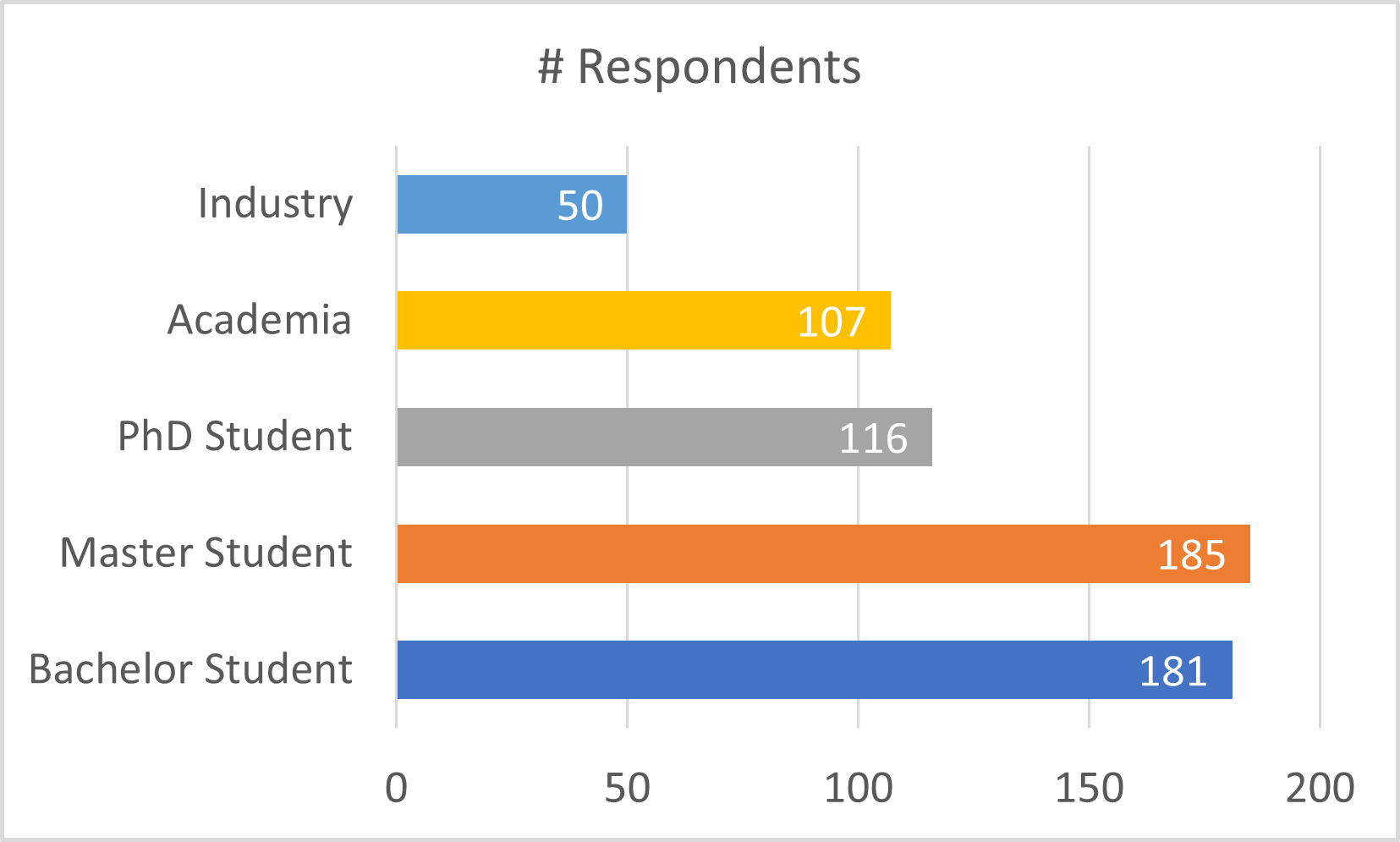}}
	\caption{Respondents current occupation. Some respondents hold more than one position (e.g. Master Student and Industry).}
	\label{fig:RespondentsPosition}
\end{figure}

Concerning their current occupation illustrated in Figure \ref{fig:RespondentsPosition}, most of our respondents were master (185) and Bachelor (181) students. Some of the respondents hold more than one position (e.g. some of the MSc students also work in industry).

We also asked our participants about their highest educational ranking achieved. We recorded the 7 answers in the \textit{Other} category: 6 of them implied holding a PhD, while the remaining one implied holding an MSc. Overall, most of our respondents declared holding a BSc or an MSc.

\begin{figure}[htb!]
	\centering
	{\includegraphics[width=0.60\columnwidth]{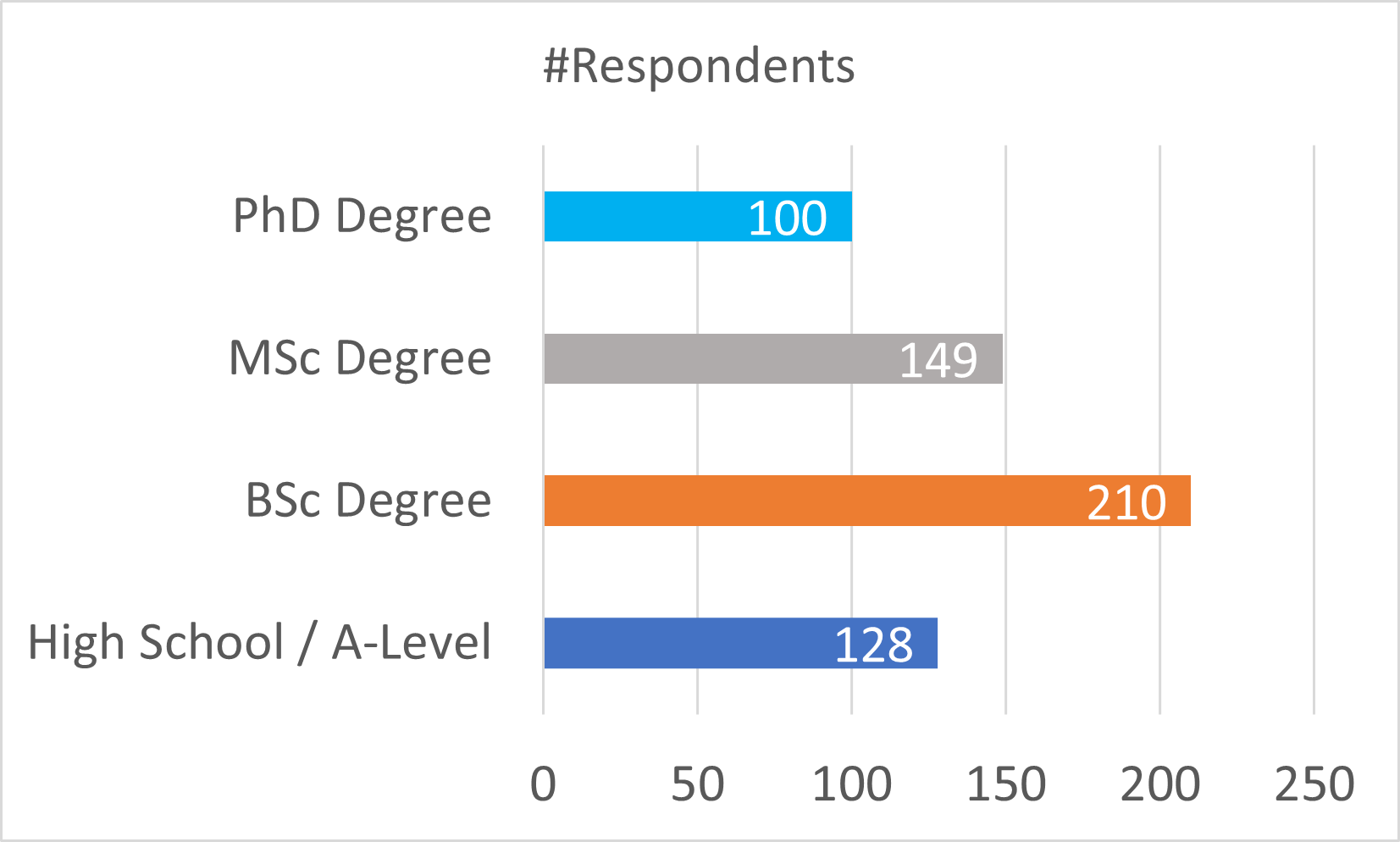}}
	\caption{Respondents highest achieved degree.}
	\label{fig:RespondentsDegree}
\end{figure}

\section{Analysis and Results} 
\label{sec:Results}
\oldtext{We performed both a qualitative analysis of the closed questions in the survey but also a quantitative analysis of the open questions. 
In this section, we} We describe the approach used for both the quantitative and qualitative analysis and the results of both types of analysis on each research question as outlined in Section~\ref{sec:introduction}.

\subsection{Approach used for Quantitative Analysis}
\label{subsec:ApproachQuantitativeAnalysis}
Our questionnaire included several closed questions for which we wanted to perform quantitative analyses, based on the frequencies of the options chosen by our respondents. \newtext{As discussed in section \ref{subsec:InstrumentDesign}, some of the variables used in our data analysis are nominal, while others are ordinal.} We started by collecting descriptive statistics on these variables. We then formulated hypotheses and performed the appropriate hypothesis testing to identify whether differences were statistically significant. \newtext{Our statistical analysis uses only \textbf{nominal} and \textbf{ordinal} data. As such, we only use \textbf{non-parametric} procedures in our data analysis which \textbf{do not require normality}.}

We used the Chi-Square test, for single-choice questions
\newtext{(i.e. nominal data). The Chi-square examines the difference between categorical variables from a random sample in order to determine whether the expected and observed results are well-fitting. In particular, we use it to test if there is a difference in proportions between several groups identified by nominal variables. For example, we use the Chi-Square test to contrast the answers from people who decided to do a PhD with those who decided against doing a PhD, concerning whether they were involved in research projects during their studies}.

We used the Mann-Whitney U test, for questions where the participant could choose their answer using a Likert-scale value\newtext{, i.e. ordinal data. The Mann-Whitney U test checks data distribution for a significant difference between two independent groups. The test applies ordinal (which is the case here) or continuous variables. It requires independent observations and the shapes of the distributions of the two groups under comparison to be roughly the same.} 

Both for Chi-Square tests and for Mann-Whitney U we used $p < .05$ for the level of significance. 

As we analyse several independent variables in the context of finding relations with the same dependent variable (whether the participant is ``PhD material'' --- has done, is doing, or plans to do a PhD --- or not), it can be argued that this implies a form of multiple comparisons so we also included post-hoc multiple comparison corrections, including the Bonferroni and the Holm-Bonferroni. The Bonferroni correction~\cite{bonferroni1936teoria} is a conservative post-hoc method that controls the Type I error rate, but significantly increases the probability of a Type II error~\cite{perneger1998s}. \newtext{The Bonferroni correction compensates for the risk of making a Type I error by dividing the significance level by the number of hypotheses. For example, for our desired $p < .05$, if we test 5 hypotheses, the corrected value is $\frac{.05}{5}=.01$, so results will be statistically significant using the Bonferroni correction if $p < .01$.} 

To achieve a more balanced control of Type I and Type II errors, we also tested whether our results are still significant after performing the Holm-Bonferroni method~\cite{holm1979simple}. \newtext{Compared to Bonferroni, the Holm-Bonferroni method increases the Type I error probability and decreases the Type II error probability. The Holm–Bonferroni method sorts the p-values from lowest to highest and compares them to nominal alpha levels of $\frac{\alpha}{m}$, where $m$ represents the number of tested hypotheses. For example, with our desired $p < 0.05$, for 5 hypotheses, the adjusted values are $\frac{.05}{5}=.01$, $\frac{.05}{4}=.0125$, $\frac{.05}{3}=.1(6)$, $\frac{.05}{2}=.025$, and $\frac{.05}{1}=.05$.} 

\oldtext{So, for} \newtext{For} all our hypothesis tests, the significance level is highlighted with different font styles, depending on whether the differences are \newtext{``just''} \textit{statistically significant}, \newtext{or remain significant when using a \underline{\textit{Holm-Bonferroni correction}} or even when using a \textbf{\underline{\textit{Bonferroni correction}}}.}
\oldtext{even when using a \underline{\textit{Holm-Bonferroni correction}} and using \textbf{\underline{\textit{Bonferroni correction.}}}}

We computed the effect size using Cramér's V ($\varphi_{c}$)~\cite{Cramer1946} for the Chi-Square tests. With one degree of freedom, the common interpretation thresholds for the effect size computed with $\varphi_{c}$ are: .10 - small effect size; .30 - medium effect size; .50 - large effect size. The Mann-Whitney U tests require a different effect sizes measure. We used the eta-square ($\eta^{2}$) effect size measure, which provides the percentage of the variance of the dependent variable (the decision to do a PhD, or not) that can be explained by the independent variable tested in the corresponding Mann-Whitney U test (e.g. the self-confidence in one's programming skills). We use the following recommended thresholds for interpretation of the effect size values computed with $\eta^{2}$: .01 - small effect size; .06 - medium effect size; .14 or higher - large effect size.

\newtext{To make our findings clearer we adopted a visual notation in our tables and text with the following symbols:}
\nopagebreak
\begin{tcolorbox}
    \begin{footnotesize}
        \textbf{\normalsize{Key for the icons used in this section}}\\
        \textbf{Participants:}\\
        \faUserGraduate~Participant who chose to do a PhD\\
        \faUser~Participant who chose not to do a PhD\\
        \faVenus~Female participant\\
        \faMars~Male participant\\
        \\
        \textbf{Descriptive statistics:}\\
        $\#$~Number of participants\\
        $\%$~Percentage of participants\\
        $\bar{x}$~mean score\\
        $\tilde{x}$~median score\\
        \textbf{Mode}~mode score\\
        
        Participants and descriptive statistics icons can be combined, e.g.:\\
        $\#$\faVenus\faUserGraduate~Number of female participants who chose to do a PhD\\
        $\%$\faMars\faUser~Percentage of male participants who chose not to do a PhD\\
        
        \textbf{Statistics tests:}\\
        \textbf{$\chi$\textsuperscript{2}(1)}~Chi square statistic\\
        \textbf{MW-U}~Mann-Whitney U\\
        \\
        \textbf{Effect sizes:}\\
        $\varphi_{c}$~Effect size (for Chi square test)\\
        $\eta^{2}$~Effect size (for Mann-Whitney U test)\\
        \\
        \textbf{p-values formatting:}\\
        .ddd~Not statistically significant\\
        \textit{.ddd}~Statistically significant\\
        {\ul\textit{.ddd}}~Statistically significant with Holm-Bonferroni Correction\\
        {\ul\textbf{\textit{.ddd}}}~Statistically significant with Bonferroni correction\\
        \\
        \textbf{Results summary:}\\
        \faChevronUp~Statistically significant difference, with a low effect size\\
        \faChevronUp\faChevronUp~Statistically significant difference, with a medium effect size\\
        \faChevronUp\faChevronUp\faChevronUp~Statistically significant difference (higher), with a high effect size\\
        (similar for lower \faChevronDown, \faChevronDown\faChevronDown, \faChevronDown\faChevronDown\faChevronDown)\\
        
        \faChevronCircleUp~Bonferroni-corrected statistically significant difference (higher), with a low effect size\\
        \faChevronCircleUp\faChevronCircleUp~Bonferroni-corrected statistically significant difference (higher), with a medium effect size\\
        \faChevronCircleUp\faChevronCircleUp\faChevronCircleUp~Bonferroni-corrected statistically significant difference (higher), with a high effect size\\
        (similar for lower \faChevronCircleDown, \faChevronCircleDown\faChevronCircleDown, \faChevronCircleDown\faChevronCircleDown\faChevronCircleDown)\\
     
        The summary of each of the results presented is preceded by a reference group identification.\\
        For example:\\ \faUserGraduate\faChevronCircleUp\faChevronCircleUp~Participants who decided to do a PhD had a statistically significantly higher value of the variable under comparison than those who decided not to do a PhD. The difference is significant even when using the Bonferroni correction. The effect size is medium.
    \end{footnotesize}
\end{tcolorbox}

We complemented the statistics tests with visualisations (divergent stacked bar charts) to help data interpretation.

\subsection{Approach used for Qualitative Analysis}
In addition to the closed questions, there are also several open questions in the questionnaire. It is also relevant to analyse their answers since data from free-text answers to open questions can provide deeper insights into the participants’ personal experiences, concerns, or motivations which the other questions in the survey may not have detected. Among 8 open questions: \newtext{question} Q7 is a basic demographics question; \newtext{questions} Q21\newtext{, Q46} and Q57 are introduced to provide space for participants to write their further comments related to previous questions; \newtext{question} Q45 aims to collect detailed information about the informative programs and activities the participants attended related to pursuing a PhD (only for the participants who answered ``yes'' for \newtext{question} Q44). All the answers to these questions were carefully read by the authors and were subsequently excluded from analysis as their contents ended up being irrelevant. \newtext{Only one comment from \newtext{question} Q46 was reported to highlight an interesting perspective of one participant where participants were asked if they have any further comments concerning PhD intentions in general.} 
We selected the remaining two open questions (Q55 and Q56) for further analysis to gain insights into the perception of the participants on the encouraging and discouraging factors for pursuing PhD studies. These two questions had a much larger number of answers, compared to other open questions, and are the most relevant open questions for this research considering the research questions. 
These two questions are as follows:
\begin{itemize}
	\item Q55 - Do you have any insights on what encourages students to start a PhD in computer science?
	\item Q56 - Do you have any insights on what discourages students from starting a PhD in computer science?
\end{itemize}

Neither of these open questions was compulsory, so the participants were free to write their comments or not. Out of 587 valid survey responses, 165 and 192 of the participants answered \newtext{question} Q55 and \newtext{question} Q56 respectively, after filtering out answers indicating that the respondent has no response to give. Note that these respondents could be considered as the group who need more information about the advantages or disadvantages of pursuing a PhD. However, for this analysis, they were removed from the data set.

We conducted a qualitative analysis employing thematic analysis \cite{braun2006using}, \cite{cunningham2017qualitative} on these free-text comments. The process involved one researcher reading all the comments and becoming familiar with both data sets: first and second question answers. Considering the recurring keywords and meanings, we identified different themes for two data sets. We assigned answers to these themes, considering their content. One answer can be assigned to multiple themes to cover the whole content of the answer. For example, we assigned the answer \textit{``I suppose being interested in the subject matter would be a start. Or maybe having fun teaching others.''} to the themes \textit{``interest and confidence in research''} and \textit{``interest in academic career and teaching''}. After assigning the themes for the answers, a second researcher checked the coding decisions randomly for 10\% of the answers for each data set. We discussed the disagreements and when necessary, made adjustments and recurring issues emerged.

\subsection{Analysis of RQ1: What main supporting factors encourage enrolment in a PhD program?}
\label{subsec:RQ1}
We organised our data analysis for RQ1 around six groups of questions: 
(i) participants' education experience through single choice questions (questions Q8-Q11), as well as whether the participant had attended any programs or activities that informed participants about PhD studies (Q44, a single choice question); (ii) participants' preferences (questions Q12-Q20); (iii) participants' intentions (questions Q22-Q23; Q25-Q30), (iv) arguments in favour of enrolling on a PhD in Computer Science (questions Q31-Q35); (v) self-reflection questions on participants' perceptions of their strengths (questions Q47-Q54); and (vi) qualitative results from the open questions identifying supporting factors for enrolling on a PhD in Computer Science. 

We consider the themes underlying all these questions, as stated in the questionnaire statements, to be plausible candidates for factors encouraging enrolment in a PhD program in Computer Science. That said, an overall disagreement concerning one of these factors may imply that something that should be an incentive (e.g. strong encouragement from the industry to study for a PhD) is, in practice, a detrimental factor (e.g. the lack of strong encouragement from the industry to study for a PhD may hinder the drive for enrolling into a PhD).

\paragraph{(i) Education experience}
Questions Q8 through Q11, and Q44, assess the participants' previous studies experience and are single choice questions, where respondents answer either \textit{Yes} or \textit{No}. 

We start by defining our hypotheses. For questions Q8 through Q11, and Q44, these are as follows:

\begin{tcolorbox}
	\begin{footnotesize}
		$H_{i-0}$: There is no statistically significant difference in \textit{factor x}, when comparing participants who choose to do a PhD (\faUserGraduate) and those who do not (\faUser).
		
		$H_{i-1}$: There is a statistically significant difference in \textit{factor x}, when comparing participants who choose to do a PhD (\faUserGraduate) and those who do not (\faUser).    
	\end{footnotesize}
\end{tcolorbox}

For example, for Q10, the corresponding hypotheses were:

\begin{tcolorbox}
	\begin{footnotesize}
		$H_{10-0}$: There is no statistically significant difference in \textit{having been involved in research projects during the participants' studies} when comparing participants who choose to do a PhD (\faUserGraduate) and those who do not (\faUser).
		
		$H_{10-1}$: There is a statistically significant difference in \textit{having been involved in research projects during the participants' studies} when comparing participants who choose to do a PhD (\faUserGraduate) and those who do not (\faUser).
	\end{footnotesize}
\end{tcolorbox}

The data collected with these questions meets the following pre-conditions: (i) the variables are categorical, (ii) all observations are independent, (iii) the cells in the contingency table (see table \ref{tab:Chi-SquareBackground}) are mutually exclusive, and (iv) \oldtext{and} the expected value of cells should be 5 or greater in at least 80\% of cells. We performed a chi-square test of independence to examine the relation between the answer to each of those questions and the choice of our respondents to do a PhD, or not.

Table \ref{tab:Chi-SquareBackground} summarises the results of the Chi-Square test for these 5 questions\newtext{. The columns represent the question identification, the answer, the number of respondents who decided not to do a PhD ($\#$\faUser), the percentage of those respondents who provided that answer ($\%$\faUser), the number ($\#$\faUserGraduate) and percentage ($\%$\faUserGraduate) of participants who decided to do a PhD that provided each answer, the Chi-Square test value (\textbf{$\chi$\textsuperscript{2}(1)}), the p-value for that test}  highlighting the statistically significant results ($p < .05$) and their corresponding effect size $\varphi_{c}$\newtext{, a pictorial representation of the statistical significance of any identified differences, their direction and effect size (\faUser vs \faUserGraduate) and total number of respondents}. 
\oldtext{We further illustrate this when contrasting the distributions of answers from participants who decided to do a PhD with those who decided against doing a PhD (see the column with heading \faUserGraduate~\textbf{\textit{vs}} \faUser).}
\newtext{In the column \faUser vs \faUserGraduate~we} \oldtext{We} represent these differences from the perspective of the PhD respondent (\faUserGraduate). 
\oldtext{The solid-up chevron (\faChevronCircleUp) represents a higher level of agreement in the answers provided by participants who decided to do a PhD in Computer Science. The number of chevrons can vary from one (small effect size) to three (large effect size). We use a lighter chevron (\faChevronUp) to represent effect sizes on differences that are \textit{statistically significant when analysed in isolation}, but not when using the \underline{\textit{Holm-Bonferroni}}, or the \textbf{\underline{\textit{Bonferroni}}} corrections.} 
\newtext{With 5 tests, the Bonferroni-corrected significance level is $p < 0.01$, while the Holm-Bonferroni-corrected significance levels progress from $p < 0.01$ to $p < 0.05$, as explained in Section \ref{subsec:ApproachQuantitativeAnalysis}. In the case of Q10, the p-value of {\ul\textbf{\textit{.000}}} (the actual value is {\ul\textbf{\textit{1.15624939375694E-16}}}) is significant even with the {\ul\textbf{\textit{Bonferroni  correction}}} (and the less strict {\ul\textit{Holm-Bonferroni correction}}).}

\begin{table}[htb]
	\caption{\oldtext{Chi-Square test for student education experience questions.} \newtext{How does the student education experience relate to the decision of pursuing (\faUserGraduate) or not (\faUser) a PhD? Note: Chi-Square test. The used Bonferroni-corrected significance level is $p<.01$.}}
	\label{tab:Chi-SquareBackground}
	\resizebox{\columnwidth}{!}{%
		\begin{tabular}{@{}lllrrrrllll@{}}
			\toprule
			\textbf{Id} & \textbf{Answer} & \textbf{$\#$\faUser} & \textbf{\%}\faUser & \textbf{$\#$\faUserGraduate} & \textbf{\%\faUserGraduate} & \textbf{$\chi$\textsuperscript{2}(1)} & \textbf{p-value} & \textbf{$\varphi_{c}$} & \faUserGraduate~\textbf{\textit{vs}} \faUser & \textbf{N}\\ \midrule 
			Q8 & No  & 154 & 41.5 & 78 & 38.4 & .519 & .471 & .030 & & 574 \\
			& Yes  & 217 & 58.5 & 125 & 61.6 & & & \\
			Q9  & No  & 164 & 44.2 & 72 & 35.3 & 4.319 & \textit{.038}& \textit{.087} &  \faUserGraduate \faChevronUp & 575\\
			& Yes & 207 & 55.8 & 132 & 64.7 & & & & &\\
			Q10 & No & 255 & 68.9 & 67 & 33.0 & 68.683 & {\ul\textbf{\textit{.000}}}& \textbf{\textit{.346}} & \faUserGraduate \faChevronCircleUp \faChevronCircleUp & 573\\
			& Yes  & 115 & 31.1 & 136 & 67.0 & & & & & \\
			Q11 & No & 191 & 51.8 & 42 & 20.7 & 52.372 & {\ul\textbf{\textit{.000}}}& \textbf{\textit{.303}} & \faUserGraduate \faChevronCircleUp \faChevronCircleUp & 572\\
			& Yes  & 178 & 48.2 & 161 & 79.3  & & & & & \\[.25cm]
			Q44 & No & 298 & 87.4 & 147 & 79.0 & 6.399 & {\ul\textit{.011}}& \textit{.110} & \faUserGraduate \faChevronUp  & 527\\
			& Yes & 43 & 12.6 & 39 & 21.0 & & & & & \\\bottomrule
		\end{tabular}%
	}
\end{table}

\begin{figure}[htb!]
	\centering
	{\includegraphics[width=1.0\linewidth]{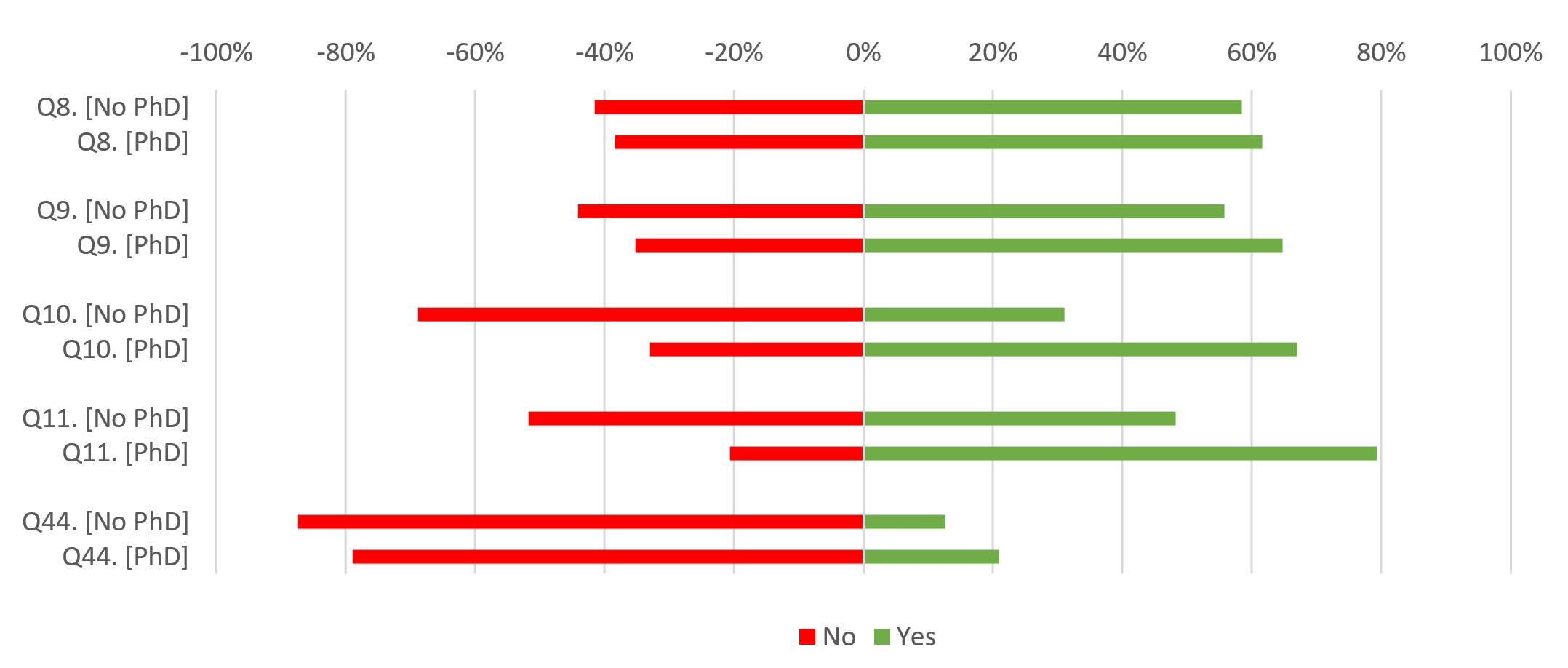}}
	\caption{Supporting factors for pursuing a PhD from single choice questions.}
	\label{fig:RQ1SupportingFactorsSingleChoice}
\end{figure}

For question Q8, the relation between these variables was not significant. The proportion of subjects who decided to do a PhD did not differ by having any computer science courses during high school (Q8). 
For questions Q9, Q10, Q11, and Q44 on the other hand, the relation between these variables was statistically significant (Q10 and Q11 even when using the Holmes-Bonferroni, or the stricter Bonferroni correction\newtext{, while Q44 remained statistically significant with the Holm-Bonferroni correction, but not the Bonferroni correction}). 
Participants who were involved in research projects during their undergraduate studies, and/or personally supervised by a research leader were more likely to enrol in a PhD, in both cases with a medium effect size (Q10 and Q11, respectively). This effect is small for participants who were personally supervised by a more junior researcher, such as a PhD student, or a teaching assistant (Q9).
Participants who attended programs or activities that informed about PhDs were more likely to enrol in a PhD, again with a small effect size (Q44). %
Figure \ref{fig:RQ1SupportingFactorsSingleChoice} summarises these results. 

\begin{insight}{Research involvement and PhD life information}
\label{insight:phd_life_info}
Participants who were involved in a research project during their undergraduate, or MSc studies, particularly when directly supervised by a senior researcher, were more likely to enrol on a PhD. Participants who attended activities aimed at informing of life as a PhD were also more likely to enrol in a PhD.
\end{insight}

\paragraph{(ii) Participants' preferences and self-confidence}

Questions Q12 through Q20 focus mostly on the preferences expressed by our respondents concerning Computer Science as an area. The overarching rationale behind these questions is that PhD students often express their enthusiasm for the research topic they are working on. As such, we hypothesised participants who choose to do a PhD would like Computer Science topics more than their peers not pursuing a PhD, and feel more confident about their skills. For example, for Q14, the hypotheses would be:

\begin{tcolorbox}
	\begin{footnotesize}
		$H_{14-0}$: There is no statistically significant difference in the \textit{perception of having good grades}, when comparing participants who choose to do a PhD (\faUserGraduate) and those who do not (\faUser).
		
		$H_{14-1}$: There is a statistically significant difference in the \textit{perception of having good grades}, when comparing participants who choose to do a PhD (\faUserGraduate) and those who do not (\faUser).
	\end{footnotesize}
\end{tcolorbox}

We created similar hypotheses for the remaining questions. Table \ref{tab:Mann-WhitneyUPhDNoPhDQ12-Q20} summarises the collected data for all these Likert-scale questions. The data collected with these questions meets the following criteria: (i) the variables are ordinal, (ii) the observations are independent, and (iii) the shapes of the distributions are roughly the same. With these pre-conditions met, we used a Mann-Whitney U test to assess these hypotheses. \newtext{The same set of preconditions for the Mann-Whitney U test were checked throughout the paper for all these tests. In particular, these preconditions were checked for the tests summarised in Tables \ref{tab:Mann-WhitneyUPhDNoPhDQ12-Q20}, \ref{tab:Mann-WhitneyUPhDNoPhDQ22-Q30}, \ref{tab:Mann-WhitneyUPhDNoPhDQ31-Q35}, \ref{tab:Mann-WhitneyUPhDNoPhDQ47-Q54}. While we present these tables separately, we computed the Holm-Bonferroni and Bonferroni corrections to the tests summarised in these tables altogether. These 38 questions lead to a Bonferroni corrected significance level of $p < .0013157895$, while the Holm-Bonferroni corrected significance level ranges from $p < .0013157895$ to $p < 0.05$.} 

Table \ref{tab:Mann-WhitneyUPhDNoPhDQ12-Q20} presents the results including the question number, 
the Asymptotic 2-tailed significance level for the Mann-Whitney U tests, the effect size ($\eta^{2}$), a visual representation of the difference between the distributions in participants who decided to do a PhD and those who did not (\faUserGraduate~\textbf{\textit{vs}} \faUser), the mean (\textbf{$\bar{x}$}), the mean for those who plan to do a PhD, are doing a PhD, or have done a PhD (\textbf{$\bar{x}$}(\faUserGraduate)), and for those who don't (\textbf{$\bar{x}$}(\faUser)), median (\textbf{$\tilde{x}$}), median for those who plan to do a PhD (\textbf{$\tilde{x}$(\faUserGraduate)}) and for those who don't (\textbf{$\tilde{x}$}(\faUser)), mode and the number of cases. Note that, as participants had the opportunity to select the option \textit{``does not apply''} for any of these questions, the number of valid responses varies from one question to the next. \oldtext{We highlight the asymptotic 2-tailed significance for those questions which are \textit{statistically significant} even when using a \underline{\textit{Holm-Bonferroni correction}} and using a \textbf{\underline{\textit{Bonferroni correction}}}.} 

\begin{table}[htb!]
	\caption{How do participants' enthusiasm and self-confidence relate to the decision of pursuing (\faUserGraduate) or not (\faUser) a PhD? \newtext{Note: Mann-Whitney U test. The used Bonferroni-corrected significance level is $p<.0013157895$.}}
	\label{tab:Mann-WhitneyUPhDNoPhDQ12-Q20}
	\resizebox{\columnwidth}{!}{%
		\begin{tabular}{@{}llllllllllll@{}}
			\toprule
			\textbf{Id} & 
            \textit{\textbf{p}} & $\eta^{2}$ & \faUserGraduate~\textbf{\textit{vs}} \faUser &\textbf{$\bar{x}$} & \textbf{$\bar{x}$}(\faUserGraduate) & \textbf{$\bar{x}$}(\faUser) & \textbf{$\tilde{x}$} & \textbf{$\tilde{x}$(\faUserGraduate)} & \textbf{$\tilde{x}$}(\faUser) & \textbf{Mode} & \textbf{N} \\ \midrule
			Q12 
   & {\ul \textit{\textbf{.000}}} & .056 & \faUserGraduate \faChevronCircleUp
			& 4.486 & 4.704 & 4.367 & 5 & 5 & 5 & 5 & 587\\
			Q13 
   & {\ul \textit{\textbf{.000}}} & .043 & \faUserGraduate \faChevronCircleUp
			& 4.255 & 4.487 & 4.134 & 4 & 5 & 4 & 5 & 577\\
			Q14 
   & {\ul \textit{\textbf{.000}}} & .072 & \faUserGraduate \faChevronCircleUp \faChevronCircleUp
			& 3.953 & 4.316 & 3.766 & 4 & 4 & 4 & 4 & 576\\
			Q15 
   & {\ul \textit{\textbf{.000}}} & .043 & \faUserGraduate \faChevronCircleUp
			& 4.038 & 4.289 & 3.905 & 4 & 4 & 4 & 4 & 580\\
			Q16 
   & {\ul \textit{\textbf{.001}}} & .021 & \faUserGraduate \faChevronCircleUp
			& 3.762 & 4.000 & 3.637 & 4 & 4 & 4 & 4 & 580\\
			Q17 
   & .799 & .000 & & 3.289 & 3.289 & 3.290 & 3 & 3 & 3 & 4 & 570\\
			Q18 
   & {\ul \textit{\textbf{.000}}} & .039 & \faUserGraduate \faChevronCircleUp
			& 3.732 & 4.030 & 3.573 & 4 & 4 & 4 & 5 & 582\\
			Q19 
   & .895 & .000 & & 4.227 & 4.216 & 4.233 & 4 & 4 &  4 & 5 & 582\\
			Q20 
   & {\ul \textit{\textbf{.002}}} & .016 & \faUserGraduate \faChevronCircleUp
			& 4.161 & 4.291 & 4.090 & 4 & 5 & 4 & 5 & 583\\ \bottomrule
		\end{tabular}%
	}
\end{table}

\begin{figure}[htb!]
	\centering
	{\includegraphics[width=1.0\columnwidth]{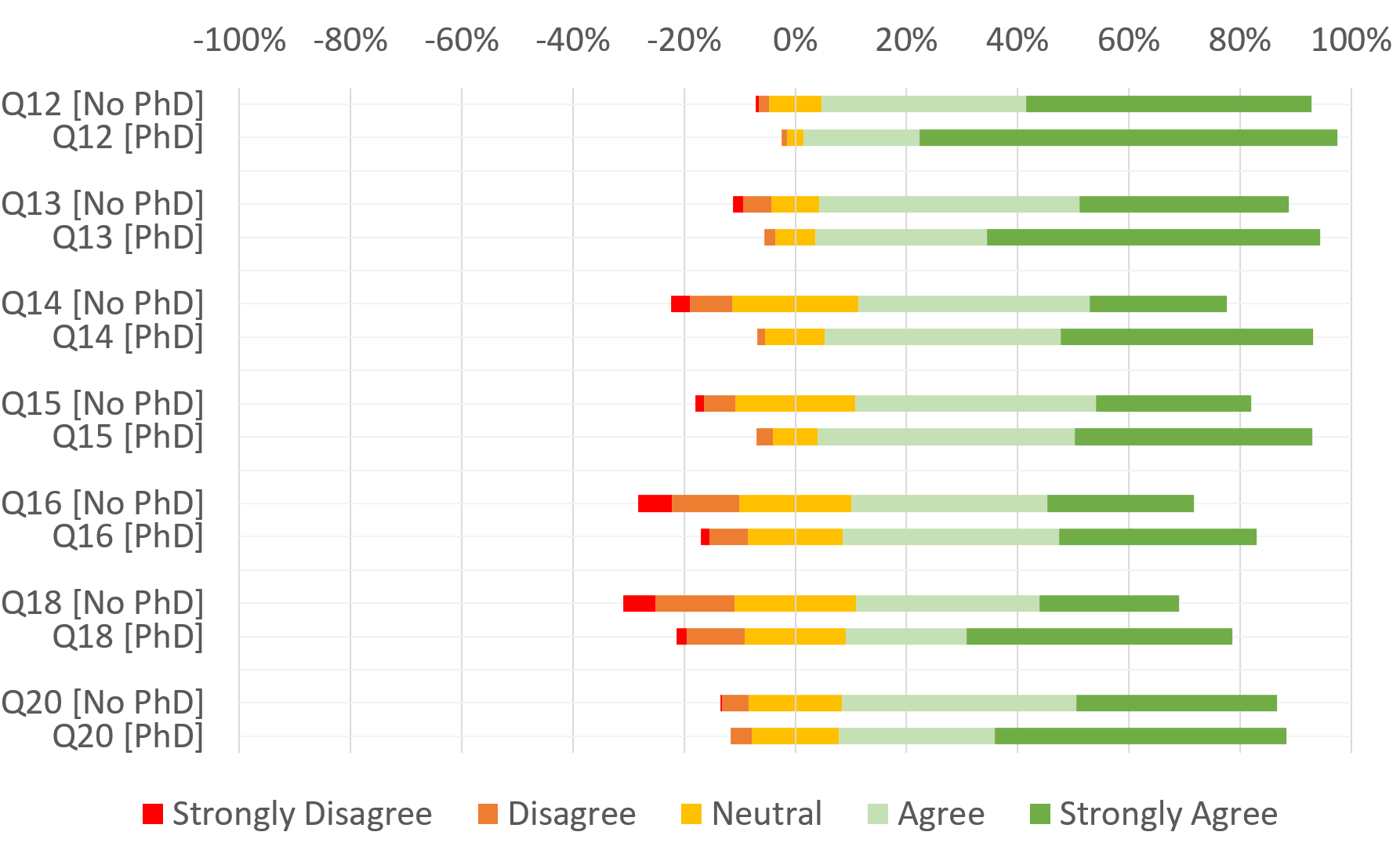}}
	\caption{The distributions of responses on enthusiasm and self-confidence, contrasting participants who decided not to do a PhD ([No PhD] \faUser) with those who decided to do one ([PhD] \faUserGraduate). \newtext{Questions for which there were no statistically significant differences were filtered out.}}
	\label{fig:RQ1_Q14_SupportingFactors}
\end{figure}

We found statistically significant differences for most questions, although effect sizes tend to be small. In all cases where we found statistically significant differences, participants who had decided to do a PhD had a higher level of agreement with the factors than those who decided not to do a PhD. Figure \ref{fig:RQ1_Q14_SupportingFactors} presents these distributions. 

The only medium effect size observed in this set of questions was on question Q14. A Mann-Whitney test indicated that the perception of having good grades was greater for participants who decided to perform a PhD than for those who didn't, with a medium effect size. There were several other statistically significant differences, all with a small effect size, and all with higher levels of agreement among participants who decided to do a PhD. These include feeling enthusiasm for Computer Science (Q12), enjoying one's studies (Q13), feeling successful in those studies (Q15), feeling self-confident in one's interactions with other students (Q16), and liking theoretical computer science (Q20) and interdisciplinary areas of computer science (Q20). In contrast, We did not find statistically significant differences in questions Q17 (whether participants were involved in study groups) and Q19 (whether participants enjoyed practical computer science subjects, such as programming).

\begin{insight}{Enthusiasm and self-confidence}
\label{insight:enthusiasm}
	Participants who decided to do a PhD in Computer Science feel more enthusiasm for Computer Science in general, and theoretical computer science and interdisciplinary studies in particular. They also enjoy their studies more than other participants. In addition, they express a higher self-confidence in the success of their studies, and their interaction with their peers.
\end{insight}

\paragraph{(iii) Participants' intentions and context}
Questions Q22, Q23 and Q25 through Q30 focused on the participant's current situation, particularly concerning the participants' views on the possibility of pursuing a PhD. We followed a similar data analysis approach to the one above  (subsection \ref{subsec:RQ1}(ii)). Our hypotheses follow the same structure as previously. 

Most of these questions had statistically significant results for the Mann-Whitney U test we performed, with effect sizes ranging from small to large, suggesting differences between participants who decided to do a PhD and those who did not. 
Only the responses to Q22 and Q28 were not statistically significant. 
Table \ref{tab:Mann-WhitneyUPhDNoPhDQ22-Q30} summarises these results.

\begin{table}[htb!]
	\caption{How do participants' intentions and other context variables relate to the decision of pursuing (\faUserGraduate) or not (\faUser) a PhD? \newtext{Note: Mann-Whitney U test. The used Bonferroni-corrected significance level is $p<.0013157895$.}}
	\label{tab:Mann-WhitneyUPhDNoPhDQ22-Q30}
	\resizebox{\columnwidth}{!}{%
		\begin{tabular}{@{}llllllllllll@{}}
			\toprule
			\textbf{Id} 
   & \textbf{\textit{p}} & $\eta^{2}$ & \faUserGraduate~\textbf{\textit{vs}} \faUser &\textbf{$\bar{x}$} & \textbf{$\bar{x}$}(\faUserGraduate) & \textbf{$\bar{x}$}(\faUser) & \textbf{$\tilde{x}$} & \textbf{$\tilde{x}$}(\faUserGraduate) & \textbf{$\tilde{x}$}(\faUser) & \textbf{Mode} & \textbf{N} \\ \midrule
			Q22 
   & .234 & .003 &
			& 4.366 & 4.404 & 4.545 & 5 & 5 & 5 & 5 & 530\\
			Q23 
   & {\ul \textit{\textbf{.000}}} & .057 & \faUserGraduate \faChevronCircleUp
			& 3.543 & 3.995 & 3.301 & 4 & 5 & 4 & 5 & 529\\
			Q25 
   & {\ul \textit{\textbf{.000}}} & .065 & \faUserGraduate \faChevronCircleUp \faChevronCircleUp
			& 3.417 & 3.849 & 3.177 & 4 & 4 & 3 & 3 & 518\\
			Q26 
   & {\ul \textit{\textbf{.000}}} & .100 & \faUserGraduate \faChevronCircleUp \faChevronCircleUp
			& 3.340 & 3.930 & 3.018 & 4 & 4 & 3 & 4 & 523\\
			Q27 
   & {\ul \textit{\textbf{.000}}} & .179 & \faUserGraduate \faChevronCircleUp \faChevronCircleUp \faChevronCircleUp
			& 3.050 & 3.877 & 2.591 & 3 & 4 & 3 & 1 & 524\\
			Q28 
   & .544 & .001 &
			& 3.052 & 3.000 & 3.080 & 3 & 3 & 3 & 3 & 498\\
			Q29 
   & {\ul \textit{\textbf{.000}}} & .101 & \faUserGraduate \faChevronCircleUp \faChevronCircleUp
			& 2.834 & 3.431 & 2.502 & 3 & 4 & 3 & 3 & 506\\
			Q30 
   & .\textit{016} & .012 & \faUserGraduate \faChevronUp
			& 1.726 & 1.900 & 1.628 & 1 & 1 & 1 & 1 &503\\
			\bottomrule
		\end{tabular}%
	}
\end{table}

\begin{figure}[htb!]
	\centering
	{\includegraphics[width=1.0\columnwidth]{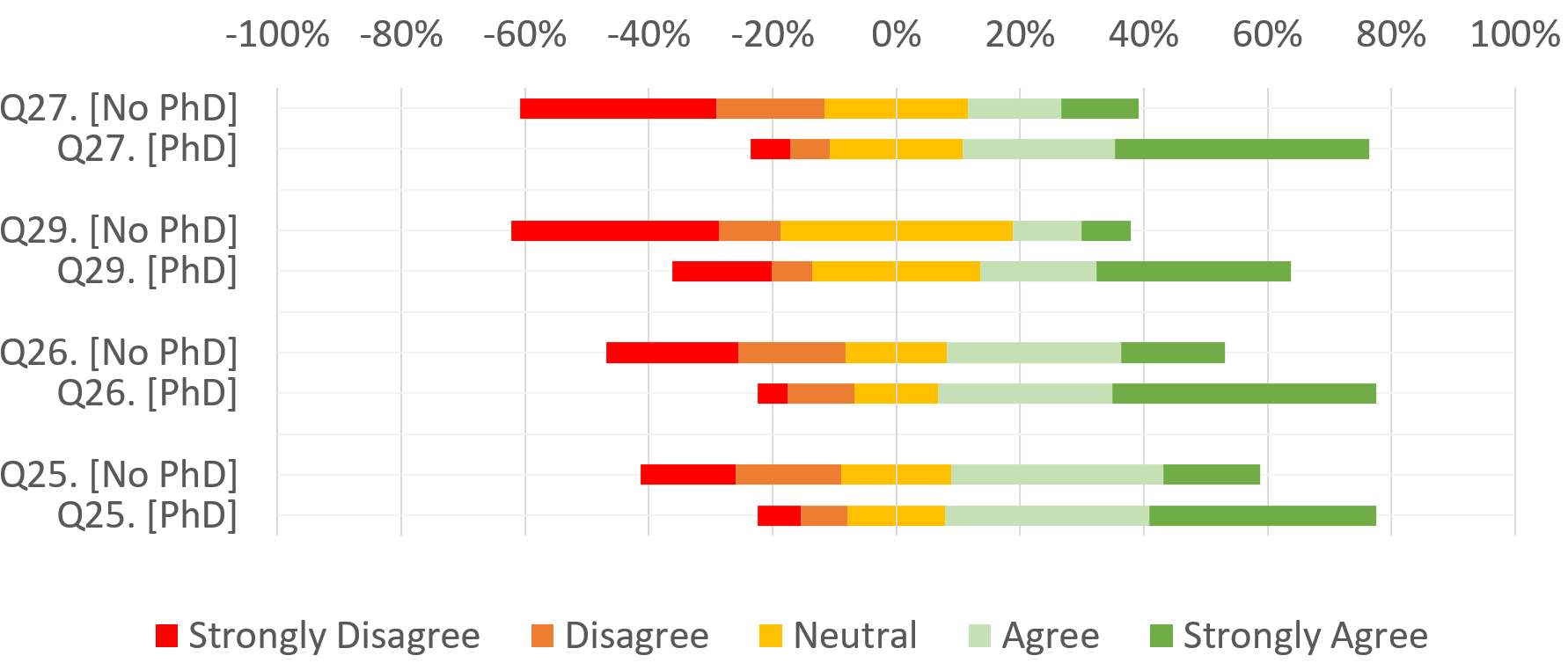}}
	\caption{The distribution of responses to questions related to participants' intent and context factors for those questions with statistically significant differences, with medium effect sizes, from those who chose to do a PhD and those who didn't.}
	\label{fig:RQ1_Q27_SupportingFactors}
\end{figure}

Participants who decided to do a PhD had a higher perception of having been encouraged to do so than the others, and this difference had a large effect size (Q27). 
A notable feature of these answers is that the mode in the group of participants that did not decide to do a PhD is 1, the lowest possible value on this scale.
We observed a similar relation concerning participants' perception of having been encouraged to do a PhD by their family, albeit with a medium effect size (Q29).
The awareness of requirements for starting a PhD, as well as of the aims and contents of PhD studies, were also higher for those who decided to start a PhD than for the other participants, with medium effect sizes (Q26 and Q25, respectively). 
Questions Q23 (having thought about doing a PhD) and Q30 (having been encouraged by industrial contacts to start a PhD) also had statistically significant results (the latter only when considered in isolation), albeit with small effect sizes. A distinguishing factor about Q30 is that it had the overall lowest agreement scores in the whole questionnaire. 
Figure \ref{fig:RQ1_Q27_SupportingFactors} illustrates this distribution. 

\begin{insight}{Encouragement and requirements}
\label{insight:encouragement_requirements}
	Participants encouraged to do a PhD were much more likely to enrol in one. Having such an incentive from one's family, or by an industrial contact is also relevant, but the latter is very uncommon. Our data also shows how many participants were \oldtext{not aware} \newtext{unaware} of the requirements for starting a PhD, and how being aware of such requirements can make a difference. 
\end{insight}

\paragraph{(iv) Arguments in favour of enrolling in a PhD}
Questions Q31 through Q35 \newtext{aim to assess} \oldtext{are aimed at assessing} our participants' perceptions concerning supporting factors that might encourage students to enrol in a PhD program. We followed the same analysis strategy \oldtext{from} \newtext{as in} the previous questions set. The hypotheses associated with each of these questions are defined \oldtext{in the same way} \newtext{similarly}.

Our test results are summarised in Table \ref{tab:Mann-WhitneyUPhDNoPhDQ31-Q35}. Participants who decided to do a PhD perceive the aims and contents suit them better (Q31), are more convinced they meet the requirements (Q32) and have a higher perception that a PhD will improve their future career (Q33) than those who decided not to start a PhD. All these differences have a large effect size.  

\begin{figure}[htb!]
	\centering
	{\includegraphics[width=1.0\columnwidth]{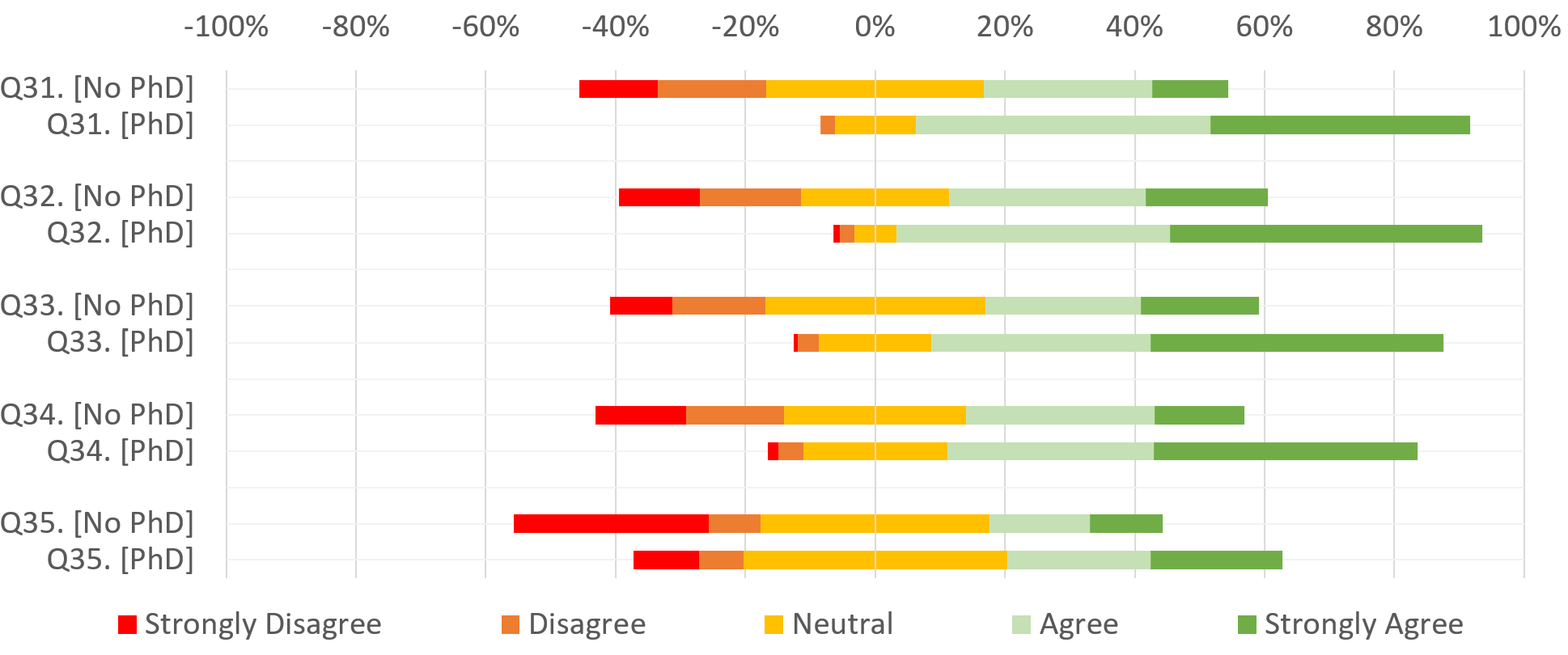}}
	\caption{The distribution of responses to questions related to participants' perceptions of supporting factors in favour of PhD studies from those who choose to do a PhD and those who didn't.}
	\label{fig:RQ31_Q35_Likert}
\end{figure}

Participants who decided to do a PhD are more likely to perceive it as fun 
than those who didn't 
with a medium effect size 
(Q34). Finally, participants who decided to do a PhD generally agree that better gender balance in Computer Science would increase their intentions of doing a PhD 
more than those who didn't, with a small effect size. 
Figure \ref{fig:RQ31_Q35_Likert} illustrates these differences.

\begin{table}[htb!]
	\caption{How do arguments in favour of starting a PhD relate to the decision of pursuing (\faUserGraduate) or not (\faUser) a PhD? \newtext{Note: Mann-Whitney U test. The used Bonferroni-corrected significance level is $p<.0013157895$.}}
	\label{tab:Mann-WhitneyUPhDNoPhDQ31-Q35}
	\resizebox{\columnwidth}{!}{%
		\begin{tabular}{@{}llllllllllll@{}}
			\toprule
			\textbf{Id} 
   & \textbf{\textit{p-value}} & $\eta^{2}$ & \faUserGraduate~\textbf{\textit{vs}} \faUser &\textbf{$\bar{x}$} & \textbf{$\bar{x}$}(\faUserGraduate) & \textbf{$\bar{x}$}(\faUser) & \textbf{$\tilde{x}$} & \textbf{$\tilde{x}$}(\faUserGraduate) & \textbf{$\tilde{x}$}(\faUser) & \textbf{Mode} & \textbf{N} \\ \midrule
			Q31 
   & {\ul \textit{\textbf{.000}}} & .234 & \faUserGraduate \faChevronCircleUp \faChevronCircleUp \faChevronCircleUp
			& 3.509 & 4.232 & 3.085 & 4 & 4 & 3 & 4 & 501\\
			Q32 
   & {\ul \textit{\textbf{.000}}} & .181 & \faUserGraduate \faChevronCircleUp \faChevronCircleUp \faChevronCircleUp
			& 3.663 & 4.341 & 3.272 & 4 & 4 & 3 & 4 & 505\\
			Q33 
   & {\ul \textit{\textbf{.000}}} & .146 & \faUserGraduate \faChevronCircleUp \faChevronCircleUp \faChevronCircleUp
			& 3.599 & 4.196 & 3.267 & 4 & 4 & 3 & 3 & 514\\
			Q34 
   & {\ul \textit{\textbf{.000}}} & .130 & \faUserGraduate \faChevronCircleUp \faChevronCircleUp 
			& 3.469 & 4.059 & 3.139 & 4 & 4 & 3 & 4 & 516\\
			Q35 
   & {\ul \textit{\textbf{.000}}} & .053 & \faUserGraduate \faChevronCircleUp
			& 2.930 & 3.360 & 2.697 & 3 & 3 & 3 & 3 & 516\\
			\bottomrule
		\end{tabular}%
	}
\end{table}

\begin{insight}{Perception of the PhD process}
\label{insight:percipation}
	Participants who chose to do a PhD were more inclined to perceive it as something that would fit their needs, improve their future careers, and be fun to do. They also felt more prepared to do so 
    and are more inclined to think that more gender balance would increase their intentions of doing a PhD.
\end{insight}

\paragraph{(v) Self-reflection on own strengths}
Questions Q47 through Q54 assess the participants' reflections on their strengths when compared to their peers. We follow the same analysis approach adopted in the previous sections. Table \ref{tab:Mann-WhitneyUPhDNoPhDQ47-Q54} summarises the results of the Mann-Whitney U tests run to test our hypotheses.

\begin{table}[htb!]
	\caption{How do participant's plans and other context variables relate to the decision of pursuing (\faUserGraduate) or not (\faUser) a PhD? \newtext{Note: Mann-Whitney U test. The used Bonferroni-corrected significance level is $p<.0013157895$.}}
	\label{tab:Mann-WhitneyUPhDNoPhDQ47-Q54}
	\resizebox{\columnwidth}{!}{%
		\begin{tabular}{@{}llllllllllll@{}}
			\toprule
			\textbf{Id} 
   & \textbf{p-value} & $\eta^{2}$ & \faUserGraduate~\textbf{\textit{vs}} \faUser &\textbf{$\bar{x}$} & \textbf{$\bar{x}$}(\faUserGraduate) & \textbf{$\bar{x}$}(\faUser) & \textbf{$\tilde{x}$} & \textbf{$\tilde{x}$}(\faUserGraduate) & \textbf{$\tilde{x}$}(\faUser) & \textbf{Mode} & \textbf{N} \\ \midrule
			Q47 
   & .152 & .004 &
			& 4.249 & 4.322 & 4.209 & 4 & 4 & 4 & 4 & 522\\
			Q48 
   & .080 & .006 &
			& 3.594 & 3.679 & 3.548 & 4 & 4 & 4 & 4 & 527\\
			Q49 
   & {\ul \textit{\textbf{.000}}} & .024 & \faUserGraduate \faChevronCircleUp
			& 4.030 & 4.212 & 3.933 & 4 & 4 & 4 & 4 & 527\\
			Q50 
   & .602 & .001 &
			& 3.640 & 3.701 & 3.607 & 4 & 4 & 4 & 4 & 525\\
			Q51 
   & .468 & .001 &
			& 4.249 & 4.271 & 4.237 & 4 & 4 & 4 & 4 & 518\\
			Q52 
   & {\ul \textit{\textbf{.000}}} & .032 & \faUserGraduate \faChevronCircleUp
			& 4.378 & 4.582 & 4.269 & 5 & 5 & 4 & 5 & 526\\
			Q53 
   & \textit{.019} & .011 & \faUserGraduate \faChevronUp
			& 4.088 & 4.202 & 4.026 & 4 & 4 & 4 & 4 & 525 \\
			Q54 
   & .068 & .006 &
			& 4.224 & 4.326 & 4.68 & 4 & 4 & 4 & 5& 523\\
			\bottomrule
		\end{tabular}%
	}
\end{table}

\begin{figure}[htb!]
	\centering
	{\includegraphics[width=1.0\columnwidth]{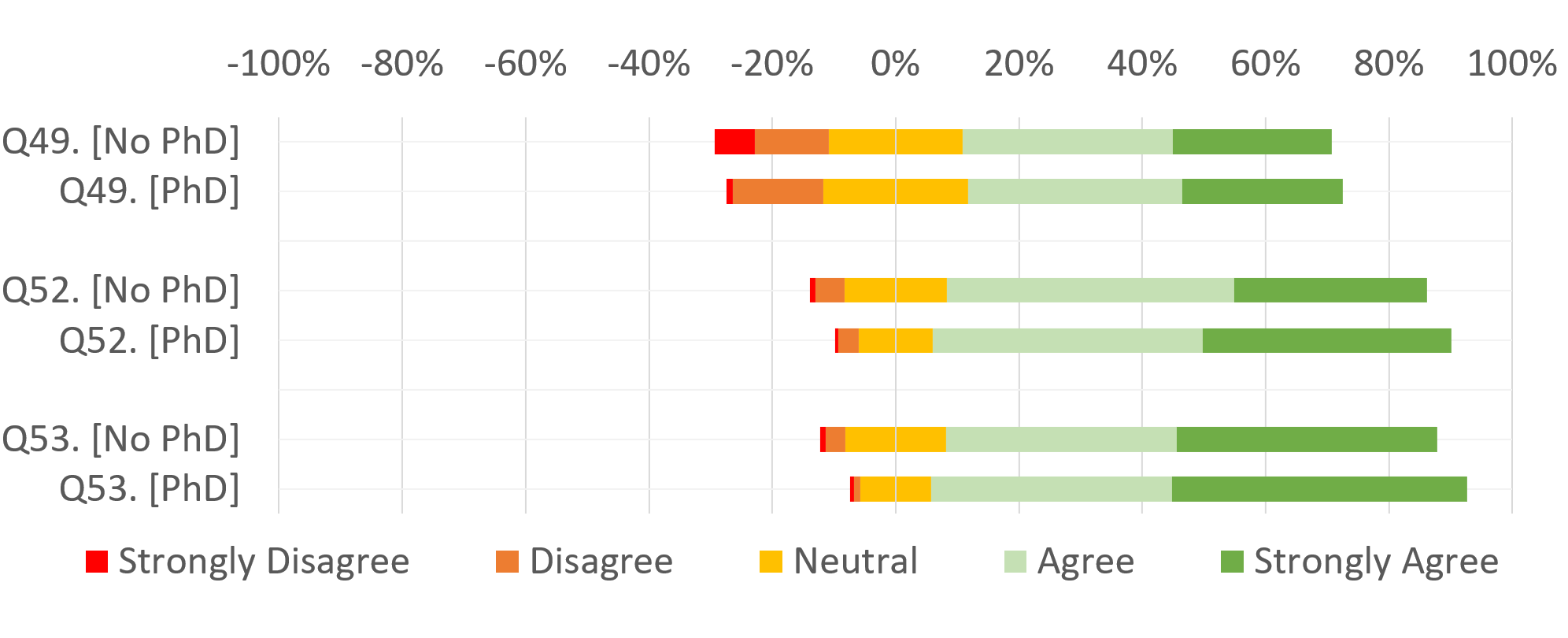}}
	\caption{The distribution of responses to questions related to participants' reflection on their strengths from those who chose to do a PhD and those who didn't. }
	\label{fig:RQ47_Q54_Likert}
\end{figure}

Participants who decided to do a PhD feel more driven by curiosity to learn (Q52), to have a better understanding of theoretical computer science 
(Q49),  
and more successful in collaborating with others 
than their peers who decided not to do a PhD 
(Q53). The results from Q53 are only statistically different when considering this question in isolation, but not when applying the Holm-Bonferroni, or the Bonferroni corrections. For the remaining questions, we found no evidence supporting an association with the decision to enrol in a PhD. Figure \ref{fig:RQ47_Q54_Likert} highlights the differences in these distributions.

\begin{insight}{Personal strengths}
\label{insight:personal_strengths}
    	Albeit with \oldtext{a} small effect size, a greater curiosity to learn, a better understanding of theoretical computer science, and the ability to collaborate with peers are perceived as more developed in people who \newtext{have} decided to do a PhD. We found no significant differences in the competencies to succeed in Computer Science, having strong programming skills, embracing problem-oriented learning, or loving computer science, as these are similarly likely to be found in participants who decided not to do a PhD. 
\end{insight}

\begin{figure}[htb!]
	\centering
	{\includegraphics[width=1\linewidth]{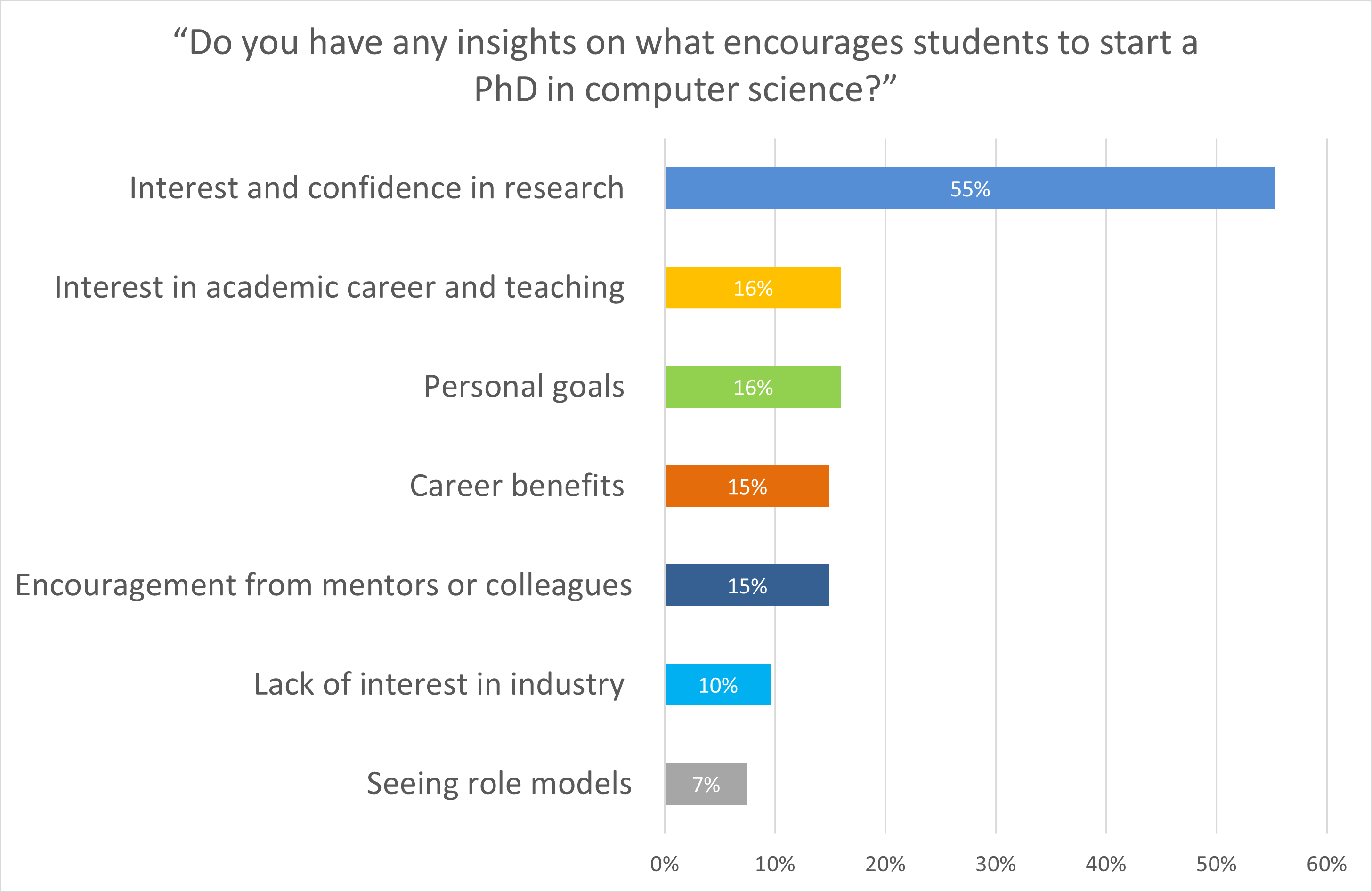}}
	\caption{Distribution of answers over themes \oldtext{for Q55} \newtext{identified in a qualitative analysis of the insights on encouraging factors for starting a PhD.}}
	\label{fig:fig1_open}
\end{figure}

\paragraph{(vi) Arguments for starting a PhD (qualitative analysis)}
The open question Q55 investigates the encouraging factors for pursuing a PhD. Table \ref{tab:themes1} presents the themes determined from the answers to this question, their explanations, a sample comment selected among responses assigned to that theme, and the total number of responses assigned to that theme. Note that, the unique survey participant ID is also reported in the sample comment column. Figure \ref{fig:fig1_open} illustrates this data in a bar chart to better visualise the distribution of the answers on this theme. \oldtext{Note that} Since a single response may have been assigned to more than one theme, the percentages are calculated by dividing the number of responses assigned for that theme by the total number of responses for that question. 

The most popular theme that was mentioned in 55\% of the responses is ``Interest and confidence in research'' as shown in Figure \ref{fig:fig1_open}. In many of these responses, this theme was mentioned as excitement, curiosity, and passion about learning and discovering new things and having a special interest in a particular research topic. Twelve respondents highlighted the positive impact of participating in research projects during bachelor or master studies in increasing interest and confidence in research. Some of these comments are:

\begin{table*}[!htbp]
	\centering
	\caption{Themes for Q55 - Do you have any insights on what encourages students to start a PhD in computer science?}
	\label{tab:themes1}
	\resizebox{\textwidth}{!}{%
		\begin{tabular}{@{}llllr@{}}
			\toprule
    \multicolumn{2}{l}{\textbf{Themes}}  & \textbf{Explanation}& \textbf{Sample Comment(s)}& \textbf{Nr. of } \\ 
			\textbf{} & \textbf{} & \textbf{} & \textbf{}& \textbf{comments} \\\midrule
			1 & Interest and confidence  & Strong interest in a particular research topic or field & P378 - \textit{“Interest in computer science  } & 104 \\
			& in research  &  of study was mentioned by many respondents as a  & \textit{in general, love for science and curiosity.”} & \\
			& & motivating factor for pursuing a PhD. This category && \\
			& &   also includes curiosity, passion for learning, and desire && \\
			& &   to solve complex problems. && \\
			2 & Personal goals & Factors such as independence, expertise, recognition, & P738 - \textit{``From my talk with 1 other student} &30 \\
			& & dignity and personal development were mentioned as &  \textit{that considered it, it was the idea of} & \\
			& &  motivations for pursuing a PhD. Some more detailed & \textit{actually doing something meaningful} & \\  
			& & personal reasons such as living abroad are also  &  \textit{with your life that meant the most to us.''} &\\
			& & categorised under this theme. & & \\
			3 & Interest in academic  & Becoming a university professor or having a career & P91 - \textit{"$\ldots$The prospect of teaching others}  &30 \\
			&  career and teaching & career in academia, and enjoying teaching   & \textit{and doing research as a career, if that }&  \\
			& & mentioned to be some of the motivations for some  &\textit{were my choice, would make me pursue } &  \\
			& & individuals to pursue a PhD. &\textit{a PhD in order to reach that further goal."} &  \\
			
			4 & Career benefits   & Some respondents mention that a PhD can have lots of& P82 - \textit{“Positive impact on career, having} &28 \\
			& and opportunities & benefits for their career, providing more independence   &\textit{ more independence on what to conduct} & \\
			& &  and variety in their choices and potentially leading to &\textit{ own research in (compared to industrial } & \\
			& &  better job opportunities.  &\textit{ jobs)”}& \\
			
			5 & Encouragement from  & Some respondents mention that  encouragement from &P232 - \textit{"A personal invitation from the } & 28 \\
			&  mentors or colleagues&  mentors or their colleagues would play a significant & \textit{professors is a big encouragement and } &\\
			& &  role in the decision to pursue a PhD.&\textit{often a deciding factor."}& \\
			6 & Lack of interest in & Some respondents mentioned that the fast-paced  and&P581 - \textit{"A friend of mine started a PhD$\ldots$ }& 17 \\
			& industry jobs &  competitive nature of industry work would influence& \textit{wasn't enthusiast of the opportunities}  &\\
			& &    their decision to pursue a PhD.  &\textit{offered by companies."} & \\
			7 & Seeing role models & Some respondents state that seeing or working with &P421 - \textit{“My professor was my role model.} & 14 \\
			& & successful role models during their studies is a big &\textit{ The way she included us to the projects}& \\
			& & motivation for them on their decisions to pursue a PhD.&\textit{  encourage me a lot.” }& \\
			\bottomrule
		\end{tabular}%
	}
\end{table*}

\begin{itemize}
	\item P320 - \textit{``Being part of real research projects in their courses and projects play a big role I think.''}
	\item P682 - \textit{``$\ldots$ A research project during bachelor or master studies could show students what it looks like and feels like to be a researcher.''}
	\item P835 - \textit{``$\ldots$ Educate them about what research and PhD are and that might help them to consider enrolling. This can be by including them in research projects, papers, providing research methodology as a course for them.''}
\end{itemize}

For some themes, the gender distributions of the respondents are noticeably unbalanced. For example, the theme ``Seeing role models'' was mentioned by only 2 male respondents, whereas all the remaining answers were given by female respondents or the respondents who identified themselves as \textit{``other''} and also the ones who did not want to disclose their gender. Similarly, 71\% of the respondents mentioned the theme \textit{``Encouragement from professor or colleagues''}, and 78\% of the respondents of the theme \textit{``Lack of interest in the industry''} were females or other genders different than males.

The responses to this open question provide valuable insights into increasing the enrolment of students pursuing their PhDs. For the encouraging factors, motivating and supporting bachelor and master students with activities that help them to feel excited and confident about doing research may have the most significant impact on encouraging students to pursue a PhD. It is also essential to show them the career benefits of having a PhD and its potential to help in personal and professional development.

\subsection{Analysis of RQ2: What main blocking factors discourage enrolment in a PhD program?} 
\label{subsec:RQ2}
For the analysis of RQ2, we included (i) the quantitative analysis of a set of closed questions (Q36 through Q43) that implied candidate arguments against starting a PhD in Computer Science, and (ii) the qualitative analysis of one open question (Q56) for participants to identify arguments against doing a PhD.

\paragraph{(i) Arguments against doing a PhD (quantitative analysis)}
We performed a similar analysis to that done in RQ1 but with a focus on participants' perceptions of blocking factors that can hinder the likelihood of prospective PhD candidates doing a PhD. The structure of the hypotheses remains the same as for RQ1. For instance, the hypothesis associated with Q39 is as follows:

\begin{tcolorbox}
	\begin{footnotesize}
		$H_{39-0}$: There is no statistically significant difference in \textit{the participant being hesitant to start a PhD due to perceiving some factors and circumstances are unclear} when comparing participants who choose to do a PhD (\faUserGraduate) and those who do not (\faUser).
		
		$H_{39-1}$: There is a statistically significant difference in \textit{the participant being hesitant to start a PhD due to perceiving some factors and circumstances are unclear} when comparing participants who choose to do a PhD (\faUserGraduate) and those who do not (\faUser).
	\end{footnotesize}
\end{tcolorbox}

We performed Mann-Whitney U
tests on the distribution of responses to each of these questions given by those who chose a PhD and those who didn't (see Table \ref{tab:DiscouragingFactorsForPhD}). 

\begin{table}[htb!]
	\caption{\oldtext{Discouraging factors for doing a PhD}\newtext{How do arguments against starting a PhD relate to the decision of pursuing (\faUserGraduate) or not (\faUser) a PhD? Note: Mann-Whitney U test. The used Bonferroni-corrected significance level is $p < .0013157895$}}
	\label{tab:DiscouragingFactorsForPhD}
	\resizebox{\columnwidth}{!}{%
		\begin{tabular}{@{}llllllllllll@{}}
			\toprule
			\textbf{Id} 
   & \textbf{\textit{p-value}} & $\eta^{2}$ & \faUserGraduate~\textbf{\textit{vs}} \faUser &\textbf{$\bar{x}$} & \textbf{$\bar{x}$}(\faUserGraduate) & \textbf{$\bar{x}$}(\faUser) & \textbf{$\tilde{x}$} & \textbf{$\tilde{x}$}(\faUserGraduate) & \textbf{$\tilde{x}$}(\faUser) & \textbf{Mode} & \textbf{N} \\ \midrule
			Q36 
   & .612 & .001 & & 2.606 & 2.628 & 2.594 & 3 & 3 & 2 & 1 & 510\\
			Q37 
   & .076 & .006 & & 2.930 & 2.790 & 3.006 & 3 & 3 & 3 & 4 & 512\\
			Q38 
   & .063 & .007 & & 3.947 & 3.820 & 4.018 & 4 & 4 & 4 & 5 & 511\\
			Q39 
   & {\ul \textit{\textbf{.000}}} & .075 & \faUserGraduate \faChevronCircleDown \faChevronCircleDown & 3.120 & 2.663 & 3.365 & 3 & 3 & 3 & 3 & 493\\
			Q40 
   & {\ul \textit{\textbf{.000}}} & .052 & \faUserGraduate \faChevronCircleDown & 3.637 & 3.262 & 3.847 & 4 & 3 & 4 & 4 & 509\\
			Q41 
   & .786 & .000 & & 3.000 & 3.022 & 2.987 & 3 & 3 & 4 & 4 & 497\\ 
			Q42 
   & \textit{.014} & .013 & \faUserGraduate \faChevronDown & 2.926 & 2.744 & 3.034 & 3 & 3 & 3 & 3 & 473\\
			Q43 
   & .626 & .001 & & 2.241 & 2.118 & 2.311 & 2 & 2 & 2 & 1 & 257\\ 
			\bottomrule
		\end{tabular}%
	}
\end{table}

\begin{figure}[htb!]
	\centering
	{\includegraphics[width=1.0\columnwidth]{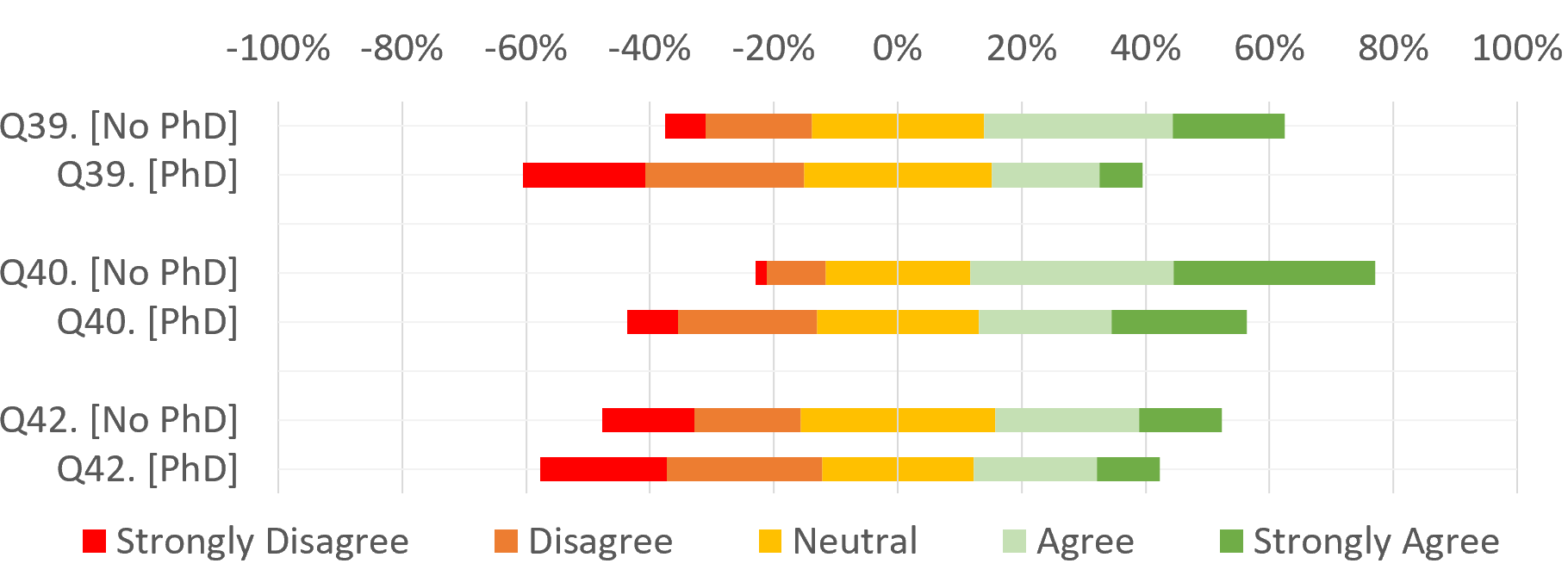}}
 \vspace{-5mm}
	\caption{The distribution of responses to questions related to participants' perceptions of discouraging factors for pursuing a PhD with statistically significant differences between those who chose to do a PhD and those who didn't.}
	\label{fig:DiscouragingFactors}
\end{figure}

Participants who decided to do a PhD feel there are less unclear factors and circumstances associated with a PhD (Q39), 
with a medium effect size. 
They also have a lower agreement with the perception that there are more flexible career opportunities in the industry 
(Q40), 
and are less afraid of making long-term job commitments for the whole duration of their PhD studies (Q42) 
then their peers who decided not to do a PhD, 
both with a small effect size. 
The results from Q42 are only statistically significant when considering this question in isolation, but not when applying the Holm-Bonferroni, or the Bonferroni corrections. We did not find evidence supporting an association with the decision to pursue a PhD for the remaining questions in this subset. 
Figure \ref{fig:DiscouragingFactors} illustrates this distribution.

\begin{insight}{Uncertainty}
\label{insight:uncertainty}
     Uncertainty about what it is like to do a PhD was the strongest discouraging factor for doing it, closely followed by a perception of lower job flexibility by participants who decided against starting a PhD. The long-term job commitment implied by starting a PhD may be a deterrent factor for some.
\end{insight}

\paragraph{(ii) Arguments against starting a PhD (qualitative analysis)}

\begin{figure}[htb]
	\centering
	{\includegraphics[width=1\columnwidth]{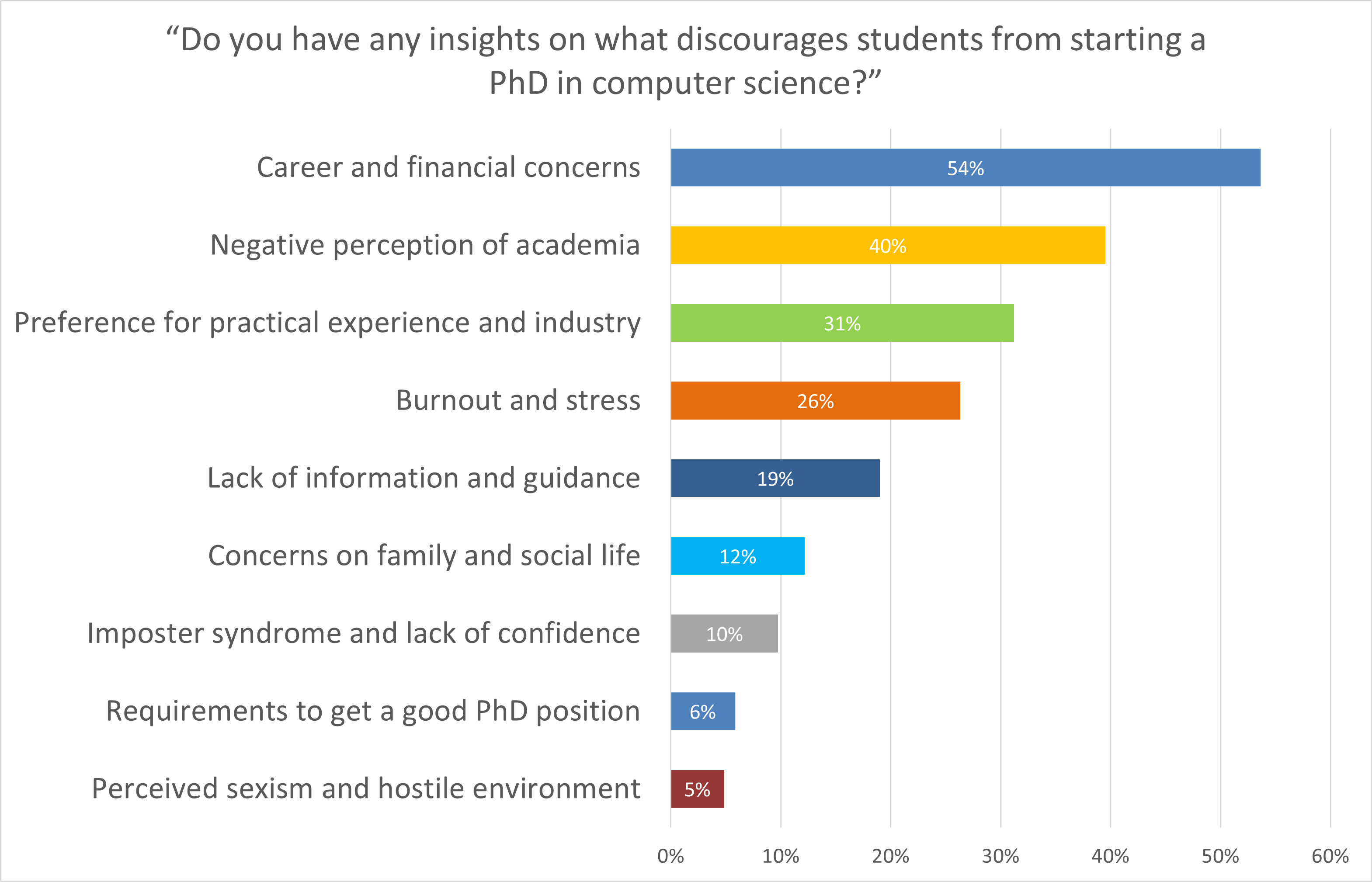}}
	\caption{Distribution of answers over themes \oldtext{for Q56} \newtext{identified in a qualitative analysis of the insights on discouraging factors for starting a PhD.}}
	\label{fig:fig2_open}
\end{figure}

The open question Q56, seeks the discouraging factors for pursuing a PhD and has more responses than the open question about encouraging factors. There is also more variety in the factors mentioned for this question. Nine different themes were identified to categorise these responses. Table \ref{tab:themes2} presents these themes and Figure \ref{fig:fig2_open} shows the distribution of the responses across these themes. 

More than half of the participants (Figure~\ref{fig:fig2_open}) who answered this question mentioned the uncertainty in the benefits of pursuing a PhD in their future careers and the low salaries during and after PhD studies compared to alternative jobs in industry. This is followed by the answers stating the negative perception of PhD studies. \oldtext{The majority of the} \newtext{Most} comments under this theme mention the long study duration and seeing this time as a ``waste of time'' for this fast-paced sector. The third most mentioned theme is the preference for practical experience where \oldtext{the majority of the} \newtext{most} respondents, instead of stating their excitement for working in the industry, mentioned their financial concerns about working in academia and the benefits of industry positions. 65\% of the respondents who mentioned their preference for practical experience also mentioned the higher salaries in industry \oldtext{as} compared to academia.

An interesting observation for this question is that the theme ``Perceived sexism and hostile environment'' was mentioned by only the female respondents or the respondents who identify themselves as ``other'' gender or who do not want to disclose their gender.

There is a comment worth mentioning regarding sexism in academia being a discouraging factor or not. Participant P47 stated their ideas in the open question Q46 as follows:
\begin{itemize}
    \item P47 - \textit{``I don't think someone with a master's in computer science would be averse to getting a PhD just because of gender imbalance. They already have a master's, they've experienced it before.''}
\end{itemize}
This argument may be one of the reasons for the low ratio of answers in the ``Perceived sexism and hostile environment'' theme. 

\begin{table*}[htbp]
	\centering
	\caption{Themes for Q56 - Do you have any insights on what discourages students from starting a PhD in computer science?}
	\label{tab:themes2}
	\resizebox{\textwidth}{!}{%
		\begin{tabular}{@{}llllr@{}}
			\toprule
    \multicolumn{2}{l}{\textbf{Themes}}  & \textbf{Explanation}& \textbf{Sample Comment(s)}& \textbf{Nr. of } \\ 
			\textbf{} & \textbf{} & \textbf{} & \textbf{}& \textbf{comments} \\\midrule
			1 & Career and financial   & Many respondents feel that pursuing a 	PhD may delay&P232 - \textit{"$\ldots$no prospects of a stable job  } &110 \\  
			& concerns  &  their entry into the job market and result in lower salaries&\textit{ in academia after finishing the PhD,}  &\\
			&  &   compared to industry positions. They mentioned the&\textit{$\ldots$, better salary and more stability }& \\ 
			&  &  attractive and better job opportunities and higher income&\textit{in industry."}& \\ 
			&  &   available in the industry. && \\
			
			2 & Negative perception  & Many respondents state a negative perception of academia, &P94 - \textit{"Not being useful."}& 81 \\
			& of academia and &  considering it slow, theoretical, and disconnected from the &P580 - \textit{"Spending 4 years minimum with
			}&  \\
			&  research &   real world. They also express doubts about the value and& \textit{very low life quality and very low}& \\
			& &  relevance of research conducted during a PhD. Long study&\textit{income."} & \\
			& &  duration and lack of funding during the PhD studies are&P684 - \textit{"It is likely a bad investment$\ldots$"} & \\
			& & also categorised under this theme.& & \\
			
			3 & Preference for   & Some respondents mention their preference for gaining&P425 - \textit{"either because students are not } &64  \\
			& practical experience&   practical experience in the industry rather than pursuing&\textit{good enough, are done studying, dislike } &  \\
			& &   a PhD, as they believe they can learn and develop&\textit{researching, or want to go into industry}& \\
			& &  skills faster in a professional environment. & \textit{(for more money)"}&  \\
			
			4 &Burnout and stress & The demanding nature of a PhD, including long working&P47 - \textit{"Burn-out. Stressful and anxiety } &54 \\
			& & hours, stressful deadlines, and high workloads can lead & \textit{inducing deadlines."}& \\
			& & to burnout and anxiety. Respondents stated the high & P380 - \textit{"Lots of writing, stories of }& \\
			& & possibility of feeling overwhelmed by the pressure and &\textit{burnouts."} & \\
			& &  struggling to maintain a healthy work-life balance. & & \\
			
			5 &Lack of information &Respondents mention a lack of clear information and guidance  &P73 - \textit{"Unclear how to apply, unclear} &39 \\
			&  and guidance &  about the PhD process,  including the application process, & \textit{what I would do after the PhD. Unclear }& \\
			& & expectations,  and career paths. They may feel uncertain &\textit{ on family and relationship in the future."}& \\
			& & about what they would do after completing their PhD and  & & \\
			& & how it would impact their job opportunities outside& & \\
			& & of academia.& &\\
			
			6 & Concerns on family  & Some of the respondents mention the difficulties of &P235 - \textit{"Possibly having to give up or} & 25 \\
			& and social life&  starting a family or spending free time on weekends&\textit{ postpone many decisions in your}&\\
			& &   without being worried about the PhD work.& \textit{personal life$\ldots$"}&  \\
			
			7 & Imposter syndrome  & Some respondents mention Imposter syndrome and low   &P39 - \textit{"being the only female, imposter's }& 20 \\
			& and lack of  & self-confidence as factors that discourage them from  & \textit{syndrome$\ldots$"} & \\
			& confidence &  pursuing a PhD. They may feel intimidated by the &  P883 - \textit{"fear of failure."}&  \\
			&  & academic environment, especially if they are the only  &&  \\
			&  & female or if they perceive a lack of respect or support. &&  \\
			
			8 & Requirements to get  & Some respondents mentioned that to obtain a good PhD &P93 - \textit{"$\ldots$high requirements on grades." }& 12 \\
			& a good PhD position &  position and grant, one should have good grades and& & \\
			&  &   "excellence" in their previous studies.& &  \\
			
			9 & Perceived sexism and  & A few respondents expressed concerns about facing  &P90 - \textit{"$\ldots$I think being perceived as female}&10 \\
			& hostile environment & sexism and unequal treatment in academia. They may  & \textit{is discouraging. I know of a friend who has}&\\
			&  & have witnessed or experienced discriminatory comments & \textit{gotten weird comments for how she presents}&  \\
			&  & or disrespectful behaviour, which can discourage them & \textit{ herself, her personality etc. by coworkers.} & \\
			&  & from pursuing a PhD. & \textit{ Or who got treated with a lot less respect} & \\
			&  & & \textit{by the students she is supervising for a} & \\
			&  & & \textit{seminar. Her male-presenting coworkers} & \\
			&  & & \textit{would never have gotten treated the way she} & \\
			&  & & \textit{has been. I dislike the thought of me being} & \\
			&  & & \textit{treated differently for being myself."} & \\
			\bottomrule
		\end{tabular}%
	}
\end{table*}

These themes and specific comments can help us better understand the factors discouraging students from continuing their studies at the PhD level and work on clarifying or providing more information to the potential PhD students to minimise the impact of these discouraging factors.
According to the conclusions of this qualitative study, it is essential to clarify the career concerns and doubts of the students. Many respondents mentioned the uncertain job opportunities after finishing their PhD. These concerns can be decreased by providing more information to these students on the wide range of options and providing advice and mentorship on how to reach them. Meeting with successful role models would also help these students change their negative perception of pursuing a PhD and may help them find answers to their doubts about PhD life, family and social life balance, and the work environment. 

\subsection{Analysis of RQ3: To what extent do these factors differ by gender?} 
Looking into the overall statistics of factors favouring or discouraging the enrolment in a Computer Science PhD can obfuscate gender-related differences within our sample of participants. On the other hand, simply contrasting the answers from participants of different genders without considering whether or not they decided to start a PhD can also obfuscate key differences, as said differences may manifest only in particular subgroups. For instance, females who choose not to do a PhD may have significant differences from males who choose not to do a PhD, while those differences are not present when considering females and males who choose to do a PhD. As such, in this section, we consider four groups: females who choose not to do a PhD (\faUser \faVenus), males who choose not to do a PhD (\faUser \faMars), females who decide to do a PhD (\faUserGraduate \faVenus), and males who decide to do a PhD (\faUserGraduate \faMars). We then compare females \textit{vs} males who decide not to do a PhD, as well as females \textit{vs} males who decide to do a PhD, to identify gender differences when the decision to do a PhD, or not, is considered. We define hypotheses similar to those in the previous research questions, but now with these more refined criteria. For instance, for Q48, the hypotheses are:

\begin{tcolorbox}
	\begin{footnotesize}
		$H_{48-0-NoPhD}$: There is no statistically significant difference in \textit{the perception of having strong programming skills} when comparing females (\faUser \faVenus) and males (\faUser \faMars) who decide not to do a PhD.
		
		$H_{48-1-NoPhD}$: There is a statistically significant difference in \textit{the perception of having strong programming skills} when comparing females (\faUser \faVenus) and males (\faUser \faMars) who decide not to do a PhD.\\

		$H_{48-0-PhD}$: There is no statistically significant difference in \textit{the perception of having strong programming skills} when comparing females (\faUserGraduate \faVenus) and males (\faUserGraduate \faMars) who decide to do a PhD.
		
		$H_{48-1-PhD}$: There is a statistically significant difference in \textit{the perception of having strong programming skills} when comparing females (\faUserGraduate \faVenus) and males (\faUserGraduate \faMars) who decide to do a PhD.
   \end{footnotesize}
\end{tcolorbox}

\paragraph{Gender differences in supporting factors for pursuing a PhD}
Here, we consider the same survey questions as in RQ1. This time we segregate responses using the gender of the respondents. We compare the answers provided by females with those provided by males. Of those questions analysed with the Chi-Square test, only one of them (Q44) had a statistically significant difference \newtext{($p = .039$), without considering a Holm-Bonferroni or a Bonferroni correction}, when contrasting females and males who did not do a PhD, with a small effect size \newtext{($\varphi_{c} = .112$)}. A slightly higher percentage of females than males had participated in a program or activities informing them about PhD studies.
Table \ref{tab:RQ3EncouragingByPhDMaterial} summarises the results for the supporting factors in which there were statistically significant differences. We present in more detail those for which this difference was significant even when considering the Holmes-Bonferroni and the Bonferroni correction, as these are also the ones with larger effect sizes, and briefly mention the remaining ones. The comparison between females and males (\faVenus \textit{vs} \faMars) is always presented from the female (\faVenus) perspective.

While the factors addressed by the questions in table \ref{tab:RQ3EncouragingByPhDMaterial} are intrinsically encouraging factors, a lower level of agreement expressed by a particular subgroup of participants with an encouraging factor suggests that this encouraging factor is less prevalent for that particular subgroup. For instance, female participants who choose not to start a PhD have a worse opinion of their programming skills than males with a medium effect size (Q48 \faUser). 
The same happens for those who choose to do a PhD, with females' opinions being worse than males' and a medium effect size (Q48 \faUserGraduate). 
Female participants who choose not to start a PhD like practical Computer Science subjects, such as programming 
less than males with a small effect size (Q19 \faUser). This difference is not significant when considering participants who choose to do a PhD (Q19 \faUserGraduate).
While there is no difference between participants of both genders who choose not to start a PhD (Q20 \faUser), for those who decide to start it, female participants like interdisciplinary areas of Computer Science more than male participants, with a medium effect size (Q20 \faUserGraduate).  
Female participants who choose not to start a PhD have lower self-confidence than males, with a small effect size (Q50 \faUser). Again, this difference is not significant for those who decide to start a PhD (Q50 \faUserGraduate).
Female participants who choose not to start a PhD have a lower self-confidence in their interaction with other students than male participants, with a small effect size (Q16 \faUser). The difference is not significant for those who decide to start a PhD (Q16 \faUserGraduate).

\begin{table}[htb]
	\label{tab:RQ3EncouragingByPhDMaterial}
	\caption{Supporting factors for a PhD, segregated by doing a PhD (\faUserGraduate), or not (\faUser). The Mann-Whitney U test compares the results of females and males in each subset of participants (\faUser\faVenus \textit{ vs} \faUser\faMars, and \faUserGraduate\faVenus \textit{ vs} \faUserGraduate\faMars, respectively). For brevity, we only present questions with statistically significant differences in at least one of the groups (\faUser, or \faUserGraduate). We sort results in descending order of the maximum effect size for each pair of questions. The differences for all remaining questions concerning supporting factors for starting a PhD were not statistically significant, suggesting no relevant gender differences. \newtext{The used Bonferroni-corrected significance level is $p < .0006756757$}}
	\resizebox{\columnwidth}{!}{%
		\begin{tabular}{@{}lllllllllllll@{}}
			\toprule
			Id & \faUser/\faUserGraduate 
   & \textbf{\textit{p}} & $\eta^{2}$ & \faVenus~\textbf{\textit{vs}} \faMars & \textbf{$\bar{x}$} & \textbf{$\bar{x}$}(\faVenus) & \textbf{$\bar{x}$}(\faMars) & \textbf{$\tilde{x}$} & \textbf{$\tilde{x}$}(\faVenus) & \textbf{$\tilde{x}$}(\faMars) & Mode & N \\ \midrule
			Q48 & \faUser 
   & {\ul \textit{\textbf{.000}}} & .093 & \faVenus \faChevronCircleDown \faChevronCircleDown & 3.548 & 3.249 & 3.885 & 4 & 3 & 4 & 4 & 334 \\
			Q48 & \faUserGraduate 
   & {\ul \textit{\textbf{.000}}} & .088 & \faVenus \faChevronCircleDown \faChevronCircleDown & 3.676 & 3.344 & 4.047 & 4 & 3 & 4& 4 & 182 \\[.2cm]
			Q19 & \faUser 
   & {\ul \textit{\textbf{.000}}} & .055 & \faVenus \faChevronCircleDown & 4.236 & 4.021 & 4.474& 4 & 4 & 5 & 5 & 369 \\
			Q19 & \faUserGraduate 
   & .062 & .017 & & 4.218 & 4.109 & 4.348 & 4 & 4 & 5 & 5 & 202 \\[.2cm]
            Q50 & \faUser 
   & {\ul \textit{\textbf{.000}}} & .053 & \faVenus \faChevronCircleDown & 3.620 & 3.364 & 3.910 & 4 & 4 & 4 & 4 & 332 \\
			Q50 & \faUserGraduate 
   & .145 & .012 & & 3.709 & 3.604 & 3.910 & 4 & 4 & 4 & 4 & 182 \\[.2cm]
			Q20 & \faUser 
   & .245 & .004 & & 4.101 & 4.149 & 4.046 & 4 & 4 & 4 & 4 & 368 \\
			Q20 & \faUserGraduate 
   & {\ul \textit{\textbf{.000}}} & .067 & \faVenus \faChevronCircleUp & 4.289 & 4.464 & 4.076 & 5 & 5 & 4 & 5 & 204 \\[.2cm]
			Q16 & \faUser 
   & {\ul \textit{\textbf{.000}}} & .034 & \faVenus \faChevronCircleDown  & 3.655 & 3.469 & 3.863 & 4 & 4 & 4 & 4 & 371 \\
			Q16 & \faUserGraduate 
   & .415 & .003 & & 4.010 & 3.954 & 4.078 & 4 & 4 & 4 & 4 & 198 \\[.2cm]
			Q35 & \faUser 
   & \textit{.002} & .031 & \faVenus \faChevronUp & 2.667 & 2.891 & 2.420 & 3 & 3 & 3 & 3 & 315 \\
			Q35 & \faUserGraduate 
   & \textit{.002} & .053 & \faVenus \faChevronUp & 3.354 & 3.611 & 3.050 & 3 & 4 & 3 & 3 & 175 \\[.2cm]
			Q12 & \faUser 
   & \textit{.003} & .024 & \faVenus \faChevronDown & 4.368 & 4.274 & 4.474 & 5 & 4 & 5 & 5 & 372 \\
			Q12 & \faUserGraduate 
   & .526 & .002 & & 4.324 & 4.723 & 4.685 & 4 & 5 & 5 & 5 & 182 \\[.2cm]
			Q54 & \faUser 
   & .526 & .002 & \faVenus \faChevronDown & 4.324 & 4.051 & 4.297 & 4 & 4 & 4 & 5 & 182 \\
			Q54 & \faUserGraduate 
   & \textit{.008} & .021 & & 4.167 &4.292 & 4.360 & 4 & 4 & 4/5 & 5 & 330 \\[.2cm]
			Q47 & \faUser 
   & \textit{.017} & .017 & \faVenus \faChevronDown & 4.221 & 4.121 & 4.333 & 4 & 4 & 4 & 4 & 330 \\
			Q47 & \faUserGraduate 
   & \textit{.010} & .037 & \faVenus \faChevronDown & 4.320 & 4.198 & 4.4.59 & 4 & 4 & 5 & 4 & 181 \\[.2cm]
			Q52 & \faUser 
   & \textit{.010} & .020 & \faVenus \faChevronDown & 4.270 & 4.170 & 4.382 & 4 & 4 & 5 & 5 & 333 \\
			Q52 & \faUserGraduate 
   & .077 & .017 & & 4.577 & 4.646 & 4.500 & 5 & 5 & 5 & 5 & 182 \\[.2cm]
			Q29 & \faUser 
   & .212 & .005 & & 2.516 & 2.613 & 2.412 & 3 & 3 & 3 & 3 & 316 \\
			Q29 & \faUserGraduate 
   & \textit{.015} & .033 & \faVenus \faChevronUp & 3.439 & 3.360 & 3.181 & 4 & 4 & 3 & 5 & 180 \\[.2cm]
			Q15 & \faUser 
   & \textit{.035} & .012 & \faVenus \faChevronDown & 3.914 & 3.832 & 4.001 & 4 & 4 & 4 & 4 & 370 \\
			Q15 & \faUserGraduate 
   & .295 & .006 & & 4.292 & 4.236 & 4.360 & 4 & 4 & 4 & 4 & 199 \\[.2cm]
			Q18 & \faUser 
   & \textit{.040} & .011 & \faVenus \faChevronUp & 3.568 & 3.687 & 3.434 & 4 & 4 & 4 & 4 & 370 \\
			Q18 & \faUserGraduate 
   & .950 & .000 & & 4.030 & 4.036 & 4.022 & 4 & 4 & 4 & 5 & 201 \\ \bottomrule
		\end{tabular}%
	}
\end{table}

All the statistically significant differences in the previous factors remain significant even when using the Holm-Bonferroni, or the Bonferroni corrections. 
In addition, we also found several statistically significant differences, when considered in isolation, all with a small effect size. Some of these differences led to higher levels of agreement from females than males. 

Regardless of their decision, females were more likely to consider that better gender balance would increase their possibilities of starting a PhD than males (Q35 \faUser\faUserGraduate).
Females who decided to do a PhD considered family encouragement as a more relevant factor than males (Q29 \faUserGraduate). This difference is not significant when considering participants who decided not to do a PhD.

Females who decide not to start a PhD have nevertheless a stronger preference for theoretical Computer Science than males (Q18 \faUser). Still, this difference is no longer statistically significant when considering participants who decided to do a PhD.

In contrast, females who decided not to start a PhD are less enthusiastic than males about computer science (Q12 \faUser). They also love computer science less than males (Q54 \faUser) and feel less successful in their studies (Q15 \faUser). None of these differences are statistically significant for those who decided to start a PhD. Regardless of their decision about starting a PhD, females express lower confidence in their competencies to succeed in computer science than males (Q47 \faUser \faUserGraduate).

\begin{insight}{Enthusiasm and self-confidence (continued)}
\label{insight:love_and_self_confidence}
	Two recurring themes in our findings are a consistently slightly lower enthusiasm for computer science and a slightly lower self-confidence in their computer science abilities expressed by females who choose not to start a PhD \newtext{compared to} \oldtext{when compared with} males. These differences disappear when considering participants who decided to start a PhD, suggesting love for and self-confidence in doing computer science are less prevalent in females than in males who decide not to start a PhD. This lack of self-confidence \newtext{when considering programming skills} is stronger, even for females who decide to start a PhD. \oldtext{when considering programming skills.} On the other hand, females who decide to start a PhD expressed a higher preference for interdisciplinary areas of computer science and valued family encouragement for starting a PhD more than males. Regardless of their decision about starting a PhD, females have a stronger belief than males that a better gender balance would increase their likelihood of starting a PhD in computer science.
\end{insight}

\begin{figure}[htb!]
	\centering
	{\includegraphics[width=1.0\linewidth]{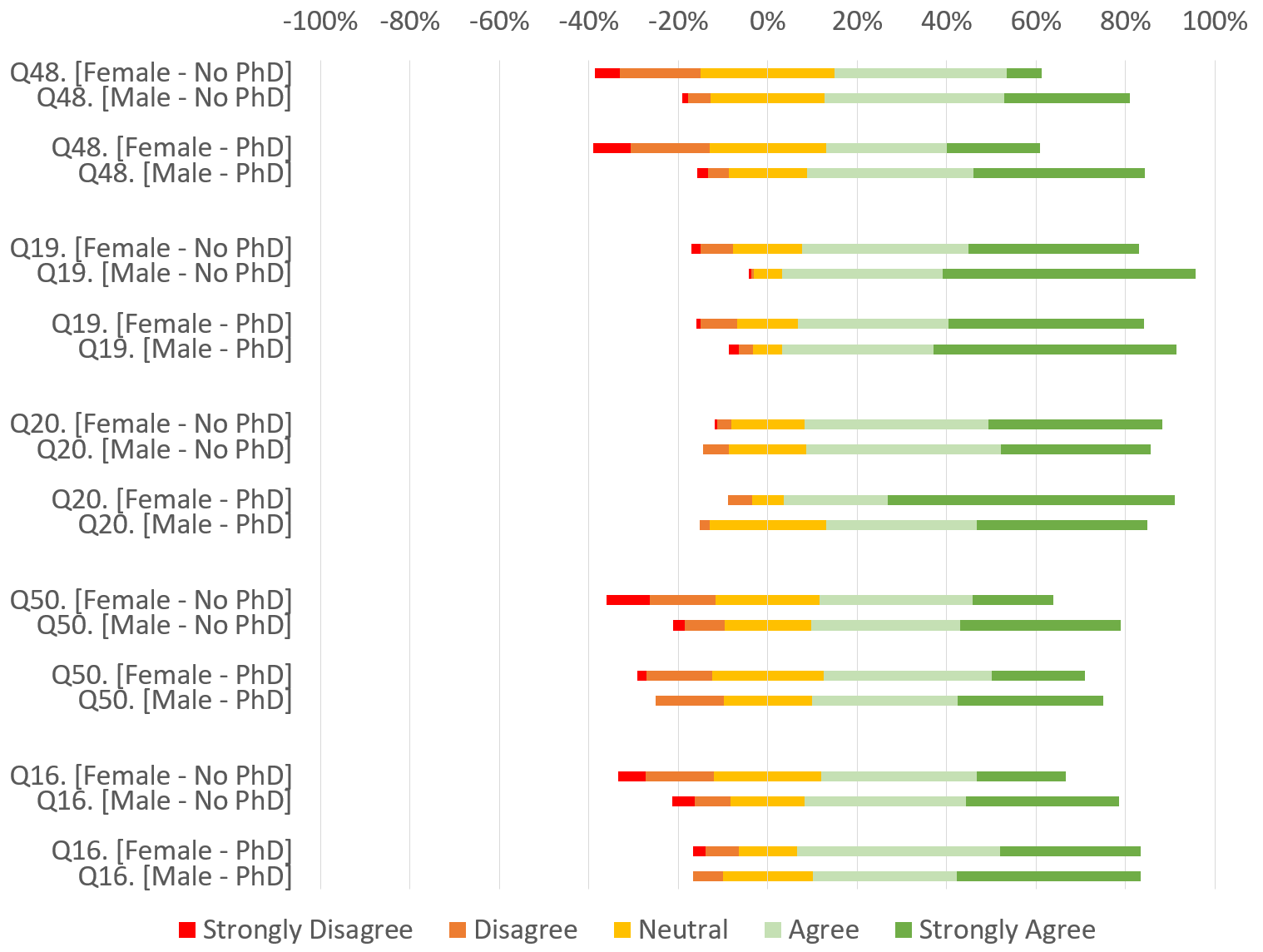}}
	\caption{The distributions of encouraging factors for pursuing a PhD grouped by gender, filtered by cases where at least one of the groups has statistically significant differences concerning another group by gender (e.g. Q48. has statistically significant differences between male and female participants who did not pursue a PhD, and another one, between male and female participants who decided to enrol in a PhD.}
	\label{fig:EncouragingFactorsByGender}
\end{figure}

\paragraph{Gender differences in factors against pursuing a PhD}
Here, we consider the same survey questions as in RQ2. This time, we segregate responses using the gender of the respondents. We compare the answers provided by females with those provided by males.
Table \ref{tab:RQ3DiscouragingByPhDMaterial} summarises the results for the discouraging factors in which there were statistically significant differences when considered in isolation with a small effect size. None of these differences are significant when considering the Holmes-Bonferroni or the Bonferroni correction. The comparison between females and males (\faVenus \textit{vs} \faMars) is always presented from the female (\faVenus) perspective.
Females who decided not to do a PhD expressed lower agreement with the idea that the industry offers better job opportunities than the University 
than males 
(Q38 \faUser). 
This was also the case for those who decided to do a PhD, with females 
also expressing a lower agreement than males 
(Q38 \faUserGraduate).
Females who decided not to do a PhD also expressed a lower agreement than males with the idea of industry offering more flexible career paths 
(Q40 \faUser).
This was also the case for those who decided to start a PhD, with females 
having a lower agreement than males 
with the perception of industry offering more flexible career opportunities, 
(Q40 \faUserGraduate).


\begin{table}[htbp]
	\label{tab:RQ3DiscouragingByPhDMaterial}
	\caption{Detrimental factors for doing a PhD, segregated by doing a PhD, or not. The Mann-Whitney U test compares the results of females and males in each subset of participants (\faUser\faVenus \textit{ vs} \faUser\faMars, and \faUserGraduate\faVenus \textit{ vs} \faUserGraduate\faMars, respectively). For the sake of brevity, we only present questions with statistically significant differences in at least one of the groups (\faUser, or \faUserGraduate). We sort results in descending order of the maximum effect size for each pair of questions. The differences for all remaining questions concerning supporting factors for starting a PhD were not statistically significant, suggesting no relevant gender differences in their results. \newtext{The used Bonferroni-corrected significance level is $p < .0006756757$}}
	\resizebox{\columnwidth}{!}{%
		\begin{tabular}{@{}lllllllllllll@{}}
			\toprule
			Id & \faUser/\faUserGraduate 
   & \textbf{\textit{p}} & $\eta^{2}$ & \faVenus~\textbf{\textit{vs}} \faMars & \textbf{$\bar{x}$} & \textbf{$\bar{x}$}(\faVenus) & \textbf{$\bar{x}$}(\faMars) & \textbf{$\tilde{x}$} & \textbf{$\tilde{x}$}(\faVenus) & \textbf{$\tilde{x}$}(\faMars) & Mode & N \\ \midrule
			Q38 & \faUser 
   & \textit{.002} & .029 & \faVenus \faChevronDown& 4.041 & 3.899 & 4.199 & 4 & 4 & 5 & 5 & 319 \\
			Q38 & \faUserGraduate 
   & .050 & .021 & \faVenus \faChevronDown & 3.818 & 3.684 & 3.965 & 4 & 4 & 4 & 5 & 181 \\[.2cm]
			Q40 & \faUser 
   & \textit{.015} & .019 & \faVenus \faChevronDown & 3.868 & 3.750 & 3.993 & 4 & 4 & 4 & 4 & 318 \\
			Q40 & \faUserGraduate 
   & \textit{.023} & .029 & \faVenus \faChevronDown & 3.260 & 3.063 & 3.477 & 3 & 3 & 4 & 3 & 181 \\
			\bottomrule
		\end{tabular}%
	}
\end{table}

\begin{figure}[htb!]
	\centering
	{\includegraphics[width=1.0\linewidth]{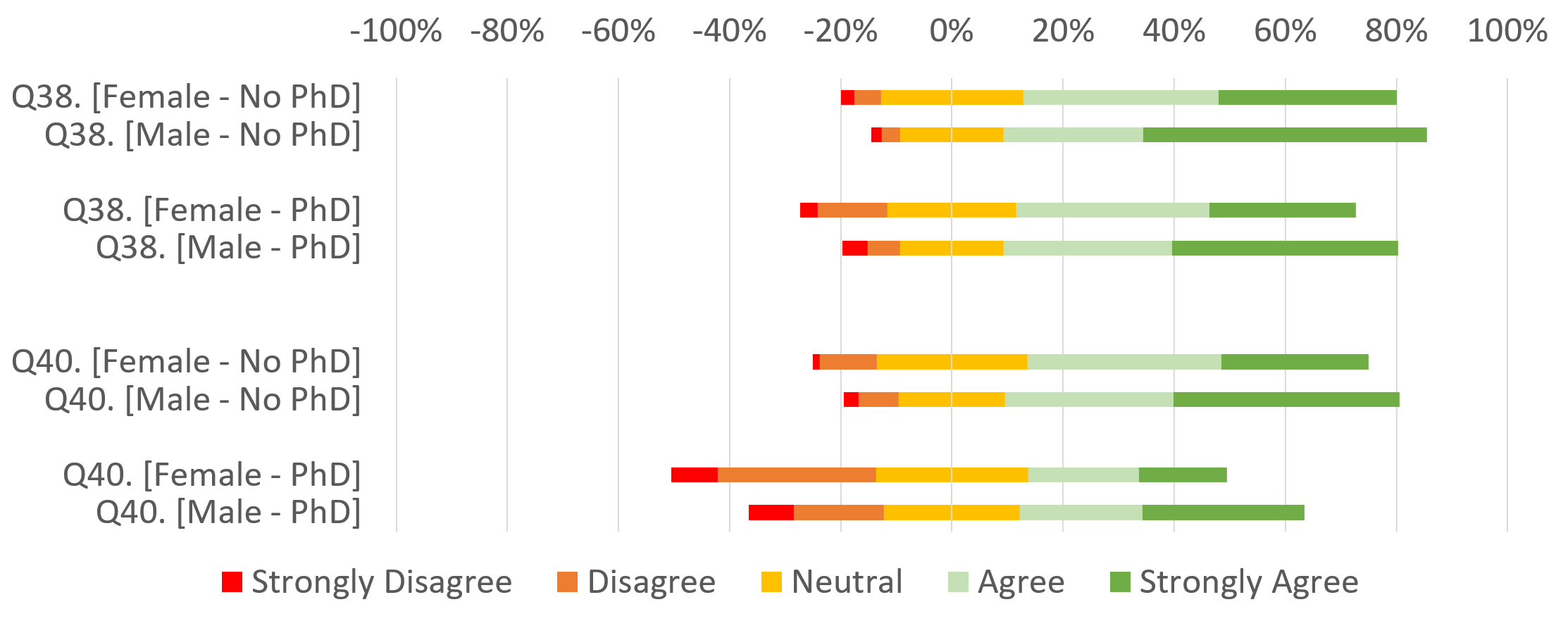}}
	\caption{The distribution of responses for questions about discouraging factors for pursuing a PhD grouped by gender. Only the questions with statistically significant differences are included.}
	\label{fig:DiscouragingFactorsByGender}
\end{figure}

Figure \ref{fig:DiscouragingFactorsByGender}  illustrates the distribution of responses for discouraging factors with statistically significant differences across males and females.

\begin{insight}{Industry offers}
\label{insight:industry_offers}
    	 Females expressed a slightly lower agreement than males concerning the perception that the industry offers better job opportunities and more flexibility than academia.
\end{insight}

%% file: 05-Discussion.tex
\section{Discussion}
This section discusses the insights obtained from the data and how they may inform specific recommendations. We then discuss the limitations of our study and threats to validity in the context of sampling and response bias, construct, measurement, conclusion and internal validity.  This section finishes with some implications of this research and some lessons learned. 
\label{sec:discussion}
\subsection{Insights}
We start with the insights obtained from the data about encouraging factors. Then, we address discouraging factors and \oldtext{, finally,} discuss the gender-based differences found in our data. We suggest some recommendations for practice as strategies to mitigate noticed issues.
\paragraph{Encouraging factors}
Involving students in research projects and having them supervised by senior researchers is a good way of making them aware of the possibility of enrolling in a PhD in computer science. Several participants also suggested involving those students in writing and submitting research papers as a beneficial strategy to encourage them to do a PhD. 
Indeed, the most cited reason for pursuing a PhD in computer science was an interest and confidence in the ability to conduct research work in this area, which significantly outnumbers any other reason expressed by participants in this way. Self-confidence and interest differences between female and male students are often pointed out as key factors for the underrepresentation of women in computer science~\cite{beyer2014women}.  

The second most mentioned motives were personal goals such as independence, expertise, recognition, or personal development. \oldtext{, among others} Several of these goals are commonly listed as desirable outcomes of a PhD~\cite{Lovitts2005,waaijer2017perceived}. The willingness to pursue an academic career received \oldtext{a} similar \oldtext{level of} attention from our participants. 

This research interest can be complemented with an awareness program designed to provide information to potential PhD candidates. Such programs are not common, but they can raise the awareness of potential PhD students about the career opportunities a PhD may enable, as well as the inherent challenges in doing a PhD and how to navigate through them. In this paper, we present one such program, in section~\ref{sec:careerlunch}.

Furthermore, encouraging students with a high potential to do a PhD can make a difference. This encouragement often comes from academic staff and peers, but it can also come from the student’s family or industry, in line with the findings of Yates and Plagnol~\cite{yates2022female}, albeit it is the male family members or industry contacts. Unfortunately, the latter is relatively uncommon. While this is problematic, it implies a clear growth opportunity, as well, by strengthening the synergies between the industry and academia, possibly through the proposal of more applied research topics.  
\
Unsurprisingly, people who decide to start a PhD in Computer Science tend to be more enthusiastic about it and more self-confident in their abilities than their peers. In particular, students who prefer theoretical computer science or interdisciplinary computer science topics tend to pursue a PhD more often than their peers. PhD candidates tend to perceive a PhD as something that can fulfil their needs (including a greater curiosity than their peers), improve their future career options, and, ultimately, be fun. They also feel a better gender balance would increase their chances of starting a PhD. 

\paragraph{Discouraging factors} The key factors against doing a PhD seem to be the uncertainty of what it is like to do a PhD, and a perception of lower job flexibility when compared to what industry offers. One of the most cited factors against doing a PhD \oldtext{that we found} is a negative perception of academia and the [lack of] value of doing a PhD for the student in the long run. These perceptions are far from exclusive to computer science students but may be, at least to some extent, counterbalanced by personal goals~\cite{waaijer2017perceived}. The program proposed in this paper can help mitigate these concerns, as obtaining a PhD may open additional opportunities for those who do it to find more intellectually appealing jobs~\cite{waaijer2017perceived}. 

The long-term commitment implied by starting a PhD is also perceived as problematic for many. This seems to be tied to some of the feedback we got from our qualitative data. Indeed, the most cited discouragement for starting a PhD is a combination of career and financial concerns, with a PhD viewed as delaying an increased quality of life improvement by at least four years. The generally low value of PhD scholarships, when compared to salaries offered in the industry significantly harms the option of doing a PhD in Computer Science, at least during the years the student is working on the PhD. Apart from structural changes leading to more competitive scholarships to bridge the salary gap, another strategy would be to provide evidence on the salary evolution for Computer Science professionals holding a PhD (not only in Academia, but particularly for those working in the Industry), as a way to support the argument against the perception that the return on the investment in doing a PhD is insufficient. For instance, in the US, professionals \oldtext{will} earn a substantially higher salary with a PhD in computer science rather than only a bachelor’s or master's degree~\cite{Steele2023}. 

\paragraph{Differences between females and males}
When contrasting the differences between responses from female and male participants, concerning encouraging factors, a noticeable feature is that, in several cases, these differences are statistically significant when one segregates participants between those who decided to do a PhD and those who didn't. \newtext{This was the case} in 10 out of the 13 questions for which we found statistically significant differences. \oldtext{, this was the case.} In 7 of those 10 questions, the differences exist for those who choose not to do a PhD, which suggests these questions provide insights on existing biases against starting a PhD. For instance, several answers highlight the ``imposter syndrome'' as \oldtext{something} more prevalent in female participants who decide not to do a PhD\newtext{, although, as in other disciplines, men are also affected by it~\cite{bravata_prevalence_2020}}. 

In contrast, we also find that factors like having multidisciplinary computer science subjects \oldtext{are more appealing} \newtext{appeal more} to females. This is in line with the observation that hybrid computer science courses including modules from other disciplines are more attractive to female undergraduate students~\cite{alvarado2012increasing}. 
Family support is also more relevant for female students, which seems consistent, for instance with the findings in~\cite{webber2021mothers}. 

Another factor we also noticed is a lower self-perception of programming skills (possibly another symptom of the already mentioned imposter syndrome), in line with a lower love for computer science in general (but not theoretical computer science in particular) expressed by female participants. This is consistent with findings of lower confidence in technological abilities and lower motivation \cite{yates2022female}.

\paragraph{Recommendations for practice}
Considering the observations on encouraging and discouraging factors and how they vary when considering gender differences, we offer  a few practical, albeit potentially challenging recommendations:
\begin{itemize}
	\item Promote programs \newtext{clarifying} \oldtext{for the clarification of what one can expect from} the process of doing a PhD, as a way of raising awareness among students about the opportunities they may have \oldtext{, but also} \newtext{and} the underlying challenges of doing a PhD. Our results suggest that \oldtext{a large percentage of the} \newtext{many} participants who decided not to do a PhD were significantly less aware of this information than those who decided to do one. This is not surprising, but it suggests that we may be missing out on good PhD candidates who never considered that possibility, among other factors, because they might not have been so aware of that possibility or what life looks like for a PhD student. While this is problematic for all genders, our data suggests it is more \oldtext{problematic} \newtext{common problem} among females. This gender gap \newtext{aligns} \oldtext{is in line} with the generally lower risk-taking nature of females compared with males. As the unknowns resulting from lack of information entail risks, providing valuable information to potential candidates mitigates some \oldtext{of those} risks and, presumably, leads to better-informed decisions about starting a PhD. This entails a double advantage: \oldtext{not only} more students may be aware and consider doing a PhD, \oldtext{but also} \newtext{and} fewer students will enrol in a PhD ``by mistake’’, potentially leading to a more effective PhD program. \oldtext{In this paper,} We propose a ``Women Career Lunch'' program that embodies this recommendation (see section \ref{sec:careerlunch}).
	\item Systematically address the imposter syndrome by recurrently providing evidence against it. Our data suggests the gender gap associated with this syndrome is noticeably mitigated among those who choose to start a PhD, although not eliminated (e.g. concerning the participants’ perceptions of their programming skills). If we are to bolster the applications of females for PhD studies, this is one of the factors we need to address. The official data from Informatics Europe \oldtext{, for instance,} provides no evidence of a generally lower performance of female students in their studies. Indeed, overall, females tend to represent a higher percentage of graduates than the percentage of females who enrol in a course, suggesting their success rate is higher than the one achieved by males at the bachelor, master, and PhD levels.
	\item Promote multi-disciplinary topics for PhDs. As females show considerable interest in \oldtext{researching multi-disciplinary topics} \newtext{multi-disciplinary research}, there seems to be a yet-to-explore potential for growth. To be clear, we do not advocate promoting these topics \textit{instead} of more ``traditional'' computer science topics, but rather \textit{in addition to} those topics.
\end{itemize}

\subsection{Limitations and threats to validity}
\oldtext{In this section,} We discuss \oldtext{the} limitations and validity threats to this survey, and our mitigation strategies to handle them.
\subsubsection{Sampling bias}
Having a random sample of people \oldtext{who are} taking or already having a degree in Computer Science \oldtext{and related} is impractical, as there are no publicly available contact lists covering our target population. This challenge is common to Software Engineering studies \oldtext{in general} \cite{Amir2018, Baltes2022sampling}. We used a combination of convenience and snowball sampling to mitigate this risk. We leveraged the contacts of the EUGAIN COST action and Informatics Europe to reach potential respondents well beyond the contacts available to the authors of this article. Naturally, convenience and snowball sampling may introduce biases in the obtained sample. We tried to mitigate this bias by using a broad and diverse \oldtext{set of} channels to reach potential respondents. The diversity of origins of our respondents suggests that, at least to some extent, we \oldtext{were able to succeed} \newtext{succeeded} in this strategy. That said, the percentage of respondents from the different countries is not aligned with the actual percentage of potential respondents in the population, with countries like Germany or Serbia being over-represented \oldtext{when} compared to other countries. This may introduce cultural-based biases in our sample.  Also, the female gender is over-represented in our sample \oldtext{, when} compared to the estimated population. In this case, we were particularly interested in having a large set of responses from females, to help us answer RQ3, so we oversampled females by targeting our survey advertisement efforts more to potential female respondents. Finally, our sample is Europe-centric by design although we did not exclude respondents from other origins, as it is not uncommon for students and academics to have mobility among continents.

\subsubsection{Response bias}
The questionnaire was anonymous and freely available through an online link, so we cannot completely rule out the possibility of someone answering the survey more than once. We cannot also rule out the possibility of having a higher response rate from participants who were particularly interested in our research, as perceived from the invitation. There is also a potential cultural effect resulting from the responses' geographical distribution. While our analysis did not consider geography as a direct factor countries from which we got more answers may have their participants’ cultural values oversampled. All that stated, we are contrasting people who decided to do a PhD with people who decided against it (RQ1 and RQ2) and females with males (RQ3), using normalised data (e.g. the percentage of people in each cohort who provided a particular answer), the effect of the different number of members in each cohort was mitigated. Also, we had a relatively high number of respondents (587), well over our target sample size (400), so these threats were mitigated.

\subsubsection{Construct validity}
When analysing the responses to our questionnaires, we emphasised that the answers reflect the perceptions of participants rather than facts. For instance, their perception of their programming skills may not match their actual programming skills. However, we should note that we were trying to establish relationships between participants’ perceptions and their decision to start a PhD, which matches what happens in the decision process, at least on the side of the candidates. Of course, deciding to do a PhD and successfully going through the PhD candidates' selection process are two different things. Our focus was on the former. Also, we need to consider that \oldtext{a large percentage of our} \newtext{many} respondents who reported that they had decided to do a PhD have already completed it or are currently enrolled in a PhD program.

\subsubsection{Measurement validity}
Most of the questions in our survey were measured with a balanced 5-point Likert scale, presented in the survey as a set of equally-distanced options. While there is some debate on the soundness of computing, for instance, the mean in the distribution of these variables as a summary variable, the balanced, equally distanced scale points, along with \newtext{many} \oldtext{a large number of} responses, mitigate that risk. That said, we complement this information with the median, and our hypotheses tests are always done using non-parametric procedures, such as the Mann-Whitney U test, which are adequate for dealing with ordinal data. Also, we always gave respondents the possibility of not answering a particular question to mitigate the risk of randomly selected answers.

\subsubsection{Conclusion validity}
When testing our hypotheses, we always highlight the extent to which differences are statistically significant, either in isolation when using the Holm-Bonferroni correction or the stricter Bonferroni correction. We also present the computed effect sizes for our hypotheses tests so that we mitigate the risk of overstating the effect of the identified relations.

\subsubsection{Internal validity}
The internal validity concerns the causal relation between a treatment and its expected results. To do so, we would have to demonstrate correlation, precedence, and the absence of other variables that might be causing identified correlations. Our statistical tests show correlations but not precedence. That said, we specifically asked participants to answer the questionnaire from \oldtext{the perspective of their perceptions} \newtext{their perceptions} when they were deciding whether to start (or not) a PhD, which preceded the decision itself. We cannot completely discard the possibility of having external variables explaining the found relationships (e.g. we have no direct information on cultural background, including religion or ethnicity, which may affect this kind of decision. We decided not to use nationalities as proxies for this kind of analysis, as we would have to rely on stereotyped visions of the cultural background of people from those countries, possibly introducing another validity threat to our data analysis.

\subsection{Implications for researchers and leaders and future work}
This work provides new insights into the factors that may be related (either as encouraging or as detrimental \oldtext{factors}) to the decision to start a PhD in Computer Science. We further contrast those factors by segregating answers by gender \oldtext{so we} \newtext{to} get insights into how these factors may differ depending on gender. The \newtext{study results} \oldtext{results of the study} have allowed us to develop a career lunch program. \oldtext{, which aims to} \newtext{The program} mitigates some of the challenges that were particularly more prevalent in female respondents. Our results also illustrate how this is a multi-factor decision, with several different factors, most with a small to moderate effect size. \oldtext{, being combined so that a person may decide whether or not to start a PhD.} While our focus in this survey was on the enrolment in a PhD, the Career Lunch program introduced in Section \ref{sec:careerlunch} is expected to have a double effect: not only will it help potential candidates to make a more informed decision concerning starting a PhD, but also, because the decision is better informed, should contribute to help to improve the retention rates in PhD programs, \textit{vs.} assuming people who do decide to start a PhD know better what to expect. 

\oldtext{The creation and adaptation of} \newtext{Creating and adapting} the Career Lunch program to several \oldtext{different} languages, with the necessary tunings reflecting the diverse cultures associated with those languages, creates ample opportunities for leaders to adopt this program in their Universities. The next crucial step is to run this program in different instances and continuously improve it to leverage the lessons learned with each new instance of the program. Along with the lessons learned, we will also need to monitor its adoption and \oldtext{the impact it has on the recruitment of} \newtext{its impact on recruiting} new PhD students. This monitoring requires the adoption of key process indicators, so the impact of adopting such a program can be assessed via a longitudinal study.

\subsection{Lessons learned}
The \newtext{survey} design, implementation, and analysis \oldtext{of the survey} were a joint effort from a large group of researchers from \oldtext{several} different countries and cultural backgrounds. This has undoubtedly helped with the location-specific advertisement of the survey (e.g. invitation emails were translated into local languages and adapted to target recipients). Also, we leveraged the members of the EUGAIN COST Action \oldtext{as well as} \newtext{and} the access to Informatics Europe members to disseminate the questionnaire, leading to a large, diverse sample of participants.
Using a research-focused survey tool (in our case, LimeSurvey) helped us understand our participants. For instance, the tool records partial answers as well as complete ones, so we were able to accurately compute the response rates for each question and filter out questionnaire participants whose responses were incomplete.

%% file: 06-CareerLunch.tex
\section{Question catalogue}
\label{sec:catalogue}
Following the survey analysis, which focused on general and gender differences between those who progress to PhD studies and those who do not, we constructed a catalogue of questions that will prime a discussion with female students to facilitate conveying the relevant information to them about PhD studies. The questions addressed topics identified as significant from our analysis, which were categorised under the following three categories: (i) basic information and practical issues, (ii) career opportunities, and (iii) emotional support.

\subsection{Basic information and practical issues}
\label{subsec:phd_process}
The category  \textit{Basic information and practical issues} includes questions that can start a discussion to provide basic information to the students about PhD studies, the \oldtext{practical} \newtext{pragmatic} issues concerning applying for a PhD position, and the everyday life of a PhD student. 
It includes the following questions:
\begin{description}
\item[Basic information] 
What are the objectives of a PhD?
How long does a PhD take?
 What are the elements of PhD research?
 What does the supervision of my work look like?
 How to choose a topic for a PhD?
How is the quality of PhD work assessed?
What distinguishes a PhD abroad from a PhD in my country?
\item[Requirements] What are the requirements for PhD admission? Do the \oldtext{requirements} \newtext{criteria} abroad differ from those in my country?
What are the requirements for finishing a PhD?
 What \newtext{do you} \oldtext{to} do during studies to qualify for a PhD position?
    \item[Finding and choosing position, topic, group, supervisor] 
Where do I find PhD positions? How do I find PhD positions abroad? 
Which institute/department suits me? What topic suits me? 
Can I choose the topic of my PhD?  
Can I get to know the group before starting my PhD?
Can I ask doctoral candidates abroad for reports on their experiences?
How do I like the atmosphere? How big is the group? 
 How do I find good supervision for my PhD? How do I choose a supervisor?
\item[Application and relevant information]  
How do I apply for PhD positions? What do I have to consider when applying for a doctorate abroad?
What does the application process look like? What is essential in the written application? What does the PhD application interview look like?
    Where can I get advice if I want to go abroad? 
    What do I have to think about administratively?
\item[Day-to-day life and expectations] 
What funding possibilities are available? How do I finance my stay abroad? 
What is the typical day-to-day of a PhD student?
What would my future tasks and possibilities be? 
Do I have to speak the local language if I go abroad?
\end{description}

\subsection{Career opportunities}
\label{subsection:career}
The category \textit{Career opportunities} includes questions that can evoke discussion on career opportunities in general and specifically in academia and industry. 
This includes the following questions:
\begin{description}
\item[Career in general]
    How does the PhD topic determine your career?
    How important is the topic of a PhD for further career opportunities?
    Is the salary higher with a PhD than with a master's degree?
    What skills will be developed during PhD study concerning research, teaching, and leadership?
    Is the doctorate obtained abroad recognised in my country (and other countries)?
\item[Academic Career]
    What does an academic career look like?
    How does the PhD prepare you for an academic career?
    What is essential to progress to the next step in an academic career?
    Who are your partners throughout an academic career path? With whom will you work together?

\item[Industry Career]
    What can a career in the industry look like?
    Is there the possibility of doing your PhD in industry? 
    Which skills are acquired during a PhD that are relevant to the industry?
    What benefits does having a PhD give towards a career in the industry?
    Can PhD research lead to a Spin-out company?
    Do domestic and foreign companies consider the merits of having a PhD differently?
\end{description}

\subsection{Emotional support}
\label{subsection:emotional_support}
The category \textit{Emotional support} includes prompting questions that address personal doubts, imposter syndrome, lack of self-confidence, gender balance issues, burnout, and sexual harassment. 
This includes the following questions:
\begin{description}
\item[Imposter syndrome and doubts] 
        Am I capable of pursuing a PhD? Am I smart enough?
    What do I do if I am unsure or have problems during my doctorate studies?
    Will I have enough innovative ideas? What if I am not able to implement my research ideas? Can I change or adjust my research focus during the PhD?
    What are \oldtext{the} typical questions and concerns students may have about their doctorate studies?
\item[Burnout]     What do I do if I "run out of breath" during the doctorate? Is it just me? 
\item[Family planning] 
    Is doing a PhD compatible with family planning?
\item[Gender issues] 
    Am I okay with being one of the few women in the group? What can I do if I receive sexist remarks? Where can I find help for sexual harassment?
\item[Finding support] 
    Who can I contact if I have problems? Where can I find advice and support?
    How are the dignity and respect rights of the PhD student applied in your University?
    What do I do if I am dissatisfied with my doctoral supervision? 
\end{description}

\subsection{Survey insights and question catalogue}
Table~\ref{tab:connections} illustrates the connection of each category of questions with the insights and themes identified in the survey.
The first row of this table associates each insight identified in section~\ref{sec:Results} with the appropriate question catalogue category, illustrating how the questions focus on the areas of insights discovered by the statistical analysis of the survey data. 
The second and third rows of the table associate the themes from the quantitative analysis identified in Tables~\ref{tab:themes1} and~\ref{tab:themes2} under the relevant question catalogue category, showing how these themes are covered in the questions. 
 
Note that the insight \ref{insight:personal_strengths} \textit{(personal strengths)} is not addressed within the catalogue as it refers to the internal characteristics of people who choose to do a PhD. 
Such characteristics are formed throughout the complete education process and cannot be influenced quickly. 
This holds for the encouraging factor 1 \textit{(interest and confidence in research)}, which is also not included in the table.  
Encouraging factor themes 5 \textit{(encouragement from mentors or colleagues)} and 7 \textit{(seeing role models)} also are not covered directly by specific questions in the catalogue but are likely to be addressed indirectly through the process of discussion that is prompted by other questions. We address these themes explicitly in the program we propose in the next section.     
\newtext{We include} all other insights and themes \oldtext{are included} in the proposed question catalogue to prompt discussion in these significant areas. 

\begin{table*}[ht!]
    \caption{Question catalogue and its connection to the insights and themes from the survey}
    \centering
	\resizebox{\textwidth}{!}{%
 \begin{tabular}{@{}llll@{}}
    \toprule
     & \textbf{Basic information and practical issues}  & \textbf{Career opportunities} & \textbf{Emotional support} \\
    \midrule
    
    \textbf{Key insights} &\ref{insight:phd_life_info} \textit{(PhD life information)}  &\ref{insight:uncertainty} \textit{(uncertainty)}& \ref{insight:enthusiasm} \& \ref{insight:love_and_self_confidence} \textit{(enthusiasm and self-confidence)} \\
    
    &  \ref{insight:encouragement_requirements} \textit{(encouragement and requirements)} & \ref{insight:industry_offers} \textit{(industry offers)} & \\
    
& \ref{insight:percipation} \textit{(perception of the PhD process)}  & & \\

    & \ref{insight:uncertainty} \textit{(uncertainty)} & & \\

     & & &  \\
    \textbf{Encouraging factors} & id 2 \textit{(personal goals)} &id 3 \textit{(interest in academic career and teaching)}& \\
    \textbf{(Table \ref{tab:themes1})}&& id 4 \textit{(career benefits and opportunities)}& \\
    &&id 6 \textit{(lack of interest in industry jobs)}&  \\

    &  & &  \\
    \textbf{Discouraging factors} & id 5 \textit{(lack of information and guidance)}  &id 1 \textit{(career and financial concerns)}&id 4 \textit{(burnout and stress)} \\
    \textbf{(Table \ref{tab:themes2})}  & id 8 \textit{(requirements to get a good PhD position)} & id 2 \textit{(negative perception of academia and research)} & id 6 \textit{(concerns on family and social life)} \\
    && id 3 \textit{(preference for practical experience)}& id 7 \textit{(imposter syndrome and lack of confidence)}\\
    &&id 5 \textit{(lack of information and guidance)}& id 9 \textit{(perceived sexism and hostile environment)}  \\

    \bottomrule
\end{tabular}
    }
    \label{tab:connections}
\end{table*}

\section{Women Career Lunch}
\label{sec:careerlunch}

The \textit{Women Career Lunch (WoCa Lunch)} program is a supporting initiative for female students that aims to acquaint female undergraduate and master's students with the possibilities for advancing their careers if they pursue PhD studies. The program aims to deliver the information identified as significant by the survey to potential female PhD students in a structured, engaging, and collaborative way. 

The program was developed by dividing the identified question catalogue (Section \ref{sec:catalogue}) into structured modules. Each module represents a cohesive unit suitable for a one-hour discussion on a topic triggered by the questions in the module.  
Also, this program includes two vital encouraging factor themes that were not explicitly handled in the question catalogue: encouragement from
mentors or colleagues and seeing role models (key insight \ref{insight:encouragement_requirements}, and encouraging factor themes 5 and 7 from Table \ref{tab:themes1}).
These aspects can be provided by the \oldtext{facilitators of the program} \newtext{the program facilitators } and guests from academia or industry who share their personal experiences. 
Participation should encourage students and enable them to make well-informed decisions about whether they should enrol in a PhD program. 

We prepared a multi-language version of the WoCa program to foster dissemination and assist in implementation. It includes guidance on the structure and format,  the practical procedures of the initiative and a comprehensive catalogue of questions, that were refined from the original list, that should be discussed within each module. This catalogue is publicly available \cite{zenodoMultiLanguageCatalog}. It is available in the following languages: English, French, German, Portuguese, Serbian, Spanish, Turkish, and Ukrainian.

\textit{WoCa Lunch} was successfully implemented at the Department of Computer Science and the Department of Mathematics at RWTH Aachen University, Germany. The following descriptions and recommendations are based on the obtained experiences, and lessons learned including participants' feedback. 

\paragraph{Overview} The program has eight modules. The initial and the last modules should be in person. The rest can be virtual. Each module takes place during lunchtime and lasts approximately one hour. One guest is invited and interviewed for each module (except for the first and last modules). The interview guests might be members of the executing institution or external guests. The goal is to address the items from the question catalogue in a familiar environment and further give insights into the career development of female role models.

\paragraph{Organising Group} The team organising the program has ideally two moderators and admin support. The moderators should have a PhD and \oldtext{should} be \oldtext{members of the} academic staff \newtext{members}. At least one moderator should be a female, and at least one moderator should be a senior researcher. 

\paragraph{Invited Guests} Each module has a guest appropriate for the module and at least one moderator. If possible, each guest should be female, and at least one should be an early-stage researcher, while at most one guest should be local to the department/institute running the event. Guests should be suitable for role models (enthusiastic, motivating, communicative, open). Guests should speak a local language, as the modules should be run in the local language. At least two weeks before their module, a guest should receive a catalogue of questions they will be asked. Knowing questions in advance enables guests to prepare well for their module. 

\paragraph{Structure} The first module presents the basic ideas of the program and what the students can expect from it. The final module wraps up the program as a whole and gathers feedback. \oldtext{about the program itself.}
The remaining modules address the most important topics related to PhD studies. Modules aim to initiate discussions and reflections, dispel uncertainties, and clarify the career path for those \newtext{pursuing} \oldtext{who choose to pursue} this path. Depending on the group, modules can be merged, or new modules can be introduced.

The modules discuss questions identified in Section \ref{sec:catalogue}. The questions are grouped around the following topics:
\begin{description}
    \item[(Module 0) Kick off] This initial session introduces the program and what the students can expect.
    \item[(Module 1) What is a PhD?] This module clarifies the objectives of a PhD, the requirements for PhD admission, and the skills that will be developed during PhD studies. Students get basic ideas about the main elements of PhD research and the requirements for finishing a PhD dissertation. Students also learn about the typical day-to-day life of a PhD student.  
    \item[(Module 2) Why a PhD for an Academic Career?]  This module clarifies the challenges of an academic career and how PhD studies prepare students for that career. Students learn about the process of selecting a research topic, research group, and supervisor for their PhD studies, and the impact of this choice on their future careers. Students should understand the evaluation processes of the quality of their work and the duration of their doctoral studies.
    \item[(Module 3) Why a PhD for a Career in Industry?] This module clarifies the benefits of obtaining a PhD for a career in industry and the skills acquired during PhD studies relevant to the career path in the industry. It presents different possibilities for doing a PhD in collaboration with industry partners, possible research topics, and requirements. It also addresses the possibility of creating a spin-out company.    
    \item[(Module 4) How to find a PhD position?] This module clarifies the formal and informal aspects of the application process, including finding a suitable PhD position, a group, and a supervisor. Students also learn about funding possibilities and how this choice influences their PhD studies. 
    \item[(Module 5) How to handle doubts or problems] \textbf{before and during a PhD?} This module clarifies typical concerns students have during their studies. This module emphasises the emotional issues that might arise and how to deal with them. Students learn where to find advice and support in case of self-doubts, burnout, and sexual harassment.  
    \item[(Module 6) Considering a PhD abroad?] This module clarifies different aspects of applying for PhD studies abroad. It addresses the application process, administrative issues, financial issues, and differences compared to a PhD in a native country.
    \item[(Module 7) Wrap-up] This last session is used to wrap up the program and gather feedback from the participants about the program, to support its continuous monitoring and evolution.
\end{description}
In addition to the described topics, each module contains personal experience sharing. Therefore, general information is also observed from a personal point of view.
Questions for personal experience sharing are identified based on each topic and its main general questions and should foster further discussion on these topics based on real situations.  
Personal questions enable getting to know the guest as a person and a role model. As an example, all the topics that should be addressed within Module 5, including the personal experience sharing, are given in \ref{appendix:module5}. 

\paragraph{Lessons Learned}
The feedback from the various implementations has been overly positive. A number of lessons have been learned from the different implementations of the WoCa Lunch that have taken place. Organisers of previous lunches found that the questions were beneficial in reducing the preparation effort and should not necessarily be considered prescriptive. 

It is crucial to establish with the students and the invited speakers that the communication approach should be informal and interactive. Invited speakers delivering slides do not have the same effect and impact as informal discussions. In certain situations, the WoCa lunch was not part of a departmental strategy and had little or no financial support. Therefore, obtaining funds for the lunch itself required additional effort.

Various experiences have shown that, typically, it is the women in the departments that organise and run these events. This reflects the experience shown with other gender inclusion initiatives where the women are those that are mostly involved~\cite{armstrong2024critical}.
Having male allies working side by side with women in these initiatives can have a positive effect, as it helps to frame these initiatives as beneficial for the whole community rather than for women alone~\cite{farrell2021you}.

%% file: 07-Conclusions.tex
\section{Conclusions and further work}
\label{sec:conclusions}
Relatively few women pursue PhD studies in computer science. This is problematic not only for equality reasons but also for the untapped potential that more women with PhDs in computer science could bring to industry and academia. The underlying goal of the research reported in this article is to increase the percentage of females in computer science, focusing on their transition to a PhD.

We conducted an extensive survey to characterise the encouraging and discouraging factors leading to the decision to start a PhD \newtext{and} \oldtext{as well as} to determine the extent to which these factors vary by gender. As expected, our results suggest this is a multi-factor decision, with several encouraging and discouraging factors, whose combination ultimately leads to a \oldtext{decision} \newtext{conclusion}.

Encouraging factors toward PhD study include interest and confidence in research, arising from research involvement during earlier studies; enthusiasm for computer science and self-confidence in addition to an interest in an academic career; encouragement from external sources, including academic staff, peers, family and the industry, although the latter is relatively uncommon; and a positive perception towards PhD studies which can involve achieving personal goals. The effect sizes of these factors are mostly medium and small.

Discouraging factors include uncertainty and lack of knowledge of the PhD process, a perception of lower job flexibility and the requirement for long-term commitment, with medium to small effect sizes.

Gender differences highlighted that female students who pursue a PhD have less confidence in their technical skills than males but a higher preference for interdisciplinary areas. Females are less inclined than males to perceive the industry as offering better job opportunities and more flexible career paths than academia, although with a small effect size.  

These insights helped develop a question catalogue to initiate discussions with potential female PhD students. The catalogue covers basic information and practical issues, career opportunities, and emotional support for PhD students.

We used the catalogue in the definition of a WoCa Lunch program as a supporting initiative to acquaint female undergraduate and master’s students with the possibilities for advancing their careers if they pursue PhD studies. We successfully conducted two pilot iterations of this program at the University of Aachen, with positive feedback from participants. We created localised versions of the WoCa Lunch program guidelines in 8 different languages: English, French, German, Portuguese, Serbian, Spanish, Turkish, and Ukrainian.

We plan to conduct the WoCa lunch program in different countries and monitor its short and long-term effects, using appropriate key process indicators, such as the evolution of the number of enrolled and graduated PhD students following their participation in the program. This will require a longitudinal study, spanning several years. We acknowledge the challenges of having the program in different languages, countries, and cultures. These may require local adaptations. Monitoring the impact of running this program in different countries will be crucial for its continuous improvement.

%% file: 08-DataAvailability.tex
\section*{Data availability}
A comprehensive replication package, including our anonymous dataset, instruments and analysis scripts, along with high-resolution files with the charts in this article, is available as a GitBook publication in {\url{https://papers-1.gitbook.io/why-do-women-pursue-a-phd-in-computer-science/}}. The dataset is also stored permanently in Zenodo~\cite{zenodoDataset}.

%% file: AppendixModule5.tex
\section{Module 5: How to handle doubts or problems before and during a PhD?}
\label{appendix:module5}

\subsection*{General questions}
\begin{itemize}
\item 
Am I capable of pursuing a PhD? Am I smart enough?

\item 
Is doing a PhD compatible with family planning?

\item 
What do I do if I am unsure or have problems during my doctorate studies?

\item 
What are typical questions and concerns students may have about their doctorate studies?

\item 
Will I have enough innovative ideas? What if I am not able to implement my research ideas? Can I change, or adjust, my research focus during the PhD?

\item 
What do I do if I "run out of breath" during the doctorate? Is it just me? (Everyone knows this feeling)

\item 
What do I do if I am dissatisfied with my doctoral supervision? (Time, support, appreciation)

\item 
Am I okay with being one of the few women in the group? What can I do if I receive sexist remarks? Where can I find help for sexual harassment?

\item 
Who can I contact if I have problems? Where can I find advice and support?

\item 
How are the dignity and respect rights of the PhD student applied in your University?
\end{itemize}

\subsection*{Personal experience sharing}

\begin{itemize}
\item 
How did you deal with any doubts you had? How did you deal with the feeling that you might not be smart enough for a PhD?
\item 
What obstacles/uncertainties did you face during your doctorate?
\item 
Who did you get support from?
\end{itemize}